\documentclass[twocolumn,tighten]{aastex63}
\pdfoutput=1
\usepackage{graphicx}
\usepackage{ulem}
\usepackage{amsmath}
\usepackage{wrapfig}
\usepackage{appendix}
\usepackage{enumitem}
\usepackage{multirow}
\usepackage{floatrow}
\usepackage{color}
\usepackage{longtable}

\begin{document}

\shorttitle{\texttt{CMC} cluster catalog}
\shortauthors{Kremer et al.}

\title{Modeling Dense Star Clusters in the Milky Way and Beyond with the \texttt{CMC} Cluster Catalog}

\author[0000-0002-4086-3180]{Kyle Kremer}
\affil{ Department of Physics \& Astronomy, Northwestern University, Evanston, IL 60208, USA}
\affil{Center for Interdisciplinary Exploration \& Research in Astrophysics (CIERA), Northwestern University, Evanston, IL 60208, USA}

\author[0000-0001-9582-881X]{Claire S. Ye}
\affil{ Department of Physics \& Astronomy, Northwestern University, Evanston, IL 60208, USA}
\affil{Center for Interdisciplinary Exploration \& Research in Astrophysics (CIERA), Northwestern University, Evanston, IL 60208, USA}

\author[0000-0002-1884-3992]{Nicholas Z. Rui}
\affiliation{Department of Astronomy, University of California at Berkeley, CA, USA 94720}

\author[0000-0002-9660-9085]{Newlin C. Weatherford}
\affil{ Department of Physics \& Astronomy, Northwestern University, Evanston, IL 60208, USA}
\affil{Center for Interdisciplinary Exploration \& Research in Astrophysics (CIERA), Northwestern University, Evanston, IL 60208, USA}

\author[0000-0002-3680-2684]{Sourav Chatterjee}
\affil{Tata Institute of Fundamental Research, Homi Bhabha Road, Mumbai 400005, India}

\author[0000-0002-7330-027X]{Giacomo Fragione}
\affil{ Department of Physics \& Astronomy, Northwestern University, Evanston, IL 60208, USA}
\affil{Center for Interdisciplinary Exploration \& Research in Astrophysics (CIERA), Northwestern University, Evanston, IL 60208, USA}

\author[0000-0003-4175-8881]{Carl L. Rodriguez}
\affil{Harvard Institute for Theory and Computation, 60 Garden St, Cambridge, MA 02138, USA}

\author[0000-0003-0930-6930]{Mario Spera}
\affil{ Department of Physics \& Astronomy, Northwestern University, Evanston, IL 60208, USA}
\affil{Center for Interdisciplinary Exploration \& Research in Astrophysics (CIERA), Northwestern University, Evanston, IL 60208, USA}
\affil{Dipartimento di Fisica e Astronomia `G. Galilei', University of Padova, Vicolo dell'Osservatorio 3, I--35122, Padova, Italy}
\affil{INFN, Sezione di Padova, Via Marzolo 8, I--35131, Padova, Italy}

\author[0000-0002-7132-418X]{Frederic A. Rasio}
\affil{ Department of Physics \& Astronomy, Northwestern University, Evanston, IL 60208, USA}
\affil{Center for Interdisciplinary Exploration \& Research in Astrophysics (CIERA), Northwestern University, Evanston, IL 60208, USA}

\begin{abstract}

We present a set of 148 independent $N$-body simulations of globular clusters (GCs) computed using the code \texttt{CMC} (\texttt{Cluster Monte Carlo}). At an age of $\sim10-13\,$Gyr, the resulting models cover nearly the full range of cluster properties exhibited by the Milky Way GCs, including total mass, core and half-light radii, metallicity, and galactocentric distance. We use our models to investigate the role that stellar-mass black holes play in the process of core collapse. Furthermore, we study how dynamical interactions affect the formation and evolution of several important types of sources in GCs, including low-mass X-ray binaries, millisecond pulsars, blue stragglers, cataclysmic variables, Type~Ia supernovae, calcium-rich transients, and merging compact binaries. While our focus here is on old, low-metallicity GCs, our \texttt{CMC} simulations follow the evolution of clusters over a Hubble time, and they include a wide range of metallicities (up to solar), so that our results can also be used to study younger and higher-metallicity star clusters. Finally, the output from these simulations is available for download at \url{https://cmc.ciera.northwestern.edu/home/}.

\vspace{1cm}

\end{abstract}

\section{Introduction}
\label{sec:intro}

Globular clusters (GCs) present rich opportunities for studying the importance of gravitational dynamics in dense stellar environments. In a GC, dynamical interactions play significant roles in both the evolution and survival of the system as a whole \citep[see, e.g.,][]{HeggieHut2003} and also in the formation of a number of exotic populations including X-ray \citep[e.g.,][]{Clark1975,Verbunt1984,Heinke2005,Ivanova2013,Giesler2018,Kremer2018a}, radio \citep[e.g.,][]{Lyne1987,Sigurdsson1995,Ransom2008,Ivanova2008,Fragione2018a,Ye2018}, and gravitational wave (GW) sources \citep[e.g.,][]{Moody2009,Banerjee2010,Bae2014,Ziosi2014,Rodriguez2015a,Rodriguez2016a,Askar2017, Banerjee2017,Hong2018,Fragione2018b,Samsing2018a,Rodriguez2018b,Zevin2018,Kremer2019b}.

About 150 GCs are known in the Milky Way (MW) \citep[e.g.,][]{Harris1996,Baumgardt2018}. 
All are sufficiently dense to have experienced significant relaxation, and many have extremely high core densities. Furthermore, all GCs appear fundamentally different from the stellar population in the Galactic field: they are generally old systems (ages  10 Gyr or more) with low metallicities (typically $Z\sim0.1Z_{\odot}$) and they also appear to have much lower binary fractions ($f_b \lesssim 5-30\%$) compared to the field \citep[$f_b \gtrsim 50\%$; e.g.,][]{Sana2012, Kroupa2018}. Over the past half century, a number of key differences amongst the MW GCs have been unveiled.
First, GCs have individual masses that span several orders of magnitude, from $\sim10^3 - 10^4M_{\odot}$ clusters that exhibit features similar to their younger open cluster counterparts, all the way to giant clusters with masses well in excess of $10^6\,M_{\odot}$ that may be linked to galactic nuclei \citep{Harris1996,Baumgardt2018}. Second, GCs in the MW exhibit a striking bimodal distribution in core radii separating the so-called ``core-collapsed'' and ``non-core-collapsed'' clusters \citep[e.g.,][]{Harris1996,McLaughlin2005}. Third, the numbers of various observed tracers of dynamical interactions -- such as X-ray binaries, radio pulsars, cataclysmic variables (CVs), and blue stragglers -- can vary dramatically from cluster to cluster \citep[e.g.,][]{Ransom2008,Heinke2010,Knigge2012, Ferraro2012}.

Much of the recent theoretical work on GC dynamics has focused on the crucial role played by stellar black hole (BH) remnants in these systems.
Several studies have shown that large numbers of stellar-mass BHs can be retained in typical GCs all the way to the present, and their dynamical interactions in the cluster core over its $\sim 12\,$Gyr of evolution provide a natural physical explanation for many of the diverse cluster features alluded to in the previous paragraph \citep[e.g.,][]{Morscher2015,Mackey2008,BreenHeggie2013,Askar2018,Kremer2019a}. The first BH candidate in a cluster was identified in the extragalactic GC NGC 4472 by \citet{Maccarone2007} through X-ray observations.
Soon thereafter, several BH candidates were identified in the MW GCs through X-ray and/or radio measurements, including M22 \citep{Strader2012}, M62 \citep{Chomiuk2013}, 47 Tuc \citep{Miller-Jones2015}, and M10 \citep{Shishkovsky2018}. Recently, the MUSE survey team have reported a BH candidate in NGC 3201, marking the first identification of a stellar-mass BH via purely dynamical measurements \citep{Giesers2018}.
Subsequent follow up has revealed two additional radial-velocity BH candidates within NGC 3201 \citep{Giesers2019}.

Computational and theoretical analyses have corroborated the recent observational evidence for BHs in GCs. In particular, it is now generally understood that a large number of BHs (100s--1000s) form through stellar evolution processes in clusters \citep[e.g.,][]{Kroupa2001,Morscher2015}. The subsequent evolution of a cluster's BH population is then governed by a number of dynamical processes: once formed, the BHs will quickly mass segregate to the center of their host cluster on a sub-Gyr timescale, assembling a BH subsystem that dominates the cluster's innermost region \citep[e.g.,][]{Spitzer1969,Kulkarni1993,Sigurdsson1993}. In this BH-dominated core, dynamically-hard BH binaries promptly form through three-body interactions \citep[e.g.,][]{Morscher2015}. As they sink to the cluster core, these binaries provide energy to passing stars in scattering interactions, a process which further hardens the binaries while energizing the rest of the cluster \citep[e.g.,][]{HeggieHut2003,BreenHeggie2013}. Furthermore, as BHs undergo these series of (binary-mediated) dynamical encounters within their host cluster's core, they frequently attain large dynamical kicks which temporarily eject them from the core. Once ejected, these BHs will rapidly mass-segregate back to the cluster's core, thereby depositing further energy into the cluster's stellar bulk. Cumulatively, these BH dynamics (binary-burning, ejection, and mass segregation) act as an energy source for their host cluster in a process we refer to as ``BH burning'' \citep[for review, see][]{Kremer2019d}. These effects are now well-understood and several in-depth studies have achieved consensus \citep{Merritt2004,Mackey2007,Mackey2008,BreenHeggie2013,Peuten2016,Wang2016,ArcaSedda2018,Kremer2018b,Zocchi2019,Kremer2019a,Antonini2019}. While a large BH population remains in a cluster, the cluster exhibits a large observed core radius due to BH burning. The observed core radius steadily shrinks as the BH population erodes, and only when the BHs are almost fully depleted, can a cluster attain a structure that would observationally be identified as ``core-collapsed.''

In \citet{Kremer2019a}, we demonstrated that the initial cluster size (set as the initial virial radius, $r_v$) is the key parameter which determines the ultimate fate of a cluster and its BH population. A cluster's half-mass relaxation time is related to its virial radius through the expression

\begin{equation}
\label{eq:relaxationtime}
t_{\rm{rh}} \sim \frac{M^{1/2}}{\langle m \rangle G^{1/2}\ln \Lambda} r_v^{3/2}
\end{equation}
\citep[Equation 2-63 of][]{Spitzer1987}, where $M$ is the total cluster mass, $\langle m \rangle$ is the mean stellar mass, and $\ln \Lambda$ is the Coulomb logarithm where $\Lambda \simeq 0.4 N$, where $N$ is the total number of particles. Thus, clusters with smaller initial $r_v$ have shorter relaxation times and are thus more dynamically evolved at their present age ($t\sim12\,$Gyr) compared to clusters born with larger initial $r_v$.

In \citet{Kremer2019a}, we employed a small set of cluster simulations with a number of initial parameters, such as total particle number and metallicity, fixed to reflect the median values of the clusters observed in the MW. In that analysis, we developed best-fit models for a set of four MW clusters (NGC 3201, M10, M22, and NGC 6752) by matching various observed features (including surface brightness and velocity-dispersion profiles). Here, we expand upon the results of \citet{Kremer2019a} and explore the effect of initial virial radii, and subsequent BH dynamics, on clusters of various masses, metallicity, and locations within the Galactic tidal field. We develop a grid of 148 independent cluster simulations, run using \texttt{CMC} (for \texttt{Cluster Monte Carlo}) which covers roughly the complete range of GCs observed at present in the MW. In Section \ref{sec:method}, we summarize the main computational methods incorporated within \texttt{CMC}, describe the choice of initial parameters for our grid of simulations, and define important quantities such that our models can be compared to observations. In Section \ref{sec:results}, we compare various features of our grid to the full population of MW clusters and demonstrate the ways that stellar-mass BHs determine cluster features. In Section \ref{sec:XRBs}, we discuss the number of BH and neutron star (NS) binaries that appear in our models at late times and discuss the implications for both radial-velocity searches for BHs and NSs in clusters as well as for X-ray binaries. We discuss the total number of pulsars in our models and compare to observations in Section \ref{sec:pulsars}. In Section \ref{sec:WDs}, we discuss white dwarf (WD) populations and applications to CVs and high-energy transient events. In Section \ref{sec:stellar_collisions}, we discuss luminous star collisions and possible applications for massive BH formation. We explore the number of blue stragglers found in our models in Section \ref{sec:BSs} and binary BH mergers plus applications to GW astronomy in Section \ref{sec:BBHs}. We discuss our results and conclude in Section \ref{sec:conclusions}.

\section{Methods}
\label{sec:method}

\subsection{Summary of \texttt{CMC}}
\label{sec:summary}

To model GCs, we use \texttt{CMC}, a H\'{e}non-type Monte Carlo code that computes the long-term evolution of GCs \citep{Henon1971a,Henon1971b,Joshi2000,Joshi2001,Fregeau2003,Chatterjee2010,Chatterjee2013,Pattabiraman2013,Rodriguez2015a}. \texttt{CMC} incorporates various physical processes relevant to both a cluster's structural evolution and the evolution of its constituent objects. A detailed description of \texttt{CMC} which will include detailed descriptions of all the latest updates as well as a comprehensive user guide will be presented in an upcoming paper \citep{Rodriguez2019c}. Here, we briefly review the methods relevant to several key physical processes.

     $\bullet\,$ \textit{Stellar and binary evolution:} we incorporate the single star and binary star evolution codes \texttt{SSE} and \texttt{BSE} \citep{Hurley2000, Hurley2002}, where we have implemented up-to-date prescriptions of compact object formation \citep{Fryer2001,Vink2001,Belczynski2002,Hobbs2005,Morscher2015}, as described below.
    
     $\bullet\,$ \textit{Neutron star formation:} we implement two scenarios for NS formation: standard iron core-collapse supernovae (CCSNe) and electron-capture supernovae (ECSNe). We adopt the ``rapid model" for stellar remnants formed through CCSNe \citep{Fryer2012}. As described in \citet{Ye2018} (and references therein), ECSNe may occur through several channels including evolution-induced collapse, accretion-induced collapse of an oxygen-neon white dwarf that accretes to the Chandrasekhar limit, or merger-induced collapse of a pair of WDs. We direct the reader to \citet{Ye2018} for more detailed discussion of each of these different formation scenarios and simply note here that we assume all NSs formed through CCSNe (ECSNe) receive natal kicks drawn from a Maxwellian with dispersion $\sigma=265\,\rm{km\,s}^{-1}$ ($20\,\rm{km\,s}^{-1}$). Additionally, we now incorporate updated prescriptions for the formation and evolution of pulsars (relating specifically to spin period and magnetic field evolution), as described in \citet{Ye2018}.
    
    $\bullet\,$ \textit{Black hole formation:} We assume BHs are formed with mass fallback \citep[again, using the ``rapid model'' of][]{Fryer2012} and calculate BH natal kicks by sampling from the same distribution as CCSN NSs but with BH kicks reduced in magnitude according to the fractional mass of fallback material: $V_{\rm{BH}} = V_{\rm{NS}}(1-f_{\rm{fb}})$, where $V_{\rm{NS}}$ is the kick velocity drawn from the \citet{Hobbs2005} distribution for CCSN NSs and $f_{\rm{fb}}$ is the fallback parameter, i.e., the fraction (from 0 to 1) of the stellar envelope that falls back upon core collapse \citep[see][for further details]{Fryer2012,Belczynski2002,Morscher2015}.
    
    We also implement prescriptions to treat pulsational-pair instabilities and pair-instability supernovae as described in \citet{Belczynski2016b}. To summarize, we assume that any star with a pre-explosion helium core mass between 45 and 65 $M_{\odot}$ will undergo pulsations that eject large amounts of mass, until the final product is at most 45 $M_{\odot}$. Assuming that $10\%$ of the final core mass is lost during the conversion from baryonic to gravitational matter at the time of collapse, a BH of mass $40.5\,M_{\odot}$ remains. We assume stars with helium core masses in excess of $65\,M_{\odot}$ are completely destroyed by a pair-instability supernova so that no remnant is formed.

\begin{figure}
\begin{center}
\includegraphics[width=0.95\linewidth]{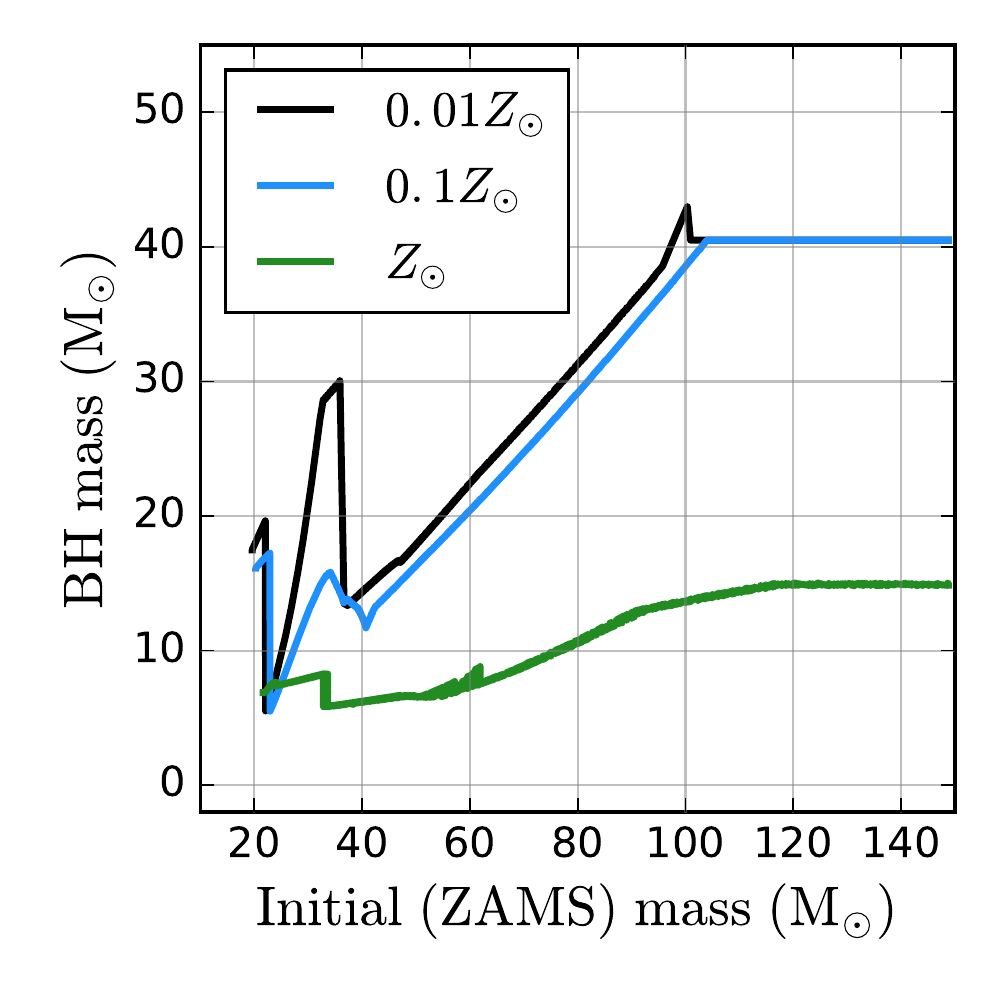}
\caption{\footnotesize \label{fig:BH_vs_ZAMS} BH mass versus initial (Zero-Age Main Sequence; ZAMS) mass for single stars computed using the stellar evolution prescriptions adopted in this study (see text for details). The three colors denote the three metallicities considered in this study.}
\end{center}
\end{figure}

    In Figure \ref{fig:BH_vs_ZAMS}, we show the initial-final mass relation for single stars (i.e., neglecting any binary or dynamical effects) and the dependence of this relation on metallicity. As the figure shows \citep[and as is discussed in detail in, e.g.,][]{Belczynski2016b}, at metallicities of $0.01$ (black curve) and $0.1Z_{\odot}$ (blue curve), the most massive stars ($M_{\rm{ZAMS}} \gtrsim 100\,M_{\odot}$) are subject to pulsational-pair instabilities, hence the visible flattening of the BH mass function at $40.5\,M_{\odot}$ for both populations. However, this is not the case at solar metallicity (green curve in Figure \ref{fig:BH_vs_ZAMS}). This is because stars at high metallicity are subject to prominent stellar-wind mass loss \citep[again, adopting the stellar wind prescriptions of][]{Vink2001} preventing the formation of helium cores above $45\,M_{\odot}$.

    $\bullet\,$ \textit{Direct integration of strong encounters:} during the evolution of a GC, binary stars will often pass sufficiently close to single stars and other binaries to undergo so-called ``strong'' encounters \citep[e.g.,][]{HeggieHut2003}. In \texttt{CMC} these binary--single and binary--binary (we neglect higher multiples) strong encounters are integrated using \texttt{Fewbody} \citep{Fregeau2004,Fregeau2007}. See \citet{Fregeau2007} for a detailed description of the use of \texttt{Fewbody} within the overall framework of \texttt{CMC}. Also, note that \texttt{Fewbody} has now been updated to include gravitational radiation reaction for all encounters involving BHs \citep[see][for more information]{Rodriguez2018a,Rodriguez2018b}.
    
    $\bullet\,$ \textit{Two-body relaxation:} two-body relaxation is the primary physical process at play in the global evolution of a GC \citep[e.g.,][]{HeggieHut2003}. We use the H\'{e}non orbit-averaged Monte Carlo method to simulate two-body relaxation \citep{Henon1971a,Henon1971b}. We direct the reader to \citet{Joshi2000} for a detailed description of how these techniques are implemented in \texttt{CMC}.
    
    $\bullet\,$ \textit{Single--single GW capture:} as described in \citet{Samsing2019_singlesingle}, binary formation can occur through GW capture of pairs of single BHs in GCs. This mechanism plays an important role in the emerging picture of how binary BHs form and merge in GCs. Here, we allow BH binaries to form through GW capture of pairs of single BHs in our simulations.
    
    $\bullet\,$ \textit{Three-body-binary formation:} as the core of a cluster collapses through gravitational instability, the innermost stars will eventually reach high enough densities to form binaries through three-body-binary (3BB) formation \citep[e.g.,][]{HeggieHut2003}. We adopt the formalism for 3BB formation described in \citet{Morscher2015}, with two small modifications. First, we allow binaries to form with $\eta \geq 2 = \eta_{\rm{min}}$, where $\eta$ is the binary hardness ratio (binary binding energy to background star kinetic energy): $\eta = (Gm_1m_2)/(r_p \langle m \rangle \sigma^2).$ Here, $m_1$ and $m_2$ are the binary component masses, $r_p$ is the separation of the objects at pericenter, and $\langle m \rangle$ and $\sigma$ are the local average mass and velocity dispersion. Note that \citet{Morscher2015} adopted $\eta \geq 5 = \eta_{\rm{min}}$, a conservative assumption that only captures a small subset of all 3BB formation events. For example, see \citet{Aarseth1976}, which showed that the probability a given three-body encounter leads to a binary scales as $\eta^{-2}$, indicating that roughly 80\% of three-body encounters result in binary formation, even for $\eta=2$, our assumed $\eta_{\rm{min}}$. Secondly, we allow 3BB formation to occur for \textit{all} stars, not only BHs, as was the case in \citet{Morscher2015}. The inclusion of these two effects leads to an overall increase in 3BB formation relative to the models of \citet{Morscher2015}.
    
    $\bullet\,$ \textit{Tidal truncation:} real GCs are not isolated systems -- they are subject to the tidal field of their host galaxy. The assumption of spherical symmetry inherent in Monte Carlo codes like \texttt{CMC} does not allow for a direct calculation of stellar loss at the tear-drop-shaped tidal boundary. Instead, we employ an effective tidal mass-loss criterion that attempts to match the tidal mass loss found in direct $N$-body simulations. In short, for each simulation, we assume a circular orbit around the Galactic center with radius, $R_{\rm{gc}}$. To calculate the tidal radius of the cluster, a logarithmic potential ($\phi [R_{\rm{gc}}] = V_{\rm{gc}}^2 \ln R_{\rm{gc}}$) for the galaxy is assumed with circular velocity $V_{\rm{gc}}=220\,\rm{km\,s}^{-1}$ (assuming a flat rotation curve such that $V_{\rm{gc}}$ is independent of $R_{\rm{gc}}$). The tidal radius of the cluster is then given by:
    
    \begin{equation}
        r_t = \Big( \frac{G M_c}{2V_{\rm{gc}}^2} \Big)^{1/3} R_{\rm{gc}}^{2/3}
    \end{equation}
    where $M_c$ is the total cluster mass \citep[see, e.g.,][]{BaumgardtMakino2003}. For a detailed description of the implementation of Galactic tides, see \citet{Chatterjee2010,Pattabiraman2013}.
    
    $\bullet\,$ \textit{Stellar collisions:} Stars in realistic clusters will frequently undergo sufficiently close passages to tidally interact. Depending on the pericenter distance ($r_p$), stars may undergo tidal captures, tidal disruptions, or physical collisions, all of which are expected to lead to distinct outcomes with distinct electromagnetic signatures. In \texttt{CMC}, we handle close encounters in the ``direct collision'' approximation, meaning pairs of stars that pass close to one another are assumed to physically collide only if $r_p \leq R_1 + R_2$, where $R_1$ and $R_2$ are the radii of the two stars of interest. Such collisions can occur through single--single encounters, as well as during binary-mediated strong encounters that are integrated by \texttt{Fewbody}. See \citet{Fregeau2007} for a detailed description of how collisions are computed within \texttt{CMC}.
    
Over the past several decades it has become clear that (nearly) all GCs host significant chemical abundance spreads, in stark contrast to the old notion that stars in GCs have the same age and chemical abundances. These star-to-star abundance variations within clusters are known as multiple populations (MPs), and the study of MPs has emerged as a pillar of research in the field of star clusters \citep[for a recent review, see][]{Bastian2018}. In particular, studies of MPs can place important constraints upon various cluster processes, especially those occurring during the earliest evolutionary phases of clusters \citep[e.g.,][]{Ventura2001,Decressin2007,deMink2009,Denissenkov2014,Gieles2018}.
Although examination of chemical abundance spreads and the implications to multiple populations are rich subjects, we do not incorporate chemical abundance variations in \texttt{CMC}. From the perspective of gravitational $N$-body dynamics, the chemical abundances of individual stars is a second-order effect; these chemical anomalies have a minimal effect on the long-term dynamical evolution of the cluster (i.e. timescales of $\sim10\,$Gyr). However, the various physical processes that may produce these abundance variations may be intimately connected to dynamics in the cluster at \textit{early} times ($t \lesssim 1\,$Gyr) \citep[e.g.,][]{Sills2010}. In \texttt{CMC}, we simply assume all stars are born with fixed metallicity at a fixed time and bypass the early phases of cluster and stellar formation from collapse of a molecular cloud \citep[for recent work which considers these processes, see e.g.,][]{Fujii2016}. 

\subsection{Selection of initial model parameters}
\label{sec:model_params}

In this paper, we present a new set of 148 independent cluster simulations run using \texttt{CMC}. We vary four initial cluster parameters in this study: the total number of particles (single stars plus binaries; $N=2\times10^5$, $4\times10^5$, $8\times10^5$, and $1.6\times10^6$), the initial cluster virial radius ($r_v/\rm{pc}=0.5,\,1,\,2,\,4$), the metallicity ($Z/Z_\odot=0.01,\,0.1,\,1$), and the galactocentric distance ($R_{\rm{gc}}/\rm{kpc}=2,\,8,\,20$).
This gives us a $4\times4\times3\times3$ grid for a total of 144 models. We also run four additional models with $N=3.2\times10^6$ particles to characterize the most massive clusters in the Milky Way. For these four models, we fix the galactocentric distance to $R_{\rm{gc}}=20\,$kpc (for simplicity) and vary metallicity ($Z=0.01Z_{\odot}$ and $Z_{\odot}$) as well as virial radius ($r_v=1$ and 2 pc).
As a whole, this complete model set shares several similarities to previous large \texttt{CMC} model sets \citep[e.g.,][]{Chatterjee2010,Chatterjee2013a,Morscher2015,Rodriguez2018a}, with the key differences being that here, we expand the range in $r_v$ and $Z$, and also decouple $Z$ and $R_{gc}$ which, e.g., \citet{Morscher2015} coupled via an assumed $Z$--$R_{\rm{gc}}$ correlation based on MW observations. These differences from our previous model sets allow a more expansive comparison to all types of clusters observed in the MW. 

A number of initial properties are fixed across all simulations. We assume all models are initially described by King profiles \citep{King1962} and adopt a fixed King concentration parameter of $W_0 = 5$. We adopt the initial mass function (IMF) of \citet{Kroupa2001} with masses in range $0.08-150\,M_{\odot}$ and assume an initial stellar binary fraction of $f_b=5\%$. To assign binaries, an appropriate number of single stars (based on $N$ and $f_b$) are randomly drawn from the IMF and assigned binary companions, with secondary masses drawn from a flat distribution in mass ratio, $q$, in the range $q \in [0.1,1]$ \citep[e.g.,][]{DuquennoyMayor1991}. 

Binary orbital periods are drawn from a distribution flat in log-scale \citep[e.g.,][]{DuquennoyMayor1991}, with the orbital separations ranging from near contact ($a \geq 5(R_1+R_2)$, where $R_1$ and $R_2$ are the stellar radii) to the hard/soft boundary, while  binary eccentricities are drawn from a thermal distribution \citep[e.g.,][]{Heggie1975}.
We evolve each simulation to a final time of 14 Gyr, unless the cluster disrupts or undergoes a collisional runaway, as discussed further in Section \ref{sec:definitions}.

In Table \ref{table:models} in the Appendix, we list all initial cluster properties and various features at the end of the simulations. The output for this set of simulations will be available for download at \url{https://cmc.ciera.northwestern.edu/}. These simulations will soon be accompanied by a release of the \texttt{CMC} source code with supplementing documentation \citep{Rodriguez2019c}.

\subsection{Key definitions}
\label{sec:definitions}

In this section, we briefly define several terms used throughout the paper that are relevant to our models. In particular, we discuss differences between ``observational'' and ``theoretical'' definitions of various terms (such as core radius) and explain our treatment of cluster dissolution and collisional runaways.

\textit{Core collapse:} From an observational perspective, the term core-collapse is traditionally used to indicate a particular cluster structure that has a central power-law surface brightness profile, as opposed to non-core-collapsed clusters with profiles that can be well-fit by a King model. Thus, observational core-collapse refers to a property of a cluster's luminous stars. This is in contrast to the definition occasionally used by theorists which refers to the collapse of a BH-dominated subsystem (we abbreviate as BHS, for ``BH subsystem''). This BHS collapse has no direct effect on the light profile of the cluster \citep{Chatterjee2017a}. However, the formation and eventual dissipation of a BHS does have an important \textit{indirect} effect on the cluster's structure (and light profile) through the ``BH burning'' process, where strong dynamical encounters within the BHS act as a energy source for the rest of the cluster \citep[see][for review]{Kremer2019d}. In this study, we use the term ``core-collapsed'' in the observational sense. We define a core-collapsed cluster as one with a luminosity profile exhibiting a prominent central cusp.

\textit{Core and half-light radii:} Using \texttt{SSE}, \texttt{CMC} calculates the bolometric luminosity and temperature of all stars as a function of time, which allows us to construct Hertzsprung-Russell diagrams (see Section \ref{sec:HRdiagrams}), as well as calculate core and half-light radii consistent with observers' definitions. We estimate the observational half-light radius, $r_{\rm{hl}}$, of each model by finding the 2D-projected radius which contains half of the cluster's total light. We use the method described in \citet{Morscher2015} and \citet{Chatterjee2017a} to estimate the observational core radius, $r_c$. Note that the observed core radius is different from the theoretical (mass-density weighted) core radius (which we denote as $r_{\rm{c,\,theoretical}}$) 
traditionally used by theorists \citep{Casertano1985}.

\textit{Disrupted Clusters:} As clusters evolve, they lose mass through a variety of processes including high-mass stellar evolution, ejection of stars through dynamical encounters and natal kicks, and mass loss through the cluster's tidal boundary. In fact, given sufficient time, all tidally bound clusters will eventually disrupt completely through mass loss as a natural consequence of relaxation \citep[see, e.g.,][]{HeggieHut2003}. The time to complete disruption depends upon the cluster's relaxation timescale as well as its initial ``overfilling'' factor, which depends on the cluster's position within the MW potential. A handful of models considered in this study undergo complete disruption before reaching the $14\,$Gyr maximum integration time.

As a cluster begins to tidally disrupt, several of the basic assumptions at the heart of our Monte Carlo approach break down, in particular spherical symmetry and the assumption that the relaxation timescale is significantly longer than the dynamical timescale (disrupting clusters can lose mass on a timescale much shorter than the relaxation time). Therefore, we assume the cluster has completely disrupted once $t_{\rm{relax}} > M/\dot{M}$, where $t_{\rm{relax}}$ and $M$ denote relaxation time and total cluster mass, respectively. In practice, our model clusters typically contain 10000 stars or less when they meet this criterion. All clusters that disrupt before 14 Gyr are labeled as such in Table \ref{table:models}.

\textit{Collisional runaway:} As pointed out in a number of recent analyses, clusters with sufficiently high initial densities may lead to large numbers of stellar collisions within the first few Myr, potentially leading to the formation of a very massive star. Such objects may have important implications for the formation of intermediate-mass BHs \citep[IMBHs; e.g.,][]{Ebisuzaki2001,PortegiesZwartMcMillan2002,Gurkan2004,Freitag2006,PortegiesZwart2010,Goswami2012}. Here, we assume a cluster has undergone collisional runaway when a star with mass in excess of $500\,M_{\odot}$ is formed, with this specific limit chosen simply in accordance with earlier work \citep[e.g.,][]{Gurkan2004}.
Treatment of the runaway process and, in particular, treatment of the various physical processes relevant when an IMBH is present is outside the computational scope of the present version of \texttt{CMC}  \citep[however, for a recent attempt at incorporating within \texttt{CMC} the various processes relevant to the presence of an IMBH, see][]{Umbreit2012}. Therefore, in the event of a runaway, we stop the integration of the model. In total, only three models meet the collisional-runaway requirement: \textsc{n16-rv0.5-rg2-z0.01}, \textsc{n16-rv0.5-rg8-z0.01}, and \textsc{n16-rv0.5-rg20-z0.01}; marked with asterisks in Table \ref{table:models}. We hope to explore this topic further in future projects.

\section{Results}
\label{sec:results}



\begin{figure*}
\begin{center}
 \gridline{\fig{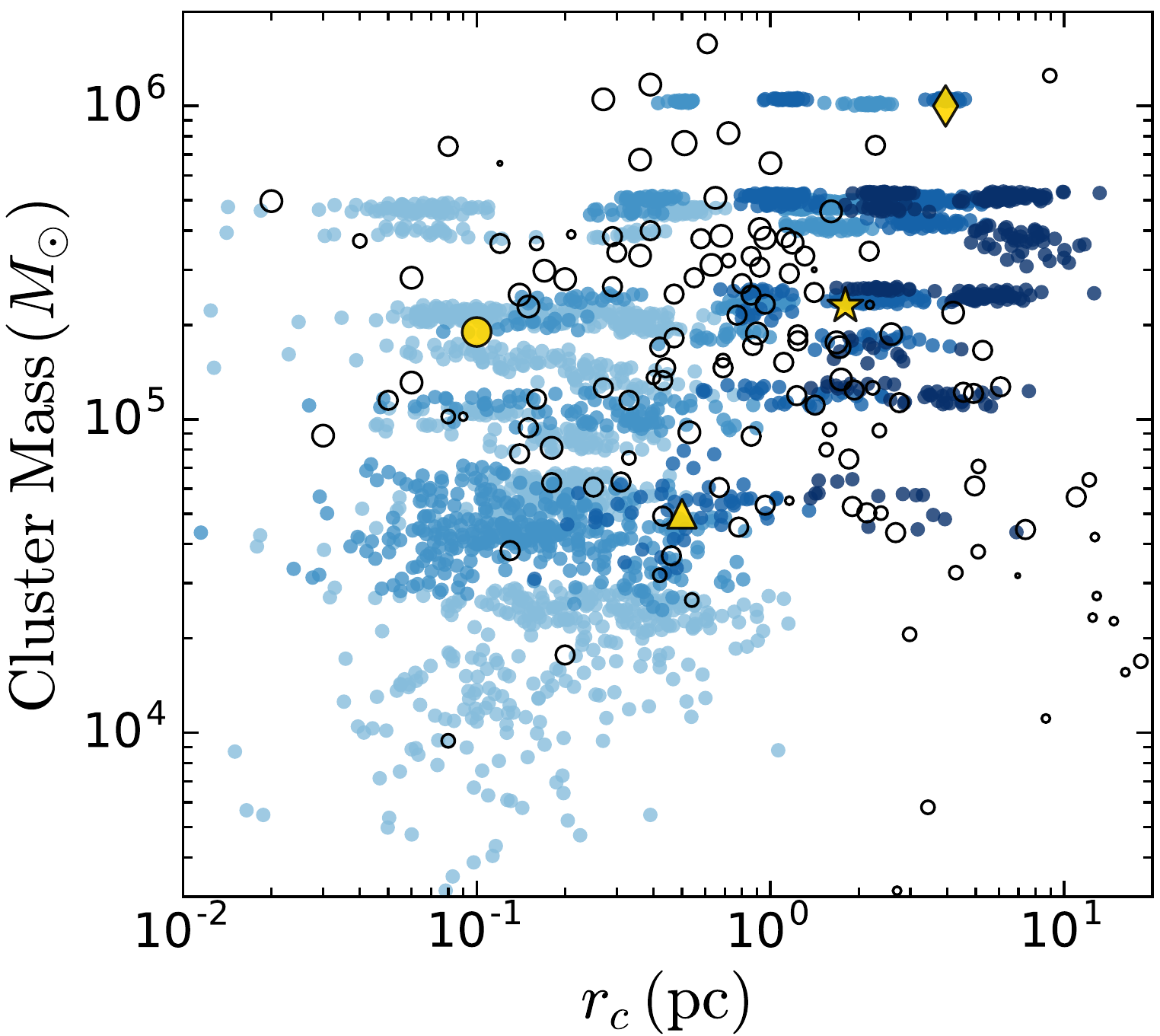}{0.36\textwidth}{(a) Total mass versus core radius}
          \fig{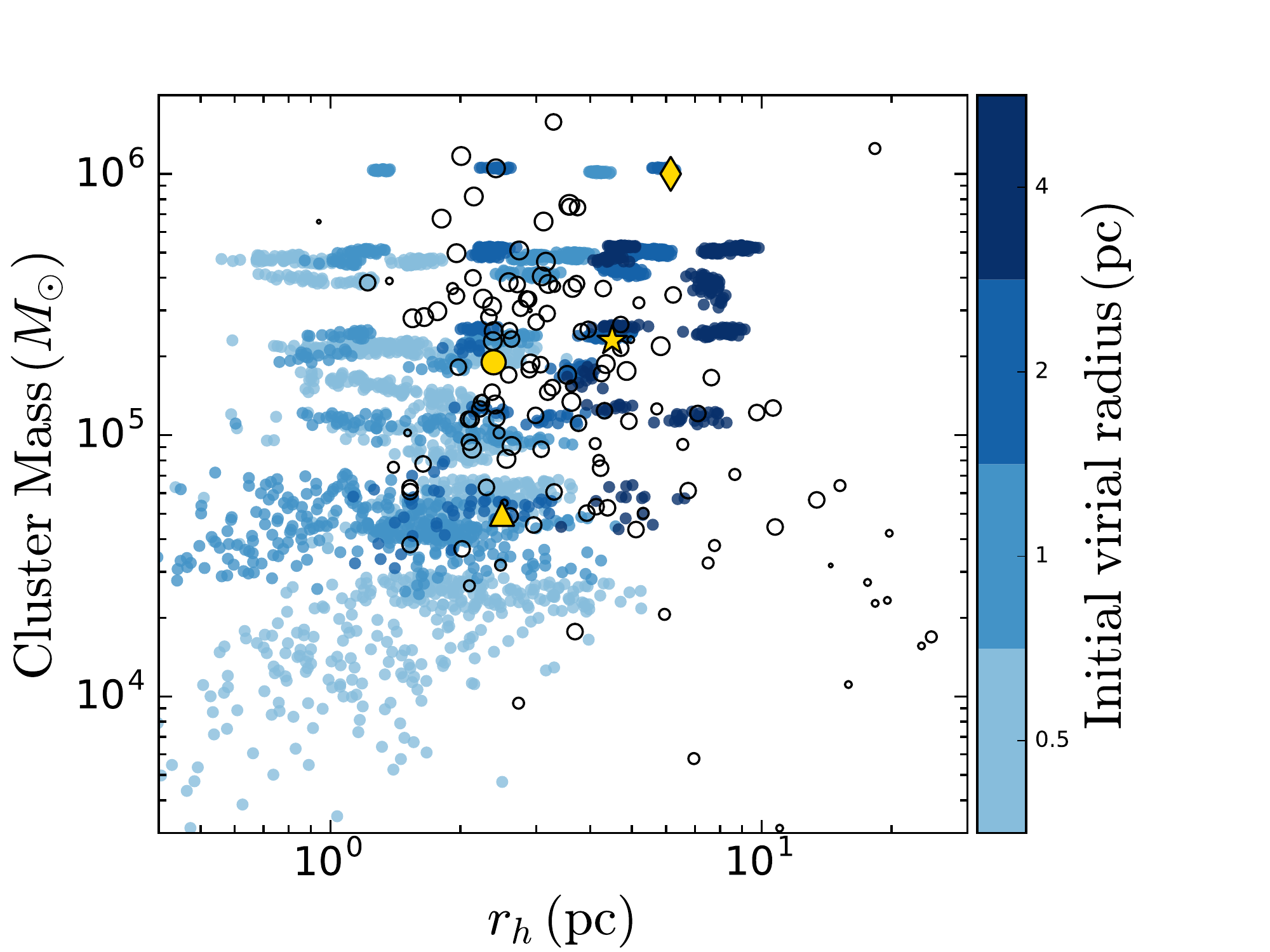}{0.427\textwidth}{(b) Total mass versus half-light radius}}
\gridline{\fig{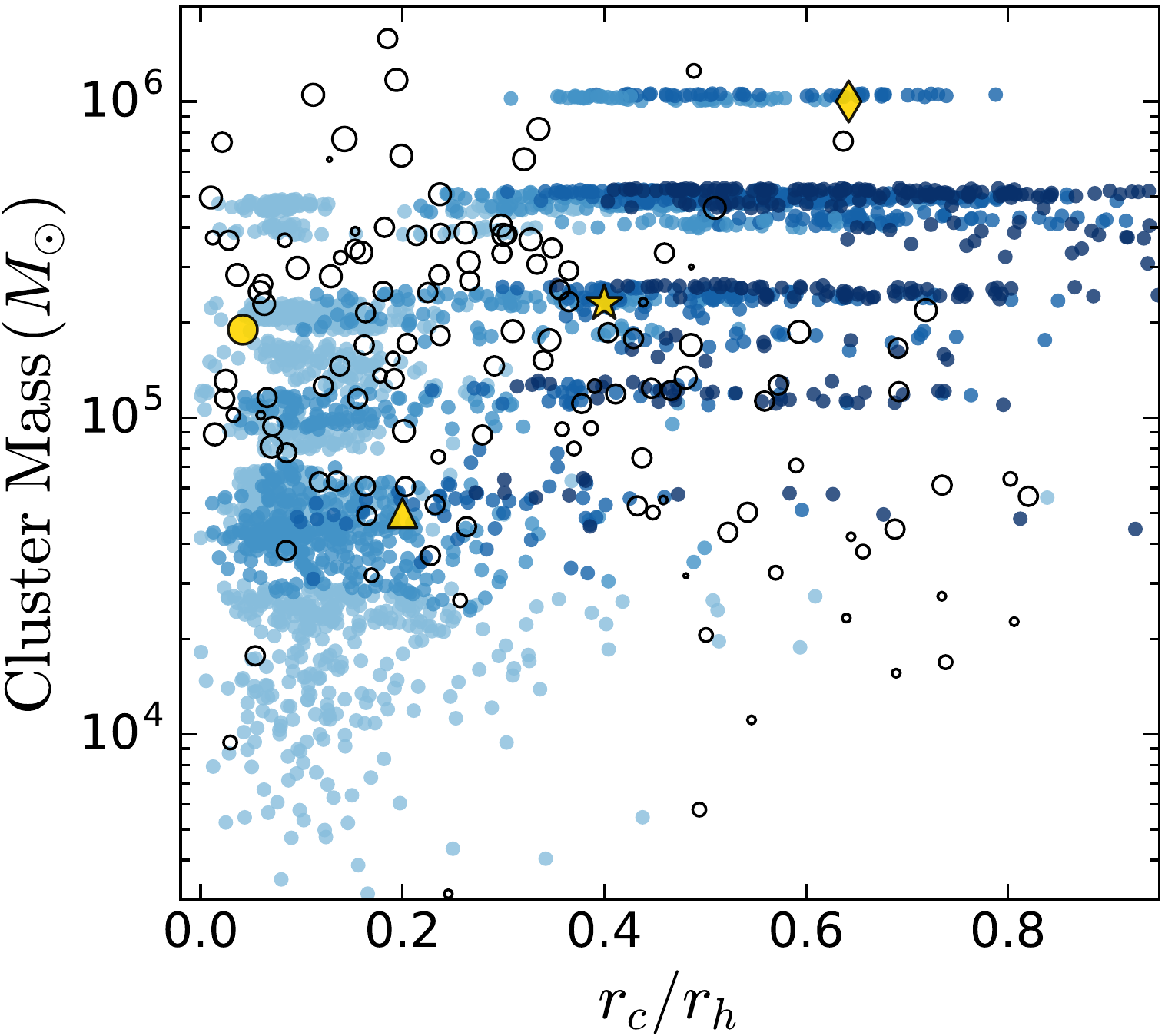}{0.36\textwidth}{(c) Total mass versus core radius/half-light radius}
          \fig{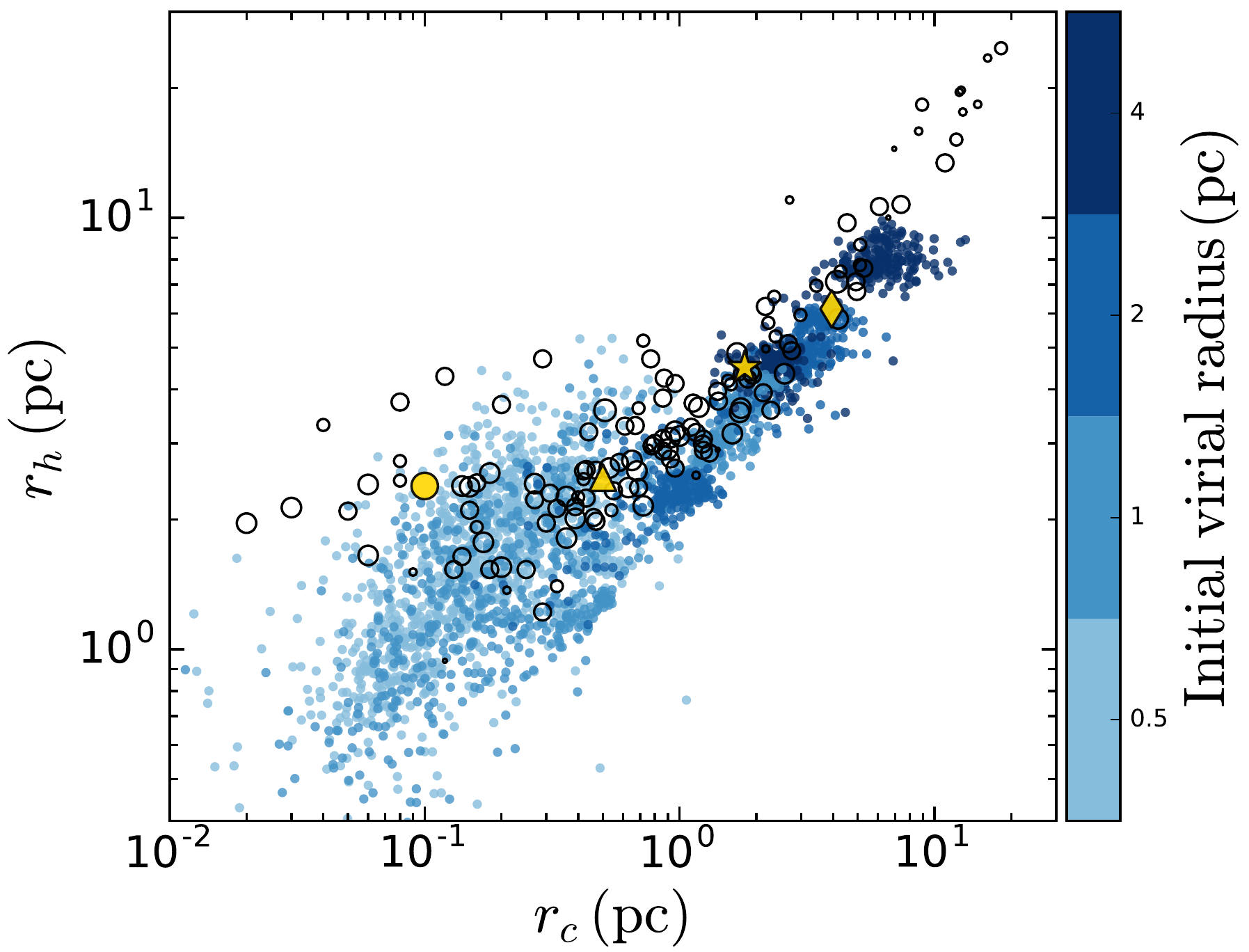}{0.427\textwidth}{(d) Half-light radius versus core radius}}
\gridline{\fig{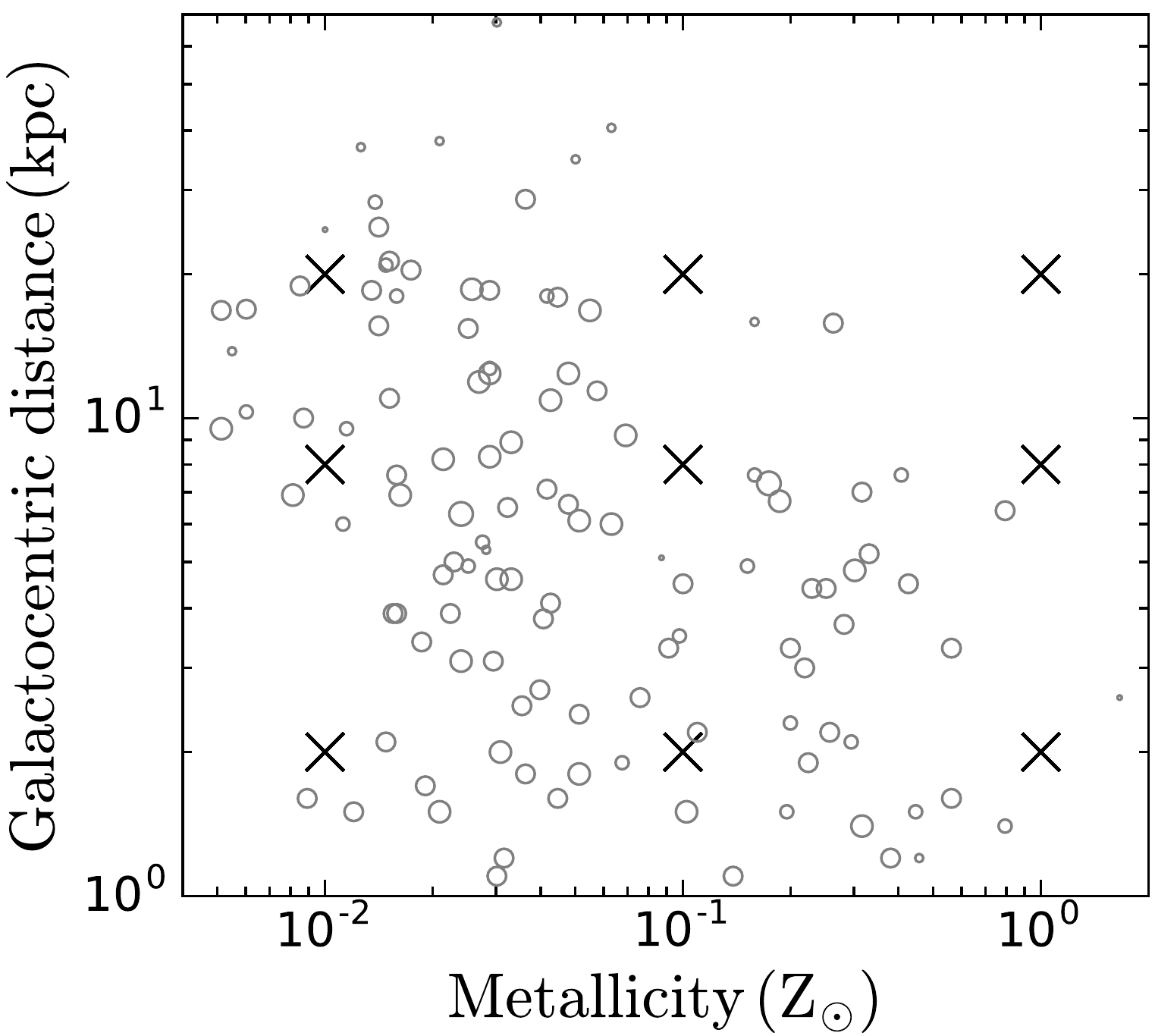}{0.36\textwidth}{(e) Galactocentric distance versus metallicity. Here X's mark initial values for model clusters}
          \fig{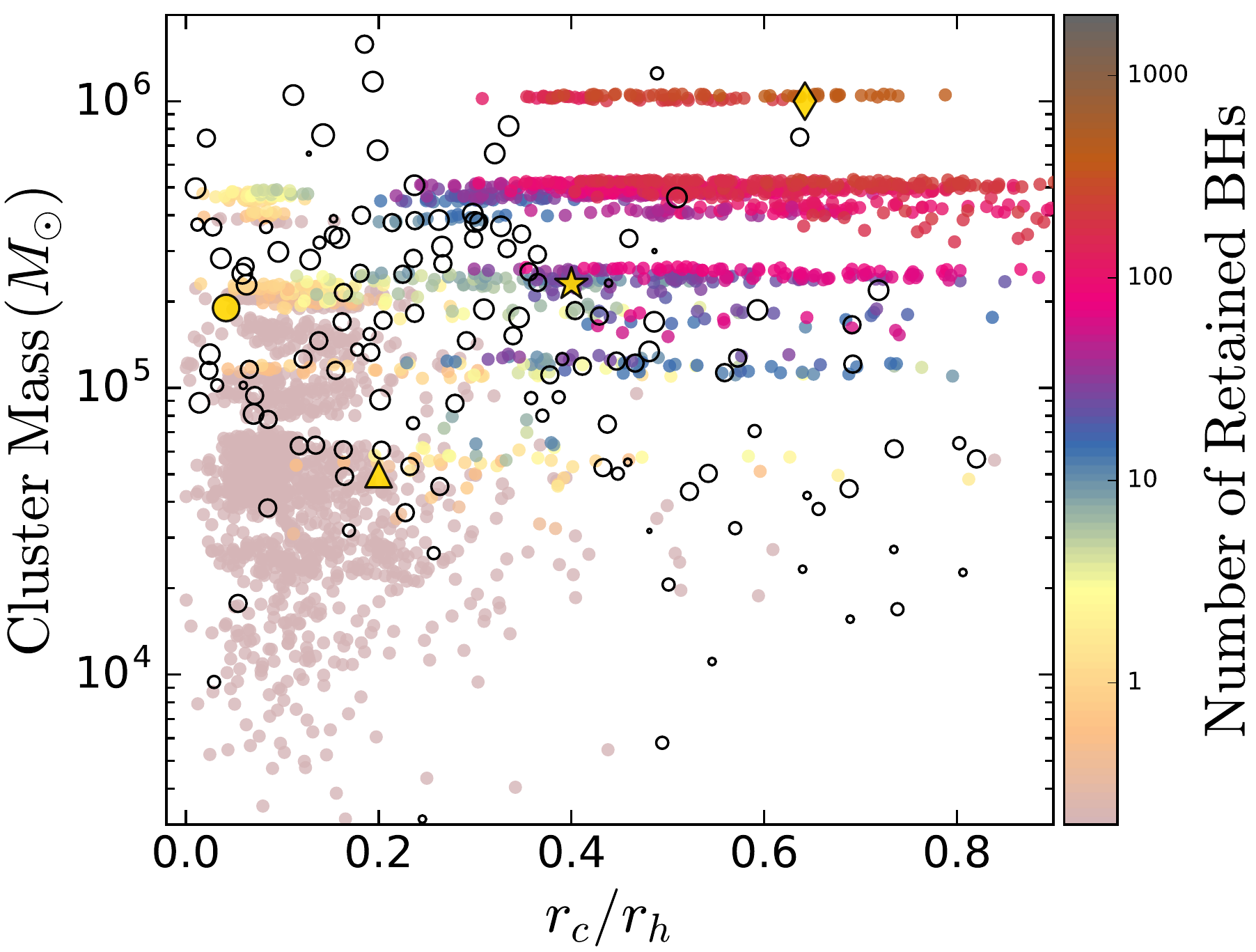}{0.427\textwidth}{(f) Same as panel (c) but with the color showing total number of BHs retained}}
\caption{\footnotesize All late time snapshots ($t=10-13\,$Gyr) for model clusters (blue points) compared to observational data for MW clusters (black points), taken from \citet{Baumgardt2018}. The size of each black point corresponds to the integrated V-band magnitude of each cluster \citep{Harris1996} such that the larger symbols denote clusters that are best observed. In panels (a)--(d), the various shades of blue show, from light to dark, models with increasing initial virial radii, from $r_v=0.5\,$pc to $4\,$pc. In panel (f), the color scheme denotes the total number of retained BHs. The panels compare various observed features, including total cluster mass, core radius ($r_c$), half-light radius ($r_h$), metallicity, and Galactocentric distance (as labeled in each of the figure subcaptions). The four gold-colored symbols in the various panels correspond to the locations of the four ``characteristic'' clusters described in the text.}
 \label{fig:Harris}
\end{center}
\end{figure*}

In Table \ref{table:models} in the Appendix, we list initial conditions and various cluster parameters at the end of each simulation for all models in this study. Models marked with a ``--'' denote clusters that disrupted before reaching the end of the simulation. In this section, we discuss a number of broad-brush features of our population of models. In Sections \ref{sec:XRBs}--\ref{sec:BBHs}, we go on to explore the formation rates of specific objects in our models.

\subsection{Comparison with the Milky Way Cluster Population}
\label{sec:comparison_to_MW}

In Figure \ref{fig:Harris}, we compare various features of our models to the MW GC data taken from \citet{Baumgardt2018}.\footnote{We also compared our models to the observed clusters in \citet{Harris1996} and found similarly good agreement. We show comparisons to \citet{Baumgardt2018} simply because that analysis quotes total cluster masses directly (as opposed to V-band magnitudes) which circumvents the need to assume a mass-to-light ratio in order to obtain cluster masses.} The size of the circle coinciding with a given MW cluster is scaled by the integrated V-band magnitude of that cluster \citep[taken from][]{Harris1996}. Thus, larger circles correspond to clusters that are more luminous in the V-band. In the top four panels (a--d), we show comparisons involving the total cluster mass as well as the ``observed'' core and half-light radii ($r_c$ and $r_h$, respectively; see definitions in Section \ref{sec:definitions}).
The color scale shows the initial virial radius of the GCs, which goes from $0.5\,$ (lightest blue) to $4\,$pc (darkest blue). To reflect the uncertainty in the ages of MW GCs, we simply show here all model snapshots with evolutionary times in the range 10--13 Gyr. Each of these separate model snapshots can be viewed as a distinct (although not necessarily statistically-independent) realization of a particular cluster.

In panel (e), we compare our models to the Galactocentric distances and metallicities of the MW clusters, with ``X''-markers indicating the discrete values of these parameters chosen for our models. Finally, in panel (f), we assign colors based on the total number of BHs retained in each model at the time of the particular snapshot shown.

The four gold-colored symbols in the various panels of Figure \ref{fig:Harris} mark the locations of four simulations that roughly characterize common cluster types observed in the MW: \textsc{n8-rv2-rg8-z0.01} (gold star), representing typical non-core-collapsed clusters with average mass ($2\times10^5\,M_{\odot}$) that have many BHs at present (e.g., NGC 3201 or M22); \textsc{n8-rv0.5-rg8-z0.01} (gold circle), representing typical core-collapsed clusters with average mass (e.g., NGC 6752); \textsc{n2-rv2-rg8-z0.01} (gold triangle), a low-mass cluster with $M=5\times10^4\,M_{\odot}$ (e.g., NGC 6144 or Terzan 3); and \textsc{n32-rv2-rg20-z0.01} (gold diamond), a massive cluster ($>10^6\,M_{\odot}$) with well over 1000 BHs at present (e.g., NGC 2808). In Sections \ref{sec:XRBs}--\ref{sec:BSs}, we discuss the numbers of various stellar sources within each of these four characteristic clusters.

As shown in Figure \ref{fig:Harris}, a clear relation exists between clusters' present-day core radii and their initial virial radii: cluster models with small initial $r_v$ (light blue scatterpoints) tend to occupy the leftmost regions of the plots shown in panels (a) and (c), while the opposite is true for models with large initial $r_v$ (dark blue scatterpoints). This is consistent with our understanding of cluster evolution. Through the natural diffusion of energy from core to halo, clusters naturally evolve toward more compact configurations \citep[$r_c$ decreases over time; see e.g.,][for an overview]{HeggieHut2003}. Clusters with smaller $r_v$ have shorter relaxation times and thus approach more compact configurations relatively quickly. Thus, by varying $r_v$ (here, from 0.5--4 pc) the full distribution of cluster core radii is naturally captured. This same result was demonstrated in \citet{Kremer2019a} for a much smaller set of models of fixed particle number and metallicity. Here, we demonstrate this result for a much broader set of cluster properties.

Furthermore, the relation between initial $r_v$ and the evolution of a cluster's core crucially depends on the cluster's BH population.
When a large number of BHs are present, the internal dynamics of the BH subsystem introduces a substantial energy source that supports the core of the cluster against its natural tendency to collapse. As the BH population is depleted through dynamical ejection of BHs in binary-mediated encounters within the core, the energy generated via this ``BH burning'' process gradually becomes less dynamically important. Ultimately, the core is no longer adequately supported and the cluster undergoes core-collapse at which point the core is supported by ``burning'' of stellar binaries \citep{Chatterjee2013a}. The depletion rate of the BHs is determined by the initial $r_v$: smaller $r_v$ corresponds to higher BH-interaction rates and therefore more rapid BH depletion \citep{Kremer2019a}. Therefore, we expect the cluster models with larger initial $r_v$ to retain more BHs at present and, as a consequence, have larger cores. This exact result is shown in panel (f) of Figure \ref{fig:Harris}. We detail the evolution of BH populations across all models in Section \ref{sec:BHpopulations}.

Figure \ref{fig:Harris} shows that our model clusters span effectively the full parameter space of the observed MW GCs. That being said, two regions of the MW cluster parameter space are visibly less well-covered than others: the most massive and most compact clusters (i.e., top-left corner of panel (c)) and the least massive and least compact clusters (i.e., bottom-right corner of panel (c)). The poor coverage in these regions can be attributed to the intentionally chosen limits in our simulation grid. More massive and compact cluster simulations are more expensive computationally. As shown in Figure \ref{fig:Harris}, the most massive clusters observed in the MW are reproduced best by simulations with initial $N=3.2\times10^6$. It is beyond the present scope of \texttt{CMC} to run such massive simulations with sufficiently small initial $r_v$ ($0.5\,$pc) to reproduce the most massive and most compact GCs observed in the MW. Not only are such simulations extremely computationally expensive, but, as discussed in Section \ref{sec:definitions}, these high density models also feature frequent collisional runaway episodes, producing massive stars (and ultimately BHs) that \texttt{CMC} in its present form is ill-equipped to handle accurately.

On the other hand, the least massive and least compact clusters (bottom-right corners of panels (a), (b), (c), and (e) and top-right corner of panel (d)) are best captured by simulations with low-$N$ and initial sizes outside the current range of our grid (i.e., $r_v=8\,$pc or larger). Although such models are relatively inexpensive computationally, we neglect to include them here simply because these diffuse low-mass clusters constitute a low fraction of the total mass of the MW GC system and also because these clusters straddle the boundary between GCs and open clusters, the latter of which are well-known to be better suited for direct $N$-body modelling \citep[e.g.,][for a review]{Aarseth2003}. With these considerations in mind, and in order to avoid fine-tuning of our simulation grid to match specific regions of the parameter space, we simply acknowledge relatively sparse coverage in the aforementioned regions and reserve more detailed studies of these regions for future work.

The total number of models (148), total final mass of the complete set (roughly $3\times 10^7\,M_{\odot}$), as well as the fraction of clusters that are core-collapsed at the end of the simulation (roughly $20\%$), are all roughly consistent with the respective values of the full population of MW GCs. These agreements between models and observations motivate the use of this model set to explore various features of the MW cluster population, as is done in Sections \ref{sec:XRBs}--\ref{sec:BSs}.

\subsection{Black hole populations}
\label{sec:BHpopulations}

\begin{figure*}
\begin{center}
\includegraphics[width=0.95\linewidth]{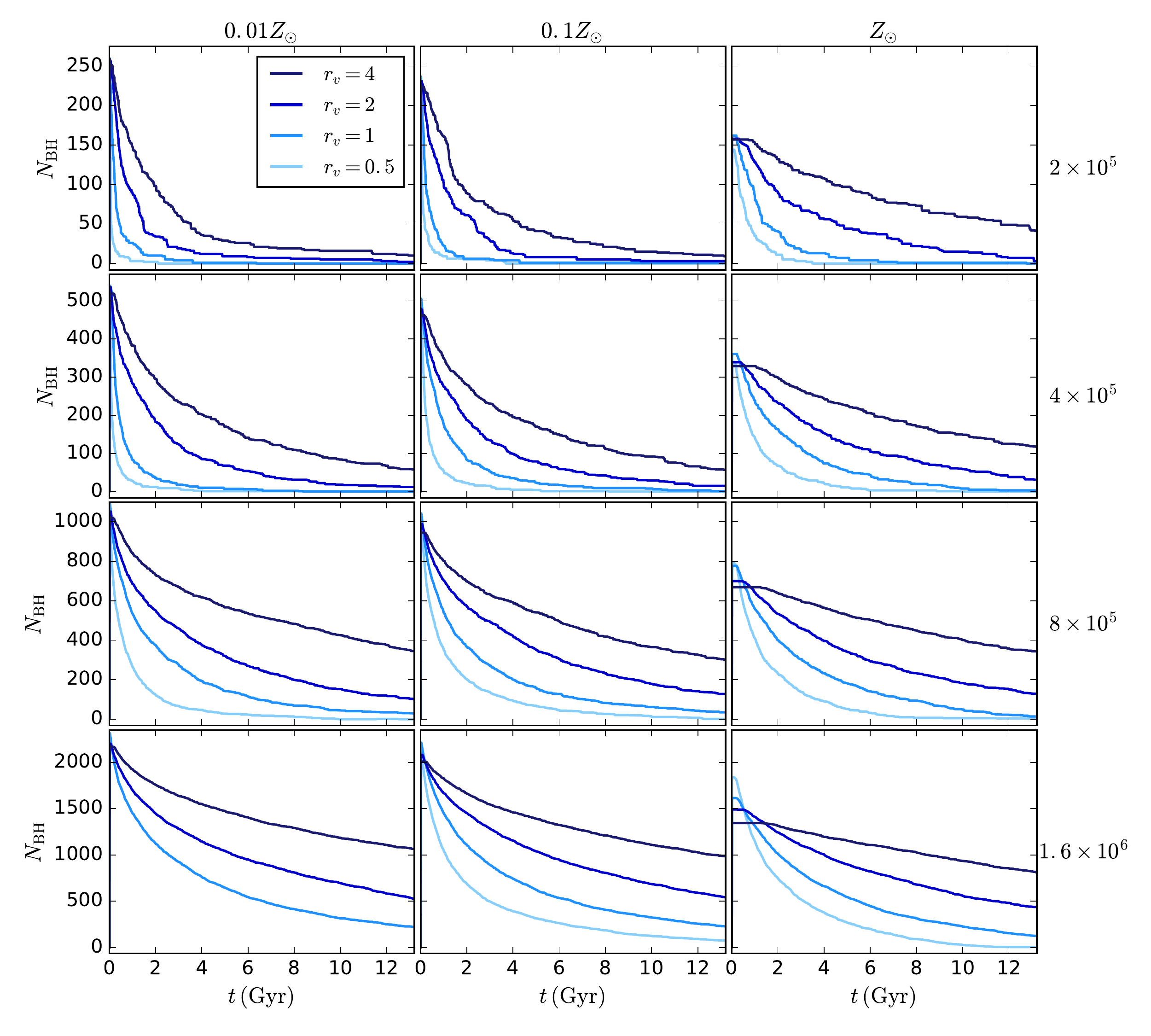}
\caption{\footnotesize \label{fig:Nbh} Total number of retained BHs versus time for all models with $R_{\rm{gc}}=20\,$kpc. The three columns show models of different metallicity (from left to right, $Z=0.01Z_{\odot}$, $0.1Z_{\odot}$, and $Z_{\odot}$) and the four rows show models of different initial particle number (from top to bottom, $N=2\times10^5$, $4\times10^5$, $8\times10^5$, and $1.6\times10^6$. As in Figure \ref{fig:Harris}, lighter to darker shades of blue indicate increasing initial $r_v$.}
\end{center}
\end{figure*}

In Figure \ref{fig:Nbh}, we show the total number of retained BHs versus time for all models with $R_{\rm{gc}}=20\,$kpc (a fixed value here for simplicity).
The four rows show, from top to bottom, models increasing in $N$, and the three columns show, from left to right, models increasing in $Z$. As in Figure \ref{fig:Harris}, curves of lighter to darker shades of blue indicate increasing $r_v$.\footnote{Note that the model with $r_v=0.5\,$pc is absent from the lower-left hand panel. As discussed in Section \ref{sec:method}, this particular model undergoes collisional runaway within the first few Myr. Treatment of this process is beyond the scope of the present version of \texttt{CMC}, so we exclude this model from our study.}

In all models, the total number of BHs decreases throughout the lifetime of the cluster. The differences in the total number of BHs formed, the total number of BHs retained at simulation end, and the depletion rate of the BH population throughout the simulation can be attributed to the modulation in $N$, $r_v$, and $Z$. We discuss each in turn below.

\textit{Initial particle number ($N$):} for models of fixed $r_v$ and $Z$, the initial number of stars determines the total number of BHs that form in the cluster through stellar evolution. Because we adopt a binary fraction of 5\% for all models in this study, stellar evolution here refers primarily to \textit{single} star evolution. As Figure \ref{fig:Nbh} shows, the total number of BHs retained in the cluster at birth scales roughly linearly with $N$: as $N$ is doubled, the initial number of retained BHs roughly doubles.

\textit{Metallicity ($Z$):} for the purposes of BH populations, metallicity determines the mass lost through stellar winds in high-mass stars, which in turn, determines the masses of the BHs at formation. As described in Section \ref{sec:method}, we adopt the wind-mass loss prescriptions of \citet{Vink2001}. In short, higher metallicity means higher line-driven winds, resulting in less massive stars just before collapse to a BH. Thus, higher-metallicity clusters yield lower-mass BHs. Because our BH formation prescriptions assume that, in general, lower mass BHs receive larger natal kicks \citep[because lower mass BHs form with less fallback; see Section \ref{sec:summary} and][]{Morscher2015}, a larger fraction of low-mass BHs formed in a higher metallicity clusters are ejected promptly at formation. Thus, higher metallicity clusters will retain fewer BHs at birth, as can be seen by comparing columns 1--3 in Figure \ref{fig:Nbh}.

\textit{Initial virial radius ($r_v$):} the initial $r_v$ affects BH retention in two ways: immediately at time of BH formation ($t \lesssim$ 10s of Myr), and subsequently over long dynamical timescale ($t\sim$Gyrs). Clusters with larger $r_v$ have shallower potential wells ($U \sim GM_{\rm{tot}}/r_v$). Thus, a larger fraction of BHs will be ejected promptly from their host cluster through natal kicks (see Section \ref{sec:method} for details of our BH natal kick prescriptions). This effect is most pronounced for higher metallicities (right-hand column of Figure \ref{fig:Nbh}, for the reasons discussed above. As a side note, the initial retention fraction of NSs also varies with $r_v$ in the same manner: more NSs are retained at birth in clusters with smaller $r_v$.

Once the population of BHs forms in a cluster through stellar evolution, the BHs rapidly mass segregate to the cluster core. As described in \citet{Kremer2019a}, smaller $r_v$ means shorter relaxation time, which means quicker mass segregation. In addition, smaller $r_v$ also leads to higher central densities, which leads to a higher rate of super-elastic strong encounters leading to quicker depletion of BHs via dynamical ejections. For fixed $N$ and $Z$, clusters with lower $r_v$ (lightly shaded blue curves in Figure \ref{fig:Nbh}) have higher initial densities and shorter relaxation times, and therefore eject their BHs relatively quickly compared to models with higher $r_v$ (darkly shaded blue curves).

\subsection{Radial profiles}

\begin{figure*}
\begin{center}
\includegraphics[width=\linewidth]{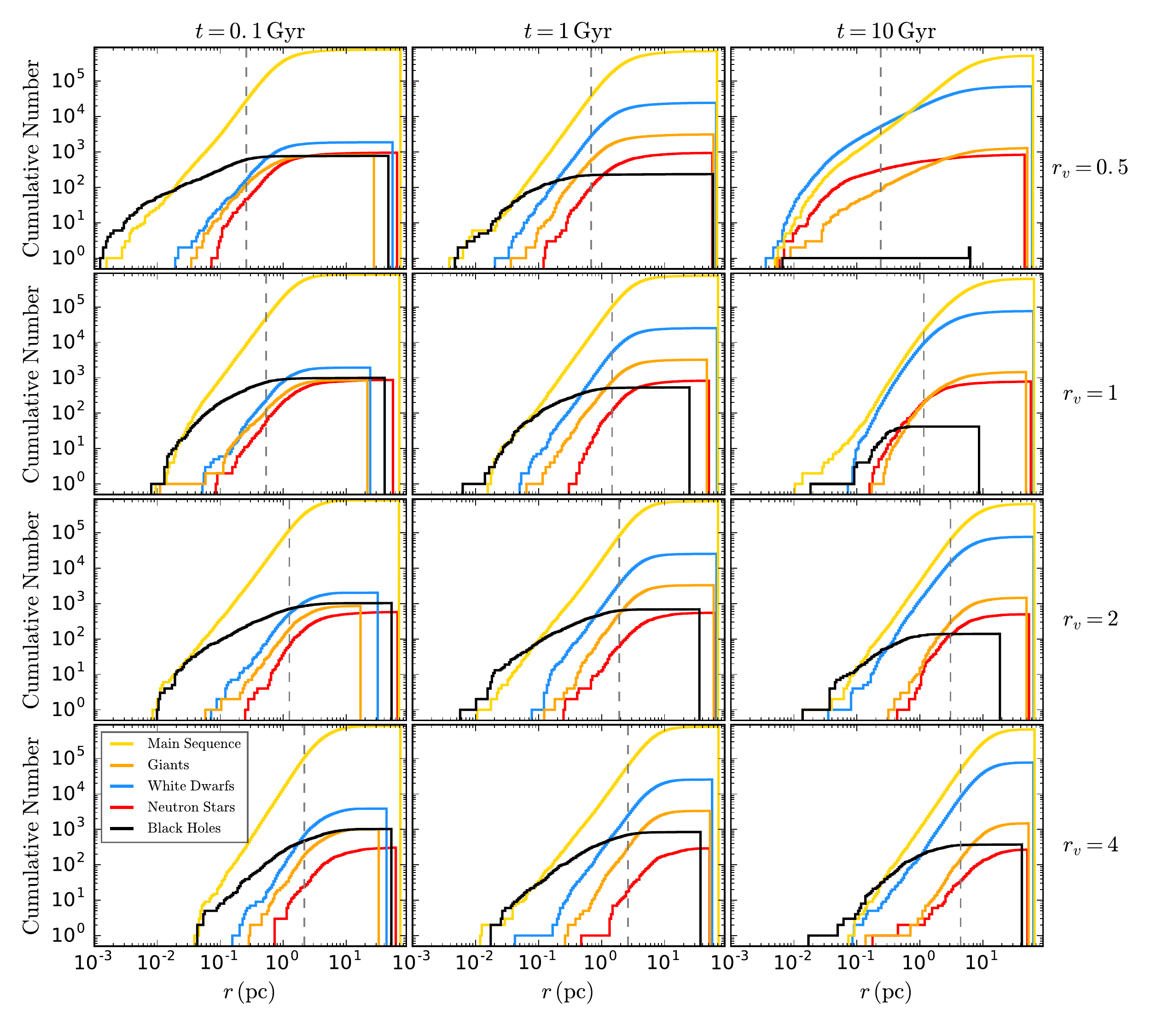}
\caption{\footnotesize Cumulative radial distributions of all stellar populations (colors denoted in legend) for three cluster snapshots in time (from left to right, $t=$0.1, 1, and 10 Gyr). From top to bottom, we show clusters of varying initial virial radius, $r_v$. All other initial cluster parameters are fixed. The vertical, dashed gray lines mark the core radius of each cluster snapshot. Note that models with smaller initial $r_v$ have fewer BHs and smaller core radii than their large $r_v$ counterparts.} \label{fig:radial_dist}
\end{center}
\end{figure*}

\begin{figure*}
\begin{center}
\includegraphics[width=\linewidth]{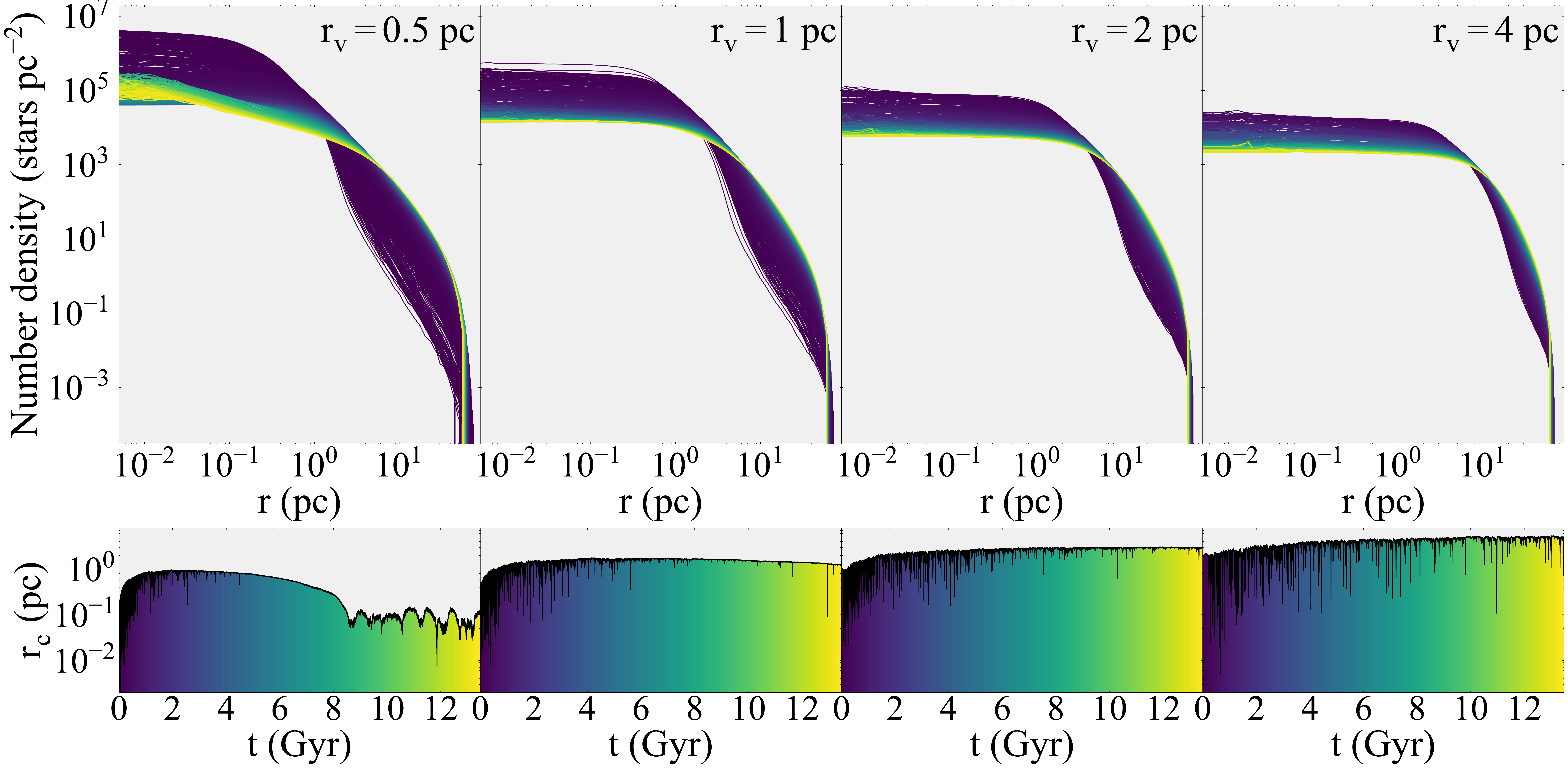}
\caption{\footnotesize \textit{Top}: Stellar number surface density profiles throughout time for models with different initial virial radii (from \textit{left} to \textit{right}: models \textsc{n8-rv0.5-rg8-z0.01}, \textsc{n8-rv1-rg8-z0.01}, \textsc{n8-rv2-rg8-z0.01}, and \textsc{n8-rv4-rg8-z0.01}).
\textit{Bottom}: Time evolution of the theoretical core radii of these models.
Model \textsc{n8-rv0.5-rg8-z0.01} is a prototypical example of a cluster which would be core-collapsed by the present day.
In contrast, models \textsc{n8-rv2-rg8-z0.01} and \textsc{n8-rv4-rg8-z0.01} represent clusters whose core-collapses have been halted.
Model \textsc{n8-rv1-rg8-z0.01} represents clusters which have only barely avoided core-collapse by the present day. Colors denote cluster age. Animated versions of this figure are available in the electronic version of the manuscript. The three panels of each animation show the number density profile, core radius, and total number of BHs, respectively, as they evolve in time.} \label{fig:lum_profiles}
\end{center}
\end{figure*}

\begin{figure*}
\includegraphics[width=0.8\linewidth]{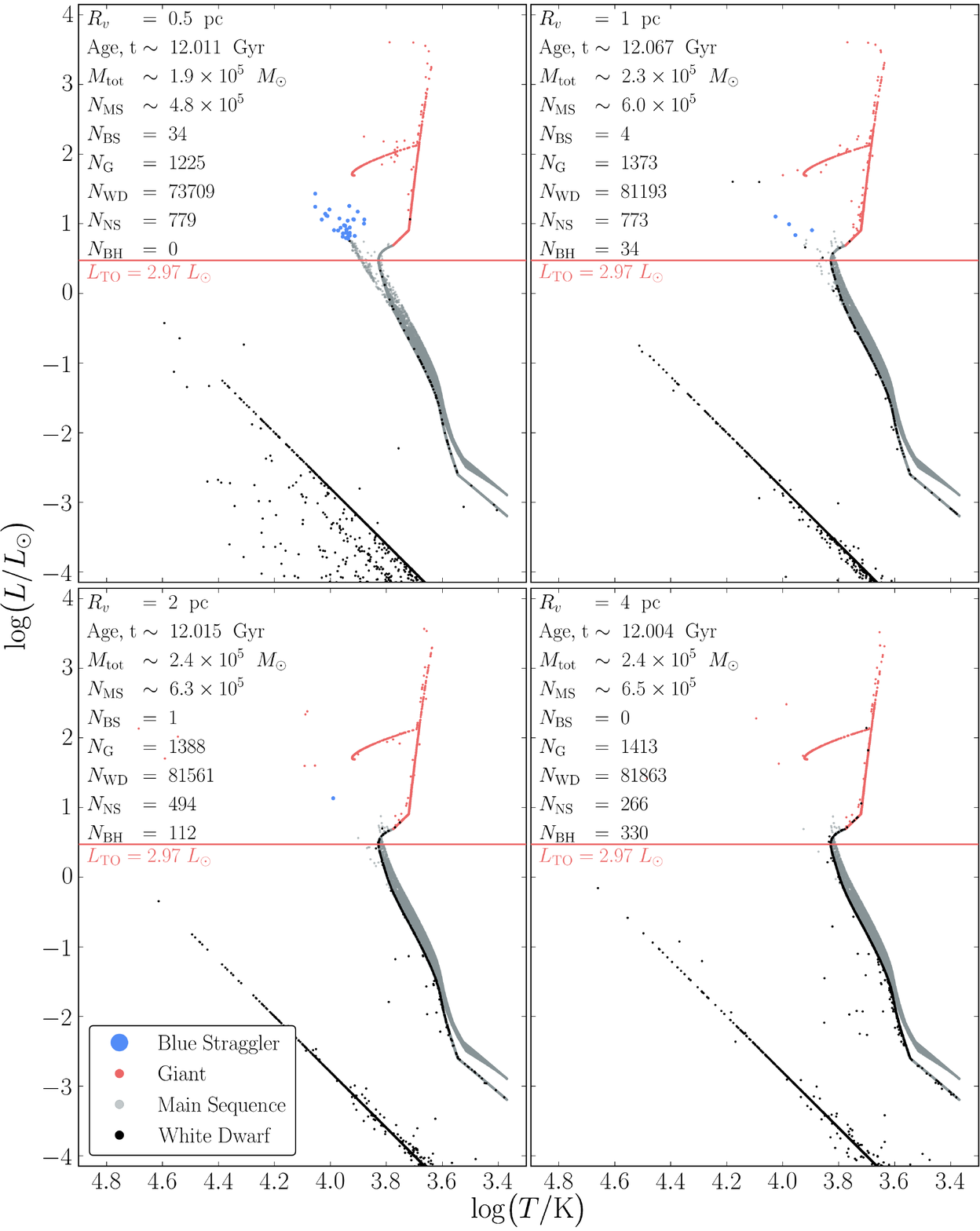}
\caption{\footnotesize Hertzsprung-Russell (HR) diagrams of the four clusters shown in Figures \ref{fig:radial_dist} and \ref{fig:lum_profiles}. Each dot represents a single or binary star (all binaries are considered unresolved), with blue stragglers (BSs) colored in blue, giants in red, the main sequence (MS) stars in gray, and any single or binary containing a white dwarf in black. Note that for binaries, the summed luminosity and luminosity-weighted mean temperature are plotted. To select BSs, the MS turnoff is first defined as the luminosity, $L_{\rm{TO}}$, corresponding to the point on the MS branch with the highest median temperature, $T_{\rm{TO}}$. $L_{\rm{TO}}$ is indicated by the red horizontal line. BSs are then defined as any MS single (or binary containing a MS star) where the above-defined luminosity and temperature exceed $L_{\rm{BS}} = 2\cdot L_{\rm{TO}}$ and $T_{\rm{TO}}$. For a more detailed view, see Figure \ref{fig:hrdiag_zoom}.}
\label{fig:hrdiag_multi}
\end{figure*}

In Figure \ref{fig:radial_dist}, we show cumulative radial distributions at three separate cluster ages ($t=$0.1, 1, and 10 Gyr). For simplicity, we limit this figure to clusters with $N=8\times10^5$, $R_{\rm{gc}}=8\,$kpc, and $Z=0.01Z_{\odot}$, but emphasize that all models exhibit similar behavior.
From top to bottom, we show models 15, 51, 87, and 123 (see Table \ref{table:models}), which have initial $r_v=0.5$, $1$, $2$, and $4\,$pc, respectively. In each panel, we show radial distributions for all different stellar populations: main sequence stars (MS; yellow), giants (orange), WDs (blue), NSs (red), and BHs (black).

The vertical, dashed gray lines mark the core radius of each cluster snapshot, as defined in Section \ref{sec:definitions}. For a fixed cluster age (i.e. a single column in Figure \ref{fig:radial_dist}), models with smaller $r_v$ typically have smaller core radii than models with larger $r_v$. This is also a direct consequence of the BH burning mechanism described previously. Furthermore, if the number of BHs (black curves) is compared across models, we see that the models with the smallest core radii at a specific time also have on average, the fewest BHs.

As seen in all panels, the innermost regions of clusters are typically dominated by the BHs (in some cases, mixed with MS stars, particularly when the MS stars are bound to BH binary companions). 
This is a consequence of mass segregation and is consistent with predictions from a number of recent analyses \citep[e.g.][]{Kulkarni1993,Sigurdsson1993,BreenHeggie2013,Morscher2015,Chatterjee2017a,Askar2018,Kremer2019a}. Notably, this is not the case for the $r_v=0.5$ and $r_v=1$ models at $t=10\,$Gyr. As described in Section \ref{sec:BHpopulations}, the dynamical clock of a cluster is determined by the initial $r_v$. Clusters with smaller initial $r_v$ are more dynamically evolved by $t=10\,$Gyr and thus have ejected a larger fraction of their BHs compared to models with larger $r_v$. For the $r_v=0.5$ and $r_v=1$ models, only 2 and 30 BHs (out of roughly 1000 total retained in the clusters at birth) remain at $t=10\,$Gyr. These BH populations no longer provide sufficient energy to the cluster through BH burning to prevent lower-mass luminous stars from entering the cluster's inner most region. The ``collapse'' of the cluster's core is most prominently seen in the $r_v=0.5$ model, as evidenced by both the core radius (dashed gray line) and the visible shift in the distribution of non-BH populations. Indeed, as discussed in Section \ref{sec:corecollapse}, this particular cluster has a surface density profile representative of a ``core-collapsed'' cluster at the conclusion of the simulation.

The same effect is apparent in the $r_v=1$ model where, at $t=10\,$Gyr, the MS stars have begun to infiltrate the cluster's innermost region, although not as prominently as in the $r_v=0.5$ model. This is expected: the transition from a BH-dominated cluster with a large core to a core-collapsed cluster is smooth (for illustration of this point, see Figure \ref{fig:lum_profiles}). This means that as the number of BHs decreases, the core becomes increasingly dominated by MS stars (the exact number of MS stars is dependent upon statistical fluctuations governed by the chaotic strong encounters in the core).
The MS star distributions shown in the four $t=10\,$Gyr panels in Figure \ref{fig:radial_dist} show clearly this transition.

Several key points can also be made regarding the NS populations in Figure \ref{fig:radial_dist} (red curves). Through the BH burning mechanism discussed previously, mass-segregation of NSs is prevented when significant numbers of BHs are present in the host cluster core \citep[e.g.,][]{Ye2018,Fragione2018a,Banerjee2018}. As a result, NSs tend to be found at relatively large radial offsets compared to the BHs. Furthermore, unlike the BHs, the total number of NSs remains roughly constant across the three snapshots in time shown for each cluster in this figure.
This is simply because NSs remain relatively inactive dynamically compared to the BHs which preferentially occupy the densest regions of their host clusters where high rates of strong encounters lead to rapid depletion of BHs via dynamical ejections.
The notable exception is the $r_v=0.5$ model at $t=10\,$Gyr (upper-right panel) where NSs occupy the cluster's innermost regions. Here, because this cluster has already ejected all but two of its BHs, NSs are the most massive stellar population remaining in the cluster and therefore have begun to efficiently mass segregate to the core. This is in line with the predictions made in \citet{Ye2018}, which noted that only in the absence of BHs (i.e., in core-collapsed clusters) will a significant fraction of NSs be found in the core of a cluster. As \citet{Ye2018} noted, this sets the stage for dynamical formation of millisecond pulsars (MSPs) and we thus expect an anti-correlation between the number of MSPs in a cluster and the total number of BHs. We return to the topic of MSPs in Section \ref{sec:pulsars}.

Several similar points can be made for the WD populations (blue curves). Unlike NSs, the number of WDs increases throughout the lifetime of the cluster, simply due to stellar evolution as less massive stars are converted to WDs at later times. Like the NSs, WDs are generally found outside the cluster's core until the BH population is sufficiently depleted. However, in core-collapsed clusters like the $r_v=0.5\,$pc model shown here, WDs can dominate the core by number. Thus, core-collapsed clusters are likely ideal factories for dynamical formation of WD binaries, Type Ia SNe, and CVs. We discuss several of these possibilities in Section \ref{sec:WDs}.

\subsection{Core-collapsed versus non-core-collapsed}
\label{sec:corecollapse}

Figure \ref{fig:lum_profiles} illustrates how initial virial radius and BH burning affect the density profile in clusters. In the top row of panels, we show the number density profile at each snapshot in time for the same four models shown in Figure \ref{fig:radial_dist}. From left to right, we show models increasing in $r_v$. In the bottom row, we show the evolution of (theoretical) core radius as a function of time. The snapshot time of each profile is colored from dark purple (early times) to yellow (late times), as indicated by the color spectrum in the lower panels.

In all cases, the core radius expands at early times ($t \lesssim 1\,$Gyr) due to mass loss associated with evolution of high mass stars as well as BH burning once a BH-core forms. The subsequent evolution varies from model to model, depending on $r_v$. For $r_v=4\,$pc, the core continues to expand throughout the cluster's evolution as a result of prolonged BH burning; as seen in Figures \ref{fig:Nbh} and \ref{fig:radial_dist}, this model retains a large population of BHs throughout its entire lifetime. This is in stark contrast to the $r_v=0.5\,$pc case, where the BH population is rapidly depleted (less than $10\%$ of the initial BH population is still retained in the cluster by $t\approx 4\,$Gyr). In this case, the core radius begins to contract relatively, only steadying out once binary burning (involving regular stellar binaries) begins at $t\approx 8\,$Gyr. Comparing the number density profiles, we see that where the $r_v=4\,$pc case retains a King profile through the entire simulation, the $r_v=0.5\,$pc model clearly reaches (and maintains) a core-collapse architecture once binary burning has begun.

In the electronic version of the manuscript, we also include animated illustrations of the four models show in Figure \ref{fig:lum_profiles}.  In the left-hand panel of these animations, we show the cluster luminosity profiles as they evolve in time. In the upper-right corner of this panel, we indicate the current cluster age as well as the total number of retained BHs and the maximum BH mass at this age. Note that the maximum BH mass decreases as the cluster evolves \citep[see, e.g.,][for detailed discussion of this point]{Morscher2015}. Note also that at early times, the maximum BH mass may exceed the $40.5\,M_{\odot}$ limit from the PPSN prescriptions described in Section \ref{sec:method}. These massive BHs are formed through either repeated BH mergers \citep[e.g.,][]{Rodriguez2018} or through stellar collisions (see Section \ref{sec:stellar_collisions}). In the middle panel of the animations, we show the theoretical core radius as it evolves over time and in the right-hand panel, we show the total number of retained BHs versus time.

\subsection{Hertzsprung-Russell diagrams}
\label{sec:HRdiagrams}

In Figure \ref{fig:hrdiag_multi}, we show Hertzsprung-Russell (HR) diagrams for the four clusters seen in Figures \ref{fig:radial_dist} and \ref{fig:lum_profiles} at $t=12\,$Gyr. We plot here the bolometric luminosity versus temperature of all stars, which are both given by \texttt{SSE} \citep[see][for further details]{Hurley2000}.

The location of the MS turnoff \citep[defined as in][]{Weatherford2018} is identical in all four panels, as expected given that these clusters have identical metallicities and ages. One noticeable difference between the four clusters here concerns the blue stragglers (blue circles), the population of stars lying leftward (hotter) and upward (brighter) of the MS turnoff \citep[e.g.,][]{Sandage1953}. We discuss blue stragglers in more detail in Section \ref{sec:BSs}.

\section{Low-mass X-ray binaries}
\label{sec:XRBs}

X-ray sources have been well-observed in GCs dating back to the 1970s \citep{Clark1975,Heinke2010}. It is understood that various dynamical processes relevant in GCs lead to formation of low-mass X-ray binaries (LMXBs) at a significantly higher rate per unit stellar mass in clusters compared to isolated binary evolution in the Galactic field \citep[e.g.,][]{Clark1975}. For several decades, the LMXBs observed in GCs came exclusively in the neutron-star-accretor variety. In fact, the conspicuous absence of BH LMXBs in clusters was traditionally used to argue that clusters have very few (if any) stellar-mass BHs at present \citep[e.g.,][]{Kulkarni1993}.

\begin{deluxetable*}{lc||ccc|ccc}
\tabletypesize{\scriptsize}
\tablewidth{0pt}
\tablecaption{Black hole and neutron star binaries in four representative clusters\label{table:XRBs}}
\tablehead{
	\colhead{} &
	\colhead{$M_{\rm{tot}}\,(M_{\odot})$} &
    \colhead{Total BHs} &
    \colhead{Det. BH binaries} &
    \colhead{Acc. BH binaries}&
    \colhead{Total NS}&
    \colhead{Det. NS binaries}&
    \colhead{Acc. NS binaries}\\
}
\startdata
Typical & $2.3\times10^5$ &105--140 & 0--5 & 0--1 & 494--498 & 0--1 & 1\\
Core-collapsed & $1.9\times10^5$ & 0--3 & 0--2 & 0 & 751--837 & 7--14 & 4--6\\
Low-mass & $5\times10^4$ &0--3 & 0--1 & 0 & 25 & 0 & 0\\
High-mass & $10^6$ &1812--2061 & 1--7 & 0-1 & 4157--4190 & 3--4 & 2\\
\enddata
\tablecomments{Ranges of total number of BH and NS binaries with luminous companions in both detached and accreting configurations for all snapshots in the range 10--13 Gyr for four representative clusters. Here, ``typical'' denotes model \textsc{n8-rv2-rg8-z0.01} (a non-core-collapsed cluster with mass, matallicity, and Galactocentric position typical for MW clusters), ``core-collapsed'' denotes model \textsc{n8-rv0.5-rg8-z0.01}, ``low-mass'' denotes model \textsc{n2-rv2-rg8-z0.01}, and ``high-mass'' denotes model \textsc{n32-rv2-rg20-z0.01}.}
\end{deluxetable*}

This picture has begun to change within the past ten years as the first stellar-mass BH candidates have been identified in both Galactic and extragalactic GCs, primarily as accreting LMXBs \citep{Maccarone2007,Strader2012,Chomiuk2013,Miller-Jones2015,Shishkovsky2018}. The discovery of these BH LMXBs, as well as the detached BH--MS binaries found in NGC 3201 by the MUSE survey \citep{Giesers2018,Giesers2019}, have motivated more detailed studies of how BH binaries may form in GCs. For example, \citet{Ivanova2010} noted that ultracompact BH LMXBs with degenerate donors \citep[similar perhaps to the X-ray source observed in 47 Tuc;][]{Bahramian2017,Church2017} may form as a result of BH--giant collisions. Later work by \citet{Kremer2018a} noted that accreting BH binaries can form through exchange encounters at rates consistent with the number of BH X-ray sources observed in clusters to date.

In Table \ref{table:XRBs_full} in the Appendix, we show the average numbers of NS and BH binaries in each model for snapshots with ages in the range 10--13 Gyr. We distinguish between various types of luminous companions (MS stars, giants, and WDs) and also between those binaries that are accreting and detached. Here, we define accreting binaries as in \texttt{BSE}, where the donor star must fill its Roche radius in the zero-eccentricity limit: $R > R_L$, where $R$ is the stellar radius and
\begin{equation}
\label{eq:RL}
R_L = a\frac{0.49q^{2/3}}{0.6q^{2/3}+log(1+q^{1/3})},
\end{equation}
where $a$ is the binary semi-major axis and $q$ is the mass ratio. In reality, an eccentricity-dependent definition for the onset of Roche-lobe overflow may be more appropriate [e.g., $R > R_L(1-e)$]. However, as shown in \citet{Kremer2018a}, such a definition is unlikely to change the results significantly in the context of forming LMXBs.

In Table \ref{table:XRBs}, we show the total numbers of BH and NS binaries (both accreting and detached) in the four simulations that characterize common cluster types observed in the MW, as described in Section \ref{sec:comparison_to_MW}.

Although we reserve a thorough study of BH/NS binary formation and their potential observability as X-ray sources for a future study dedicated specifically to the topic, we comment here on two general trends. First, the number of accreting BH binaries scales very weakly with cluster properties, consistent with several recent analyses \citep[e.g.,][]{Chatterjee2017b,Kremer2018a}. In order for BHs and luminous stars to form binaries through dynamical encounters, the two populations must overlap within a ``mixing zone'' in the cluster's core. The dynamical encounter rate between BHs and luminous stars within this zone is approximately $n_{\rm{LC}} \Sigma v_\infty N_{\rm{BH}}$, where $n_{\rm{LC}}$ is the typical number density of luminous companions and $v_\infty$ is the typical relative velocity at infinity in the mixing zone. $\Sigma$ is the cross section for encounters and $N_{\rm{BH}}$ is the total number of BHs in the mixing zone. As discussed in Section \ref{sec:results}, $N_{\rm{BH}}$ determines the density of luminous stars within clusters through BH burning; as $N_{\rm{BH}}$ increases, $n_{\rm{LC}}$ decreases. Thus we expect the total formation rate of BH--LC  binaries to remain roughly constant with $N_{\rm{BH}}$, as suggested by Table \ref{table:XRBs} and as described in detail in \citet{Kremer2018a}.

However, no such self-regulating dynamical process is expected for NS binaries; formation of these systems is expected to depend simply on the density of the cluster core, and thus, on $N_{\rm{BH}}$. Among clusters with comparable mass, those with fewer BHs have relatively dense cores (see Section \ref{sec:results}) and therefore form more NS binaries. This is analogous to the results of \citet{Ye2018}, which showed that MSP formation also anticorrelates with the total BH population size. We further discuss pulsars in the following section.

\section{Pulsars}
\label{sec:pulsars}

\begin{figure*}
\begin{center}
\includegraphics[width=1.0\textwidth]{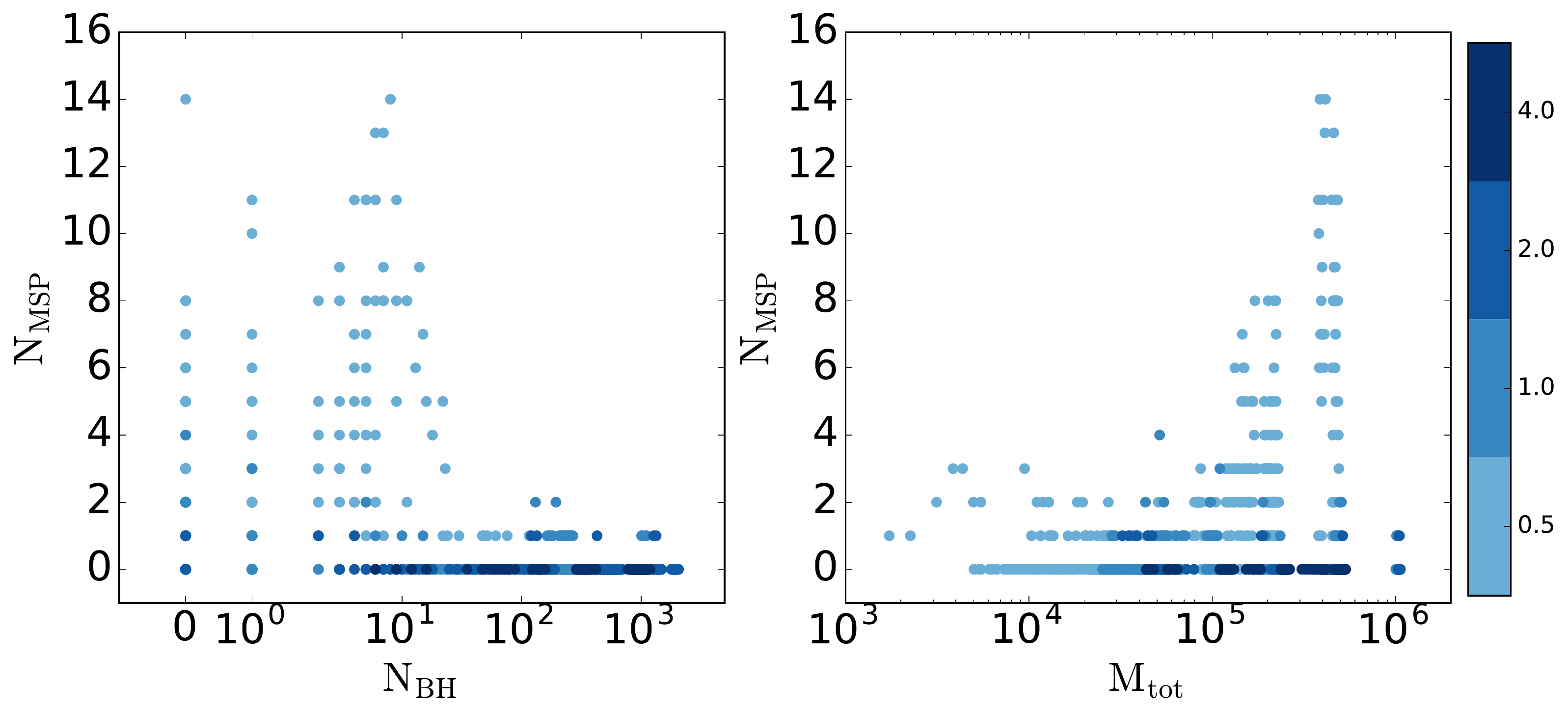}
\caption{\footnotesize \label{fig:pulsars}The number of MSPs versus the number of BHs (left panel) and the total cluster mass (right panel) in all model snapshots in the range 10--13 Gyr.}
\end{center}
\end{figure*}

In excess of 150 millisecond radio pulsars have been observed in various GCs in the MW \citep[for a recent review, see][]{Ransom2008}. MSPs are generally thought to form when an old, slowly-spinning NS is spun up through mass transfer from a Roche-lobe-filling binary companion \citep[e.g.,][]{Rappaport1995,Tauris2012}. Thus, MSPs are likely intimately linked to NS LMXBs (see Section \ref{sec:XRBs}), with the former being direct descendants of the latter. As with LMXBs, the MSP formation rate is expected to be more pronounced in GCs, due to dynamical processes, relative to isolated binary evolution \citep[e.g.,][]{Clark1975,Hut1992,Bahramian2013,Ye2018}.

Recently, \citet{Ye2018} showed that MSP formation also directly relates to a cluster's stellar-mass BH retention. When a large population of BHs is present, BH burning heats the cluster's core, delaying core-collapse, regulating the central density, and, of relevance to the formation of MSPs, limiting the dynamical encounter rate for NSs. As a cluster's BH populations becomes depleted, NSs grow more dynamically active, ultimately increasing the MSP formation rate \citep[see also][]{Fragione2018a}. Hence, as shown in \citet{Ye2018}, we expect an anti-correlation between BH number and MSP number in clusters of equal mass. GCs with the fewest BHs (i.e., core-collapsed clusters), are expected to host the largest numbers of MSPs.

\citet{Ye2018} controlled BH retention in cluster models by varying the magnitude of BH natal kicks, a proxy for more physically-motivated processes that may determine BH retention such as the cluster's initial $r_v$. In this analysis, we expand upon the results of \citet{Ye2018} by modulating BH retention by varying $r_v$ while fixing the BH natal kick physics, and also by exploring pulsar formation in cluster models with a wider range in particle number and metallicity. All physics relevant to the formation of pulsars and MSPs is the same here as in \citet{Ye2018}. Briefly, we randomly select NS spin periods and magnetic field strengths using values consistent with the young pulsars observed in the MW ($B\sim10^{12}\,$G and $P>30\,$ms). All NSs form as young pulsars in our models. MSPs can form through stable mass transfer in binaries. We assume the spin period evolves through dipole radiation and that the magnetic fields of single or detached pulsars decay exponentially with time \citep{Kiel2008}.  For pulsars that have been through mass transfer, we assume the ``magnetic field burying'' scenario and lower the pulsars' magnetic fields according to how much material is accreted \citep{Kiel2008}. These pulsars exhibit faster spins due to angular momentum transfer \citep{Hurley2002}.

In the left panel of Figure \ref{fig:pulsars}, we show the number of retained MSPs versus the total number of retained BHs for all model snapshots with age in the range 10--13 Gyr. Each scatter point indicates a distinct cluster snapshot. As in earlier figures, shades of blue denote the cluster's initial $r_v$, with lighter to darker shades indicating increasing $r_v$. For reference, the average numbers of pulsars and MSPs retained at late times in each model are also listed in Table \ref{table:WD-pulsar-BS-full}.

Comparing specifically to Figure 4 of \citet{Ye2018}, we see a similar trend: clusters with smaller BH populations are capable of producing larger numbers of MSPs (here, up to 14) while clusters with large numbers of BHs are unlikely to contain more than 1--2 MSPs. As Figure \ref{fig:pulsars} shows, this relation is determined primarily by the initial $r_v$. Models with $r_v=0.5\,$pc (light blue) that have the highest central densities at late times (see Figures \ref{fig:radial_dist} and \ref{fig:lum_profiles}) generally produce more MSPs than models with larger values of $r_v$.

For models with few BHs, we predict a slightly smaller number of MSPs compared to \citet{Ye2018}. For models with similar $N$ ($8\times10^5$) and metallicity ($0.1Z_{\odot}$), we find up to roughly 7 MSPs in models with 0--1 BHs compared to up to roughly 16 for similar models in \citet{Ye2018}. This makes sense given the differences in the two methods. In the limiting case of \citet{Ye2018}, nearly all BHs are ejected from the cluster at birth through natal kicks. For such a cluster, BH burning is nonexistent from the outset and the cluster's core quickly collapses. This allows dynamical processes relevant to pulsar formation to operate over essentially the full lifetime of the cluster, yielding many more opportunities for pulsar and MSP formation. However, when the BH natal kick physics is fixed, as in this analysis, the NSs must first wait for the BH population to be sufficiently depleted before the core can collapse. As shown in Figure \ref{fig:lum_profiles}, for $N=8\times10^5$, this takes roughly 8 Gyr. Because clusters with smaller total mass (total $N$) have shorter relaxation times (Equation \ref{eq:relaxationtime}), models with smaller $N$ will tend to core-collapse sooner; however, even small-$N$ clusters must wait some characteristic length of time before NSs become dynamically active in the core.
Hence, models in the present study spend a much shorter portion of their life in a core-collapsed state relative to those in \citet{Ye2018}, limiting the number of pulsars compared to those earlier models. However, we do stress that this effect amounts to only a factor of roughly 2. Most importantly, the general trend between MSP number and BH number is preserved.

\begin{deluxetable}{l|c|ccc}
\tabletypesize{\scriptsize}
\tablewidth{0pt}
\tablecaption{Pulsars in four representative clusters\label{table:PSRs}}
\tablehead{
	\colhead{} &
	\colhead{$M_{\rm{tot}}\,(M_{\odot})$} &
    \colhead{Total NS}&
    \colhead{Pulsars}&
    \colhead{MSPs}\\
}
\startdata
Typical & $2.3\times10^5$ & 494--498 & 1 & 1\\
Core-collapsed & $1.9\times10^5$ & 751--837 & 4--13 & 4--7\\
Low-mass & $5\times10^4$ & 25 & 0 & 0\\
High-mass & $10^6$ & 4157--4190 & 3--4 & 3\\
\enddata
\tablecomments{Ranges of total number of NSs, pulsars and millisecond pulsars for all snapshots in range 10-13 Gyr for four representative clusters, as defined in Table \ref{table:XRBs}.
}
\end{deluxetable}

In Table \ref{table:PSRs}, we show the range in total numbers of pulsars and MSPs in the same four characteristic clusters shown in Table \ref{table:XRBs}. Generally, the numbers quoted in this table are in rough agreement with observations of clusters of comparable total mass \citep[reviewed in][]{Ransom2008}. We predict up to 13 pulsars in core-collapsed clusters with total mass comparable to the characteristic core-collapsed model shown in Table \ref{table:PSRs} ($M_{\rm{tot}} \approx 2\times10^5$), e.g., NGC 6752 (5 observed pulsars).
In our four massive simulations, we identify up to 8 pulsars, which admittedly is less than the number, for example, in 47 Tuc, which has comparable total mass to our massive models. However, 47 Tuc has a very high central density and previous work has predicted this cluster likely retains a fairly small population of BHs \citep{Weatherford2018}. As described in Section \ref{sec:method}, the most massive branch of our parameter space (initial $N=3.2\times10^6$) does not extend to sufficiently small initial $r_v$ to produce a core-collapsed cluster at late times, simply due to computational limitations. However, \citet{Ye2018} modelled a single massive cluster that reached central densities comparable to 47 Tuc at late times (by assuming all BHs are ejected at birth through natal kicks as a computationally inexpensive proxy for efficient BH ejection associated with small $r_v$) and showed that, in this limiting case, the number of model pulsars are roughly consistent with observed number in a cluster like 47 Tuc. We direct the reader to \citet{Ye2018} for more detailed discussion on this topic.

\section{White Dwarfs}
\label{sec:WDs}

\begin{deluxetable*}{lc||ccc|ccc}
\tabletypesize{\scriptsize}
\tablewidth{0pt}
\tablecaption{White dwarf properties in four representative clusters\label{table:WDs}}
\tablehead{
	\colhead{} &
	\colhead{$M_{\rm{tot}}\,(M_{\odot})$} &
    \colhead{Total WDs ($\times10^4$)} &
    \colhead{Det. WD binaries} &
    \colhead{Acc. WD binaries}&
    \colhead{WD--WD coll.}&
    \colhead{WD--NS coll.}&
    \colhead{WD--BH coll.}\\
}
\startdata
\rm{Typical} & $2.3\times10^5$ & 7.7--8.4 & 400--463 & 32--35 & 0 & 0 & 0\\
\rm{Core-collapsed} & $1.9\times10^5$ & 7.1--7.5 & 85--125 & 10--18 & 66 & 17 & 0\\
\rm{Low-mass} & $5\times10^4$ &1.8--1.9 & 321--336 & 44--51 & 0 & 0 & 0\\
\rm{High-mass} & $10^6$ & 32--35 & 858--1029 & 145--181 & 2 & 1 & 0\\
\enddata
\tablecomments{Ranges of total numbers of WDs, WD binaries, and WD collisions with other compact remnants across the four characteristic models defined as in Table \ref{table:XRBs}.}
\end{deluxetable*}

WDs are expected to be abundant in GCs and, as shown in Figure \ref{fig:radial_dist}, may even be the dominant stellar population within the cores of some clusters at late times. Unlike BH and NS populations, which are expected to form early in the evolution of their host cluster ($t_{\rm{formation}} \lesssim 100\,$Myr) and then slowly decrease throughout the remainder of the cluster lifetime (see, e.g., Figure \ref{fig:radial_dist}), WDs continue to form throughout the full lifetime of their host cluster such that the number of WDs increases with time.

To handle WD formation and evolution, we adopt the treatment implemented in \texttt{BSE} \citep[for details, we direct the reader to][]{Hurley2002}. In column 12 of Table \ref{table:models}, we show the total number of WDs retained in each simulation at simulation end. The number of WDs at late times generally varies from roughly $10^4$ to over $10^5$, depending on the various simulation parameters. As can be seen from Table \ref{table:models}, the WD number depends most sensitively upon the total number of stars ($N$), with weaker dependence upon initial $r_v$ and metallicity.

When WDs interact with other stars, either through binary evolution or various dynamical processes, WDs have been associated with a number of high-energy astrophysical phenomena such as CVs \citep{Knigge2012,Ivanova2006} and Type Ia SNe \citep[e.g.,][]{SharaHurley2002}. In the following subsections, we briefly explore the processes leading to such events in our model set and discuss various implications.

\subsection{Accreting white dwarf binaries}
\label{sec:accreting_WDs}

WDs that are stably accreting material from a binary companion constitute a number of different astrophysical sources. One type of accreting WD binary is the CV, a system in which a WD accretes material from a donor on the MS \citep[e.g.,][]{Warner1995,Knigge2011}. As their name suggests, CVs are variable stars characterized by novae or nova-like outbursts. For some CVs, where the WD accretor has a strong magnetic field, magnetic activity leads to specific classes of novae. Closely related to the CVs are the AM CVn systems, compact binaries (orbital periods $\lesssim 1\,$hour) where a WD accretes hydrogen-poor material from either a He WD or naked He star \citep[e.g.,][]{Paczynski1967,Marsh2004,Nelemans2005}. Depending on the various features of the system (i.e., accretion rate, mass ratio, etc), some AM CVn may be observed as soft X-ray sources \citep[e.g.,][]{Nelemans2004} and, for mHz GW detectors like LISA, as GW sources \citep[e.g.,][]{Nelemans2004,Kremer2017}. 

Accreting WD binaries are valuable tools for studying important aspects of binary evolution, including mass-transfer processes, super-Eddington accretion, and common envelope physics. In the coming decades, WD binaries will play an emerging role in GW astronomy as WD binaries will be the most abundant source for the upcoming low-frequency GW observatory, LISA \citep{LISA2017}. GW plus electromagnetic observations of accreting WD systems will shed new light on the mass-transfer and tidal physics at work in these systems \citep[e.g.,][]{Breivik2018}. Additionally, some accreting WD binaries are expected to be the progenitors of Type Ia SNe \citep[e.g.,][]{Webbink1984,SharaHurley2002}.

All massive GCs are expected to host populations of accreting WD binaries. Populations of CVs, in particular, are well-observed in many GCs \citep[see, e.g.,][]{Knigge2012}. As with cluster X-ray binaries, some fraction of CVs in GCs are expected to be formed through dynamical processes, thus opening up alternative channels for CV formation beyond the channels relevant to binary stellar evolution alone. A number of analyses have explored the various ways accreting WD binaries may form dynamically in GCs \citep[e.g.,][]{Grindlay1995,Ivanova2006,Belloni2016,Belloni2019}.

Table \ref{table:WDs} lists the total number of WDs as well as total number of (detached and accreting) WD binaries for the four representative cluster models shown in Tables \ref{table:XRBs} and \ref{table:PSRs}. In Table \ref{table:WD-pulsar-BS-full}, we list more expansive information on the WD binaries found in every simulation.

As discussed in Section \ref{sec:XRBs}, the number of NS binaries formed in a cluster correlates in an intuitive way with various cluster properties, especially the cluster mass and density; the denser the cluster, the more dynamical encounters a NS undergoes, presenting more opportunities to form both detached and accreting NS binaries. Given that the most massive WDs in a cluster ($\approx1.2-1.4\,M_{\odot}$) have masses comparable to the typical NS, one may expect similar trends to hold for WD binaries. However, inspection of Tables \ref{table:XRBs} and \ref{table:WDs} reveals the opposite: unlike NS binaries the WD-binary population \textit{decreases} with increasing core density.

This stems from the fact the WD binary (WD--MS, WD--G, WD--WD; both detached and accreting) population derives primarily from the primordial binary population. Thus, for WD binaries, dynamical encounters are in fact more likely to \textit{destroy} binaries destined to become accreting/detached WDs as opposed to forming them. As a consquence, with increasing cluster density (i.e., as a cluster approaches core collapse), the WD binary population is depleted due to enhanced ionization and dynamical ejections. In this context, \citet{Davies1997} noted that the destruction of CV progenitors may be particularly relevant in the cores of dense clusters. The idea that dynamical interactions may commonly lead to destruction of WD binaries was also explored in e.g., \citet{Knigge2012}. This is in contrast to NS binaries and BH binaries (see Sections \ref{sec:XRBs}, \ref{sec:pulsars}, and \ref{sec:BBHs}), which mostly are formed dynamically, suggesting that the formation processes for NS/BH binaries become more efficient with increasing density.

In this case, the specific numbers of accreting/detached WD binaries with different donor types depend sensitively upon assumptions about binary evolution, especially concerning the critical mass ratio for stable mass transfer \citep[for example, see][]{Hurley2002}, as well as common envelope physics. A careful study of the role these parameters play is beyond the scope of the current study. However, for recent work on the subject see, e.g., \citet{Belloni2017}. For simplicity, we adopt here the default assumptions in \texttt{BSE} and reserve a detailed study of the interplay between the dynamical and binary evolution processes for later work.

\subsection{White dwarfs collisions -- connecting to high-energy transients}
\label{sec:WDcollisions}

\begin{figure}
\begin{center}
\includegraphics[width=\columnwidth]{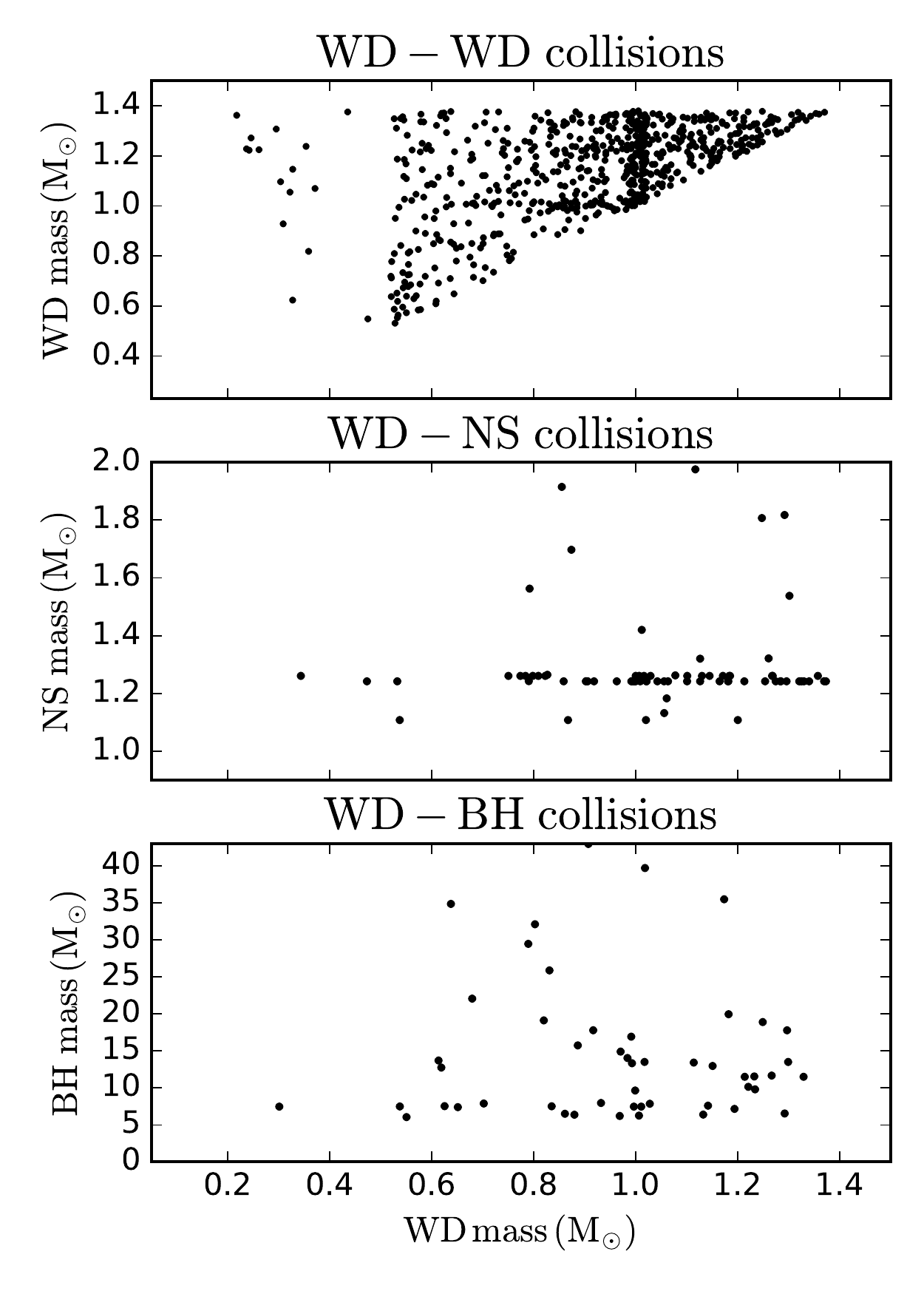}
\caption{\footnotesize \label{fig:WDcollisions} All dynamically-mediated WD collisions with various compact remnants identified in our models. In the top panel, we show WD--WD collisions; middle, WD--NS; and bottom, WD--BH. These collisions may be associated with a number of high-energy transients, as discussed in the text.}
\end{center}
\end{figure}

\begin{figure}
\begin{center}
\includegraphics[width=\columnwidth]{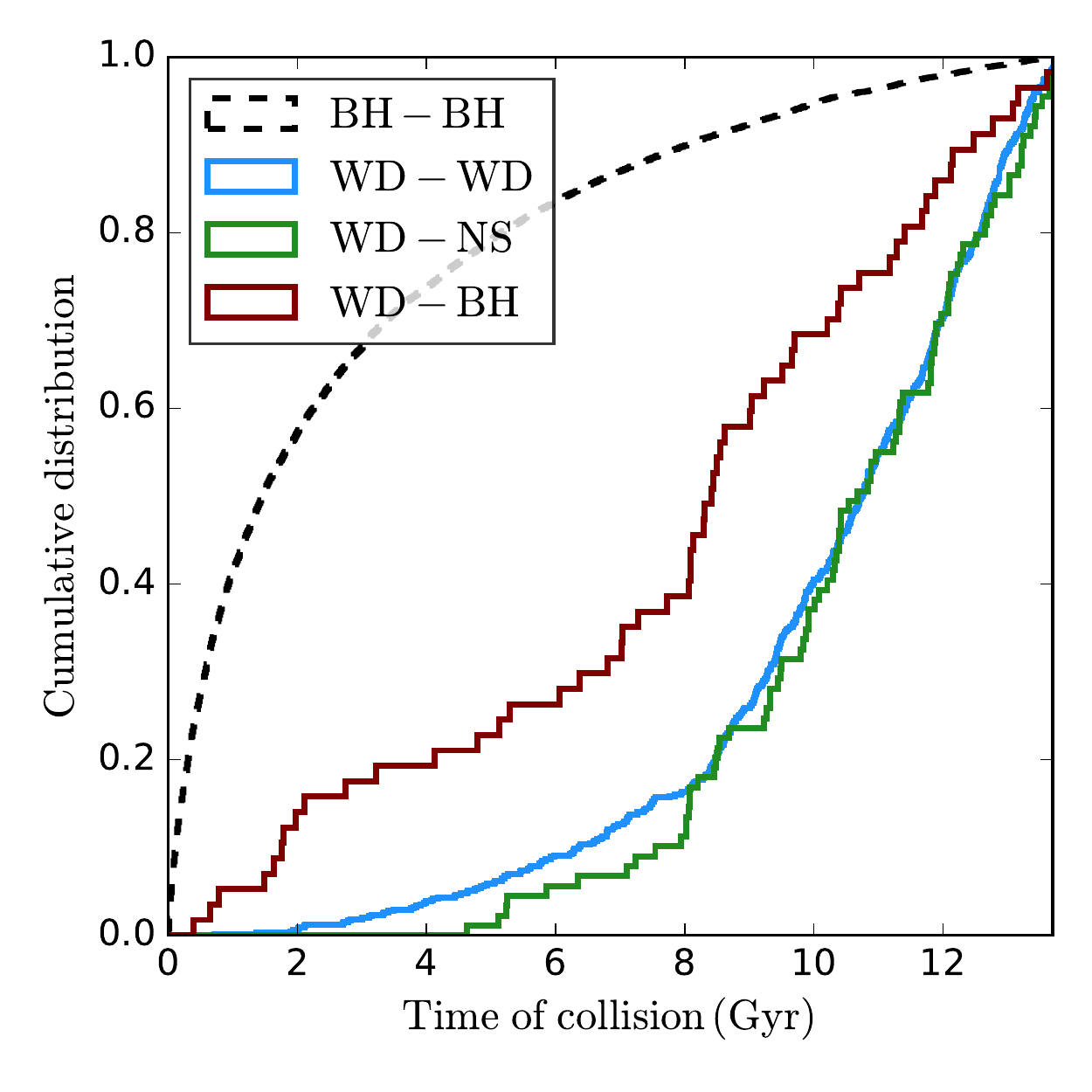}
\caption{\footnotesize \label{fig:WD_collision_times} Cumulative distribution of collision times (relative to birth time of cluster) for all WD--WD (blue), WD--NS (green), and WD--BH (red) collisions, compared to the distribution for binary BH mergers (dashed black). We show here distributions for all collisions/mergers occurring in our complete set of models.}
\end{center}
\end{figure}

In addition to their application to accreting systems such as CVs and AM CVn, WDs are also associated with a number of high-energy events.
In GCs specifically, dynamical collisions of WDs with other stellar populations (particularly other stellar remnants) have been linked to various transients. For example, \citet{SharaHurley2002} showed that dynamical interactions amplify the rates of WD--WD mergers and collisions that may lead to Type Ia SNe. \citet{Ivanova2006} demonstrated similar results regarding the contribution of cluster WDs to the Type Ia SNe rate. WD--NS collisions in GCs have been proposed as a possible mechanism of calcium-rich gap transients observed at high radial off-sets in metal-poor galaxies \citep{Kasliwal2012,Lunnan2017,De2018,Shen2019}. Additionally, tidal disruptions of WDs by BHs (especially IMBHs) in GCs have been examined in several analyses \citep{Rosswog2009,MacLeod2016,Fragleiginskoc2018}. It has been proposed that such WD TDEs serve as viable mechanisms for a number of observed high-energy events \citep[e.g.,][]{Krolik2011,Jonker2013}. In this section, we examine WD--remnant collisions in our models and discuss the implications of these events for a number of possible transients.

In columns 6--8 of Table \ref{table:WDs}, we list the total number of WD--WD, WD--NS, and WD--BH collisions occurring during dynamical encounters (e.g., single--single or binary-mediated encounters) in the characteristic cluster models from before. Because we are focusing here on only those WD collisions that occur dynamically (unlike, e.g., the CVs discussed in previous subsection, which are sensitive to binary evolution processes that complicate potential correlations with various cluster parameters), we can explain the relative rates of these events in the models shown in Table \ref{table:WDs} through simple dynamical arguments. As discussed in Section \ref{sec:results} (see especially Figure \ref{fig:radial_dist}), when large numbers of BHs are present in a cluster, dynamical interactions in less massive stellar populations (e.g., WDs), are less frequent simply because the less massive populations tend to be driven out to larger radial positions due to BH burning. Only when the BH population decreases sufficiently do less massive populations like WDs become dynamically active. Of the four clusters shown in Table \ref{table:WDs}, only the core-collapsed model (\textsc{n8-rv0.5-rg8-z0.01}) has a dynamically insignificant population of BHs (which, in turn, is what allowed this cluster to undergo core-collapse in the first place). As expected, it is in this model that we see the most WD collisions.
Indeed, the radial profiles of various stellar populations are shown for this particular model in the top panel of Figure \ref{fig:radial_dist}. We see clearly from that figure that at late times, the WDs are actually the dominant stellar population by number in the cluster's innermost regions. Thus, we expect core-collapsed clusters (e.g., M15 and NGC 6752) to be the most likely candidates for WD collisions and the  transient electromagnetic signatures associated with these events.

In our full set of 148 simulations, we identify 672 WD--WD collisions (of which 153, 356, and 163 occur through single--single, binary--single, and binary--binary encounters, respectively); 92 WD--NSs (of which 10, 47, and 35 occur through single--single, binary--single, and binary--binary encounters, respectively); and 57 WD--BHs (of which 10, 27, and 20 occur through single--single, binary--single, and binary--binary encounters, respectively). The exact number of single--single versus binary-mediated events depends upon the assumed initial binary fraction. Here, the initial binary fraction is fixed to $5\%$. We reserve for future studies a detailed exploration of the effect of binary fraction on the total number of WD collisions of various channels.

As suggested by Table \ref{table:WDs}, these dynamically-mediated collisions occur most frequently in those models with initial $r_v=0.5\,$pc that undergo core-collapse. Indeed, roughly $70\%$ of all WD collisions occur in the 36 simulations with $r_v=0.5\,$pc, while roughly $25\%$ occur in the 36 simulations with $r_v=1$ and $5\%$ in the remaining simulationss with $r_v \geq 2\,$pc. In Figure \ref{fig:WDcollisions}, we show masses for all collisions of WDs with compact remnant targets.

Furthermore, because dynamically-mediated WD collisions occur most frequently only after a cluster has undergone core-collapse (which typically takes $\sim$Gyrs; see Figure \ref{fig:lum_profiles}), we find WD collisions occur primarily at late times. This is in contrast to, for example, binary BH mergers, which begin to occur as soon as the BHs mass segregate and become dynamically active in the BH-dominated core. In Figure \ref{fig:WD_collision_times}, we show the cumulative distribution of collision times for all types of WD collisions compared to the distribution of merger times for binary BH mergers (which will be discussed in more detail in Section \ref{sec:BBHs}). Here we simply show the distribution of collision times relative to the birth time of the cluster. Cluster birth times are of course not fixed, and a more realistic estimate of the time distribution of WD collisions would need to incorporate a cluster age distribution. Indeed, the time distributions shown in Figure \ref{fig:WD_collision_times} imply that for old clusters with present-day ages $\lesssim6\,$Gyr, WD collisions may have never occurred. We reserve this more detailed analysis for future work and simply emphasize here that, in general, WD collisions preferentially occur in old GCs.

Grouping all models together, we estimate WD--WD collisions occur at a rate of roughly $10^{-7}$ per year in old ($t\gtrsim6\,$Gyr) GCs in a MW-like galaxy (assuming 150 total clusters), while the rates for WD--NS and WD--BH collisions are roughly $10^{-8}$ per year.

As described in Section \ref{sec:method}, we record collisions during close dynamical encounters in the ``physical collision'' limit: in order for two objects to undergo a collision, they must satisfy the requirement $r_p < R_1 + R_2$. However, more distant close encounters that fall in the tidal disruption or even tidal capture regime may also ultimately lead to a ``collision-like'' event.\footnote{Recent work \citep[e.g.,][]{Samsing2018a,Rodriguez2018b} has also shown that GW emission during close fly-bys during fewbody encounters play an important role in the formation of BH binaries and mergers in GCs. It is essential to include post-Newtonian corrections in these resonant encounters to capture this merger channel. However, post-Newtonian effects do not have an analogous effect for WDs, because the characteristic distance for post-Newtonian effects to become a dynamically significant source of energy dissipation for WDs lies within the WDs tidal disruption radius.} The minimum pericenter distance that leads to tidal capture for WD encounters is likely a few times the physical radius \citep[e.g.,][]{SamsingLeighTrani2018}. Hence, the collision rates reported in the above paragraph may underestimate the true number of dynamically-mediated collisions by a factor of a few to perhaps an order-of-magnitude, relative to the ideal where tidal disruptions/captures are included. Even so, these rates are several orders-of-magnitude lower than the observed rates of Type Ia SNe and Ca-rich transients, which are both estimated to occur at rates of roughly a few$\times 10^{-3}$ per year for MW-like galaxies \citep[e.g.,][]{Frohmaier2018}.

In addition to WD collisions arising through dynamical interactions, WDs may also be driven to merger with other compact remnants through binary evolution processes, in particular gravitational inspiral \citep[e.g.,][]{Kremer2015,Shen2015}. These inspiral WD binaries may form either through dynamical exchange encounters or through binary evolution processes (where, for example, a common envelope event brings the WD components close enough to inspiral within a Hubble time). Like the collision events discussed to this point, these WD mergers may similarly contribute to the total rate of high-energy transients \citep[e.g.,][]{SharaHurley2002,Shen2019}. These WD merger events may occur inside the host cluster or outside of the cluster if the WD binary is ejected through dynamical recoil associated with dynamical formation or through natal kicks associated with the formation of a NS/BH companion. Indeed, these various formation channels for WD merger events (collisions during dynamical encounters, in-cluster mergers, and ejected mergers) are analogous to the merger channels that have been discussed at length in various recent papers in the context of binary BH mergers \citep[e.g.,][]{Samsing2018a,D'Orazio2018,Rodriguez2018b,Zevin2018,Kremer2019b}, which we discuss in detail in Section \ref{sec:BBHs}. However, unlike binary BH mergers in GCs, which are essentially driven entirely by dynamical processes \citep[primordial BH mergers constitute $\lesssim5\%$ of all mergers;][]{Rodriguez2018a} and are thus largely insensitive to the initial binary properties assumed, the evolution of WD binaries are influenced less significantly by dynamical interactions, as discussed in Section \ref{sec:accreting_WDs}. This indicates that the exact numbers of these events depends sensitively upon both the assumed primordial binary fraction as well as binary evolution physics that governs the formation of compact WD binaries, especially common envelope physics \citep[e.g.,][]{SharaHurley2002}. We reserve a detailed exploration of the effect these various processes have on WD mergers for a more focused study.

\section{Stellar Collisions}
\label{sec:stellar_collisions}

\begin{figure*}
\begin{center}
 \gridline{\fig{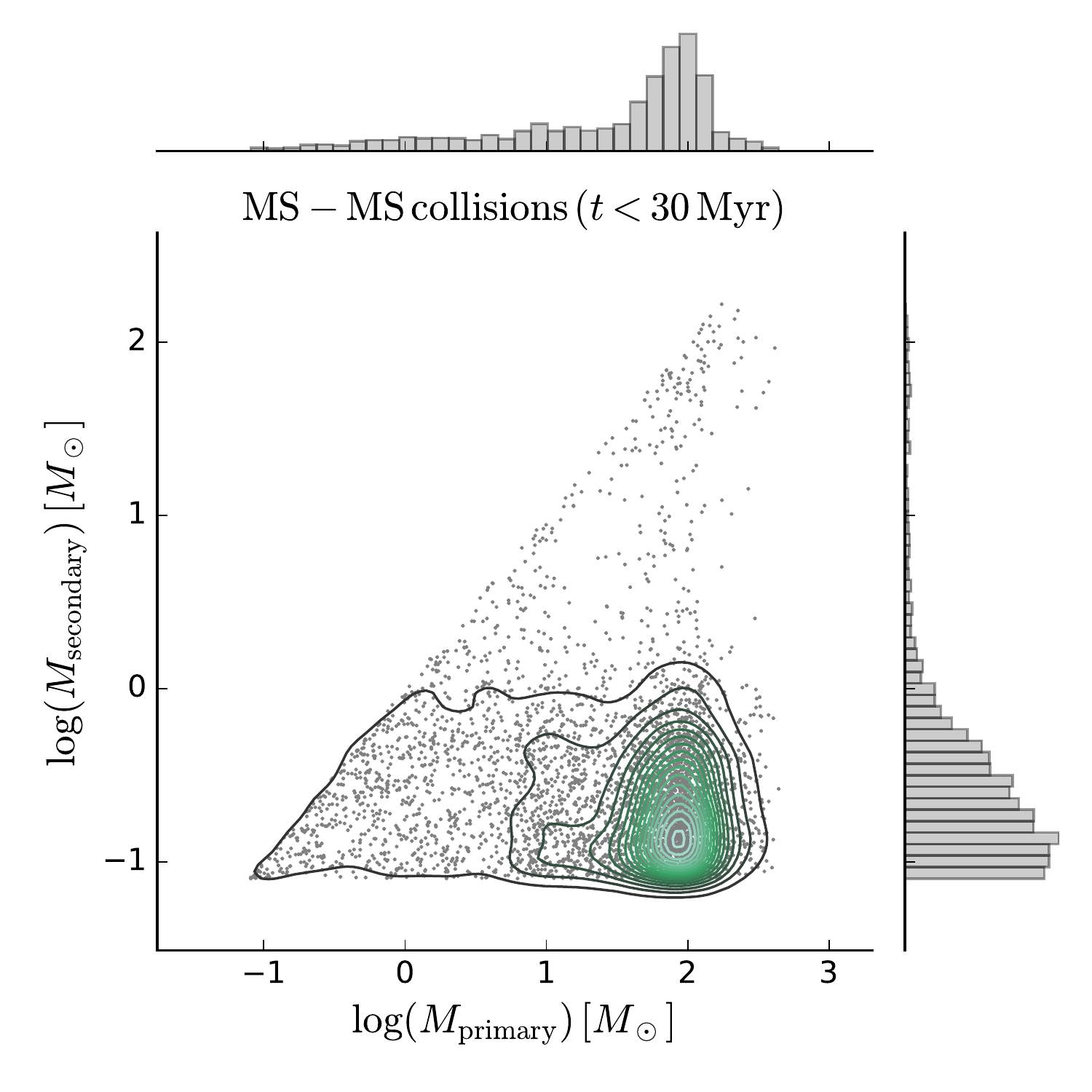}{0.44\textwidth}{(a) MS--MS collisions at early times}
          \fig{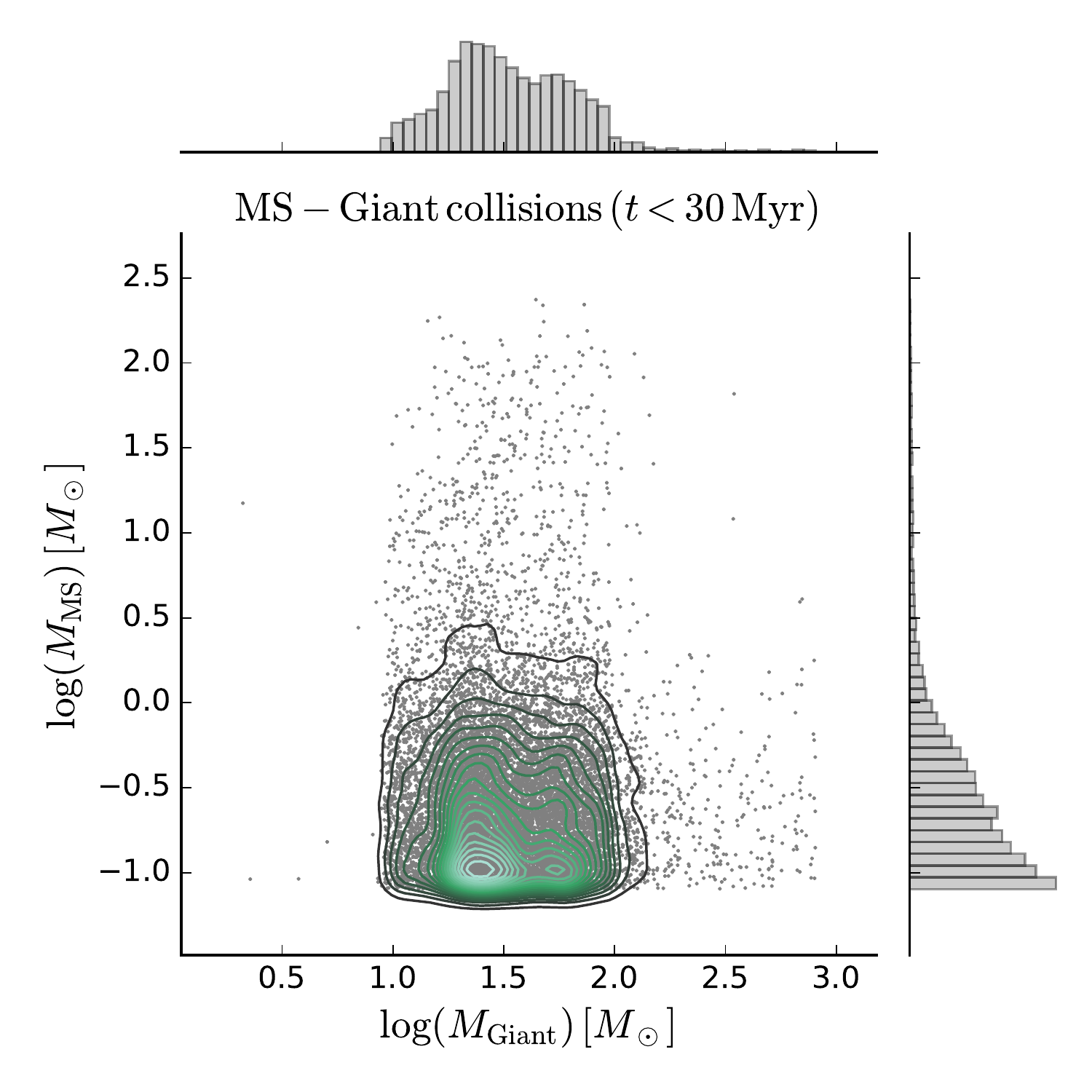}{0.44\textwidth}{(b) MS--giant collisions at early times}}
\gridline{\fig{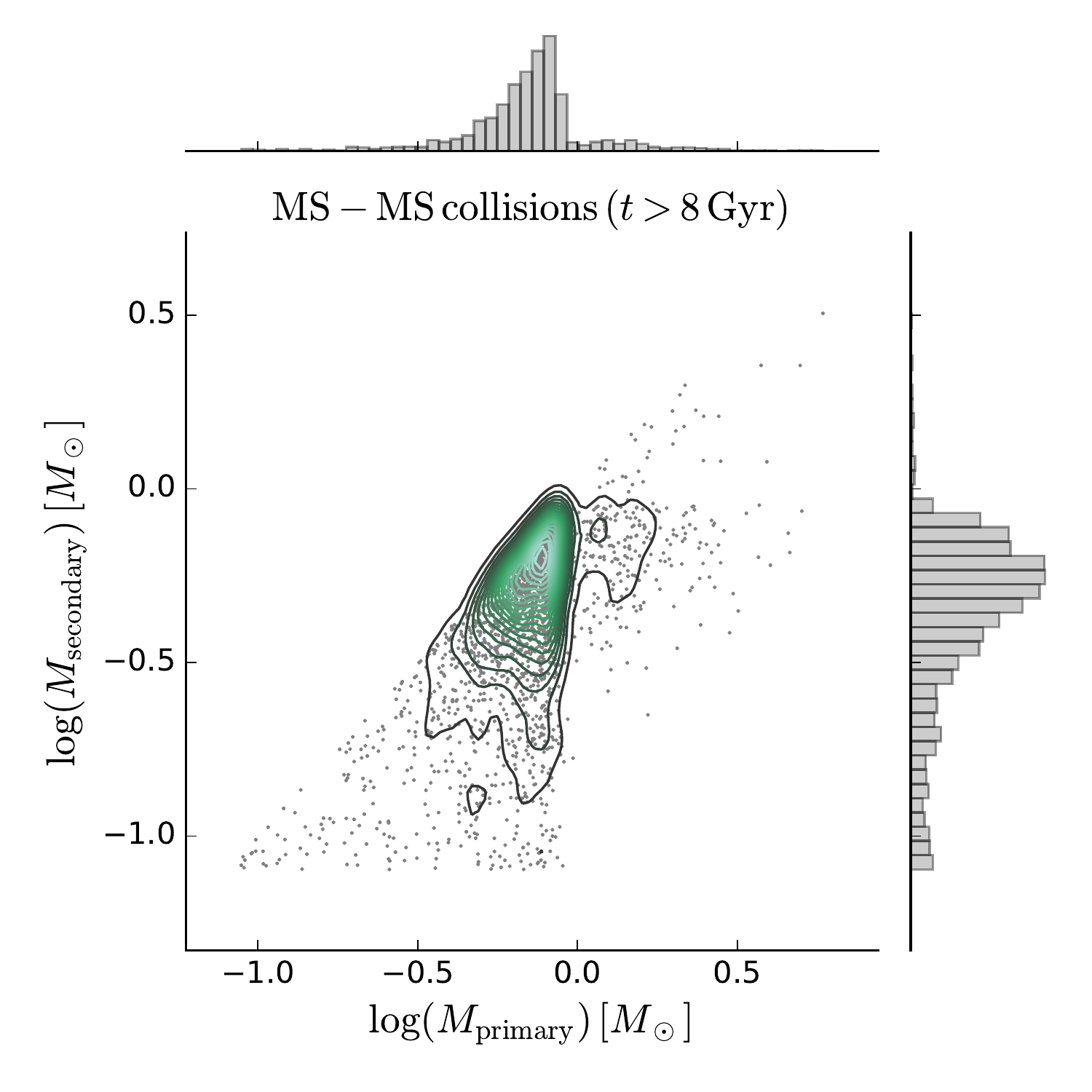}{0.44\textwidth}{(c) MS--MS collisions at late times}
          \fig{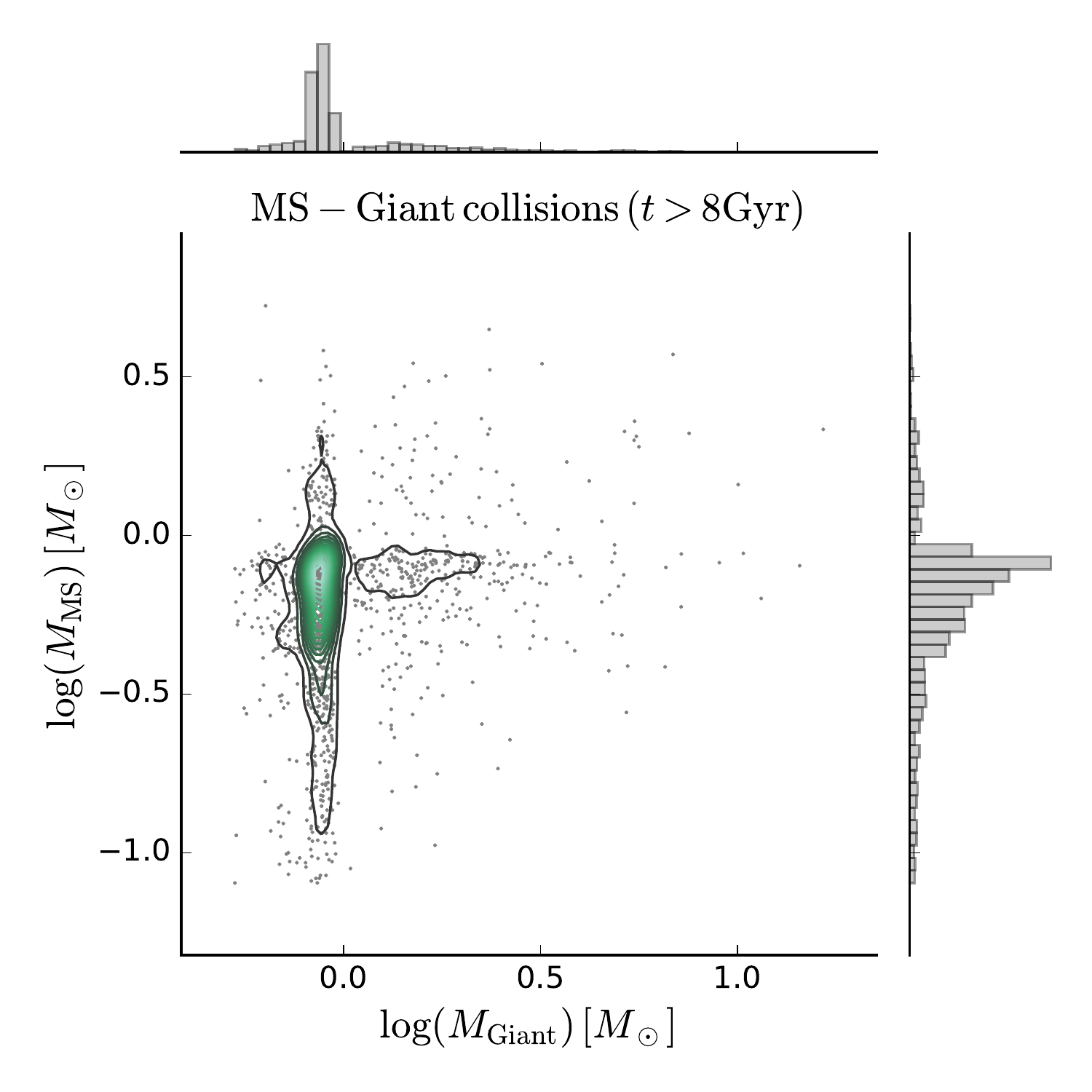}{0.44\textwidth}{(d) MS--giant collisions at late times}}
\caption{\footnotesize All luminous star collisions occurring in our simulations separated by collision component types and by time. Here, `early times' refers to $t<30\,$Myr after cluster formation while `late times' refers to $t>8\,$Gyr.}
\label{fig:stellar_collisions}
\end{center}
\end{figure*}

As hinted at in the previous section in the context of WDs, stellar collisions can occur at significant rates in dense systems like GCs. In addition to the WD collisions discussed in the previous section, which may lead to Type Ia SNe or Ca-rich transients, stellar collisions involving BHs and NSs may lead to a variety of distinct transients with implications for a number of astrophysical phenomena. For example, collisions of BHs/NSs with MS stars may lead to luminous flares and possibly ultra-long gamma ray bursts \citep[e.g.,][]{Perets2016,Kremer2019c,Fragleipernkoc2019} and collisions of BHs/NSs with giants may lead to ultra-compact X-ray binaries with WD donors \citep[e.g.,][]{Ivanova2005,Ivanova2010,Kremer2019c}.

Although collisions between compact objects and luminous stars may be most exciting for the study of high-energy transients, these types of collisions tend to be rare simply because compact objects, particularly BHs and NSs, constitute only a small fraction of the total population in a typical cluster (see Figure \ref{fig:radial_dist} and Table \ref{table:models}). More common are collisions where both objects are luminous stars (i.e. MS--MS collisions or MS--giant collisions), events that have been well-studied theoretically \citep{HillsDay1976,Bacon1996,Lombardi2002,Fregeau2007,Leigh2011,Antognini2016}. In particular, collisions between luminous stars may have important implications for the formation of blue stragglers (to be discussed in detail in Section \ref{sec:BSs}). In some cases, the post-collision evolution of these events may be observed as optical transients \citep[e.g.][]{Tylenda2011,MacLeod2017,Metzger2017}. In Table \ref{table:stellar_collisions} in the Appendix, we list the total number of various stellar collision types (i.e., MS--MS, MS--giant, MS--BH, etc.) occurring in each simulation of this study.

In Figure \ref{fig:stellar_collisions}, we show masses of all collisions between luminous stars occurring in our simulations. In the top two panels (a and b), we show those collisions that occur early in the host cluster's evolution ($t<30\,$Myr), while in the bottom panels (c and d), we show late-time collisions ($t>8\,$Gyr). The left panels show MS--MS collisions and the right panels show MS--giant collisions.

At early times, when massive stars are still present in the cluster, $100\,M_{\odot}$--$0.1\,M_{\odot}$ MS--MS collisions and $30\,M_{\odot}$--$0.1\,M_{\odot}$ giant--MS collisions are most typical. These typical values are determined by a combination of the stellar IMF and gravitational focusing. Stars of some specific mass $M_1$ and radius $R_1$ will undergo collisions with ``target'' stars of mass $M_2$ and radius $R_2$ at a rate given by:

\begin{equation}
\label{eq:coll_rate}
    \Gamma_{\rm{coll}} = n_2 \pi r_p^2 v_\infty \Bigg(1+\frac{2G(M_1+M_2)}{r_p v_\infty ^2} \Bigg) N_1,
\end{equation}

where $n_2$ is the number density of the targets, $r_p=R_1+R_2$ is the minimum pericenter distance that leads to a collision, $v_\infty$ is the relative velocity of the pair of objects at infinity, and $N_1$ is the total number of stars having the selected mass of interest, $M_1$.

As described in Section \ref{sec:method}, we adopt the IMF of \citet{Kroupa2001}, which peaks at roughly $0.1M_{\odot}$, a typical M-dwarf. For this IMF, M-dwarfs dominate over $100\,M_{\odot}$ stars by a factor of roughly 1000. By using Equation \ref{eq:coll_rate} and adopting parameters typical of young massive clusters, we can estimate the relative rates of various collision types to make sense of the trends exhibited in Figure \ref{fig:stellar_collisions}. 

We assume a typical cluster core radius of 1 pc, a typical $v_\infty=10\,\rm{km/s}$, and assume that $0.1\,M_{\odot}$ and $100\,M_{\odot}$ stars have radii of roughly $0.1R_{\odot}$ and $30R_{\odot}$, respectively. We also assume that, for a typical cluster containing $8\times10^5$ stars at birth, roughly $10^5$ stars are M-dwarfs ($M\sim0.1M_{\odot}$) while roughly 100 are high-mass stars with $M\sim 100M_{\odot}$. In this case, from Equation \ref{eq:coll_rate}, the rates of $0.1\,M_{\odot}$--$0.1\,M_{\odot}$, $0.1\,M_{\odot}$--$100\,M_{\odot}$, and $100\,M_{\odot}$--$100\,M_{\odot}$ collisions in young massive clusters are roughly $0.02\,\rm{Myr}^{-1}$, $2\,\rm{Myr}^{-1}$, and $0.007\,\rm{Myr}^{-1}$, respectively. Thus, as shown in the top panels of Figure \ref{fig:stellar_collisions}, $0.1\,M_{\odot}$--$100\,M_{\odot}$ are most common. Although the collision cross section for $100\,M_{\odot}$--$100\,M_{\odot}$ is higher, these events are limited by the relatively low number of objects. Also, though the number of possible targets is largest for $0.1\,M_{\odot}$--$0.1\,M_{\odot}$ collisions, the rate here is limited by the relatively small cross section and the decreased effect of the gravitational focusing term.

In our simulations, we assume no primordial mass segregation. Thus at $t=0$, all stars of all masses are equally mixed within the cluster radially. However, a number of recent studies \citep[e.g.,][]{Baumgardt2008,Subr2008,Habibi2013,Pang2013,Pavlik2019}, suggest that primordial mass segregation may be a more appropriate initial condition. In this case, because the most massive stars would be preferentially found closer to the cluster's center where densities are higher, collisions where both components are massive stars may dominate the overall rate. To this point, as shown in the bottom two panels of Figure \ref{fig:stellar_collisions}, at late times ($t>8\,$Gyr), when the clusters have evolved sufficiently toward a mass-segregated configuration, collisions of equal mass components become most common. In this case, the most massive stars (for $t\sim10\,$Gyr, the MS turnoff mass is roughly $0.8M_{\odot}$) are preferentially found in the dense core of the cluster and are therefore more likely to undergo collisions through dynamical encounters.

We note that a detailed, incompleteness-corrected, comparison of mass segregation ($\Delta$) in all our models to $\Delta$ observed in 50 MW GCs shows that our simulations accurately reproduce the $\Delta$ distribution in GCs \citep[see Figure 3 of][]{Weatherford2019}, suggesting that primordial mass segregation may be unnecessary. We reserve a more detailed examination of the effects of primordial mass segregation for a later study.

As discussed in e.g., \citet{Spera2019,Banerjee2019}, massive stellar mergers occurring through binary evolution may have important consequences for the formation of massive BHs, specifically BHs lying within the so-called upper mass-gap expected from (pulsational) pair instability SNe \citep[e.g.,][]{Belczynski2016b,Woosley2016,Spera2017}. In principle, dynamically-mediated stellar collisions of massive stars in GCs may have similar implications for BH formation. Indeed, if appropriate conditions are met (i.e., if a cluster is sufficiently dense at early times), stellar collisions may lead to a runaway scenario resulting in the formation of a very massive star, and ultimately, an IMBH \citep[e.g.,][]{PortegiesZwart2004,Freitag2006,Giersz2015,Mapelli2016}. For now, we simply note that, as motivated by the collision rates demonstrated in Table \ref{table:stellar_collisions} and Figure \ref{fig:stellar_collisions}, collisions may indeed play a role in massive star evolution, with specific applications to BH formation. We will more fully explore the implications of stellar collisions for BH formation in an upcoming paper.

Finally, as shown in Equation \ref{eq:coll_rate}, the overall collision rate depends on the number density of objects, which is specified in our simulations at early times by the initial $r_v$.  Thus, we expect that clusters with smaller $r_v$ (that are more likely to undergo core-collapse by the present day, as discussed in Section \ref{sec:corecollapse}) will feature more stellar collisions than models with higher initial $r_v$.
To quantify, in the $r_v=0.5\,$pc simulation, \textsc{n8-rv0.5-rg8-z0.01}, which has undergone core-collapse by $t=12\,$Gyr (see Figure \ref{fig:lum_profiles}), we identify 811 MS--MS and 3044 MS--giant collisions at early times and 886 MS--MS and 121 MS--giant collisions at late times. In contrast, in the $r_v=2\,$pc model, \textsc{n8-rv2-rg8-z0.01} -- which still retains a large population of BHs at late times and does not undergo core-collapse -- we identify only 3 MS--MS and 45 MS--giant collisions at early times, but 48 MS--MS and 16 MS--giant collisions at late times. Hence, if stellar-collisions in young clusters indeed play a role in BH formation, we expect this effect to be most pronounced in the clusters that are most dense initially and ultimately undergo core-collapse. Similarly, because these clusters also exhibit higher stellar collision rates at late times, we expect clusters with smaller $r_v$ to feature an increased number of blue straggler stars. We discuss the specific application to blue stragglers in the following section.

\section{Blue stragglers}
\label{sec:BSs}

\begin{figure}
\includegraphics[width=0.9\columnwidth]{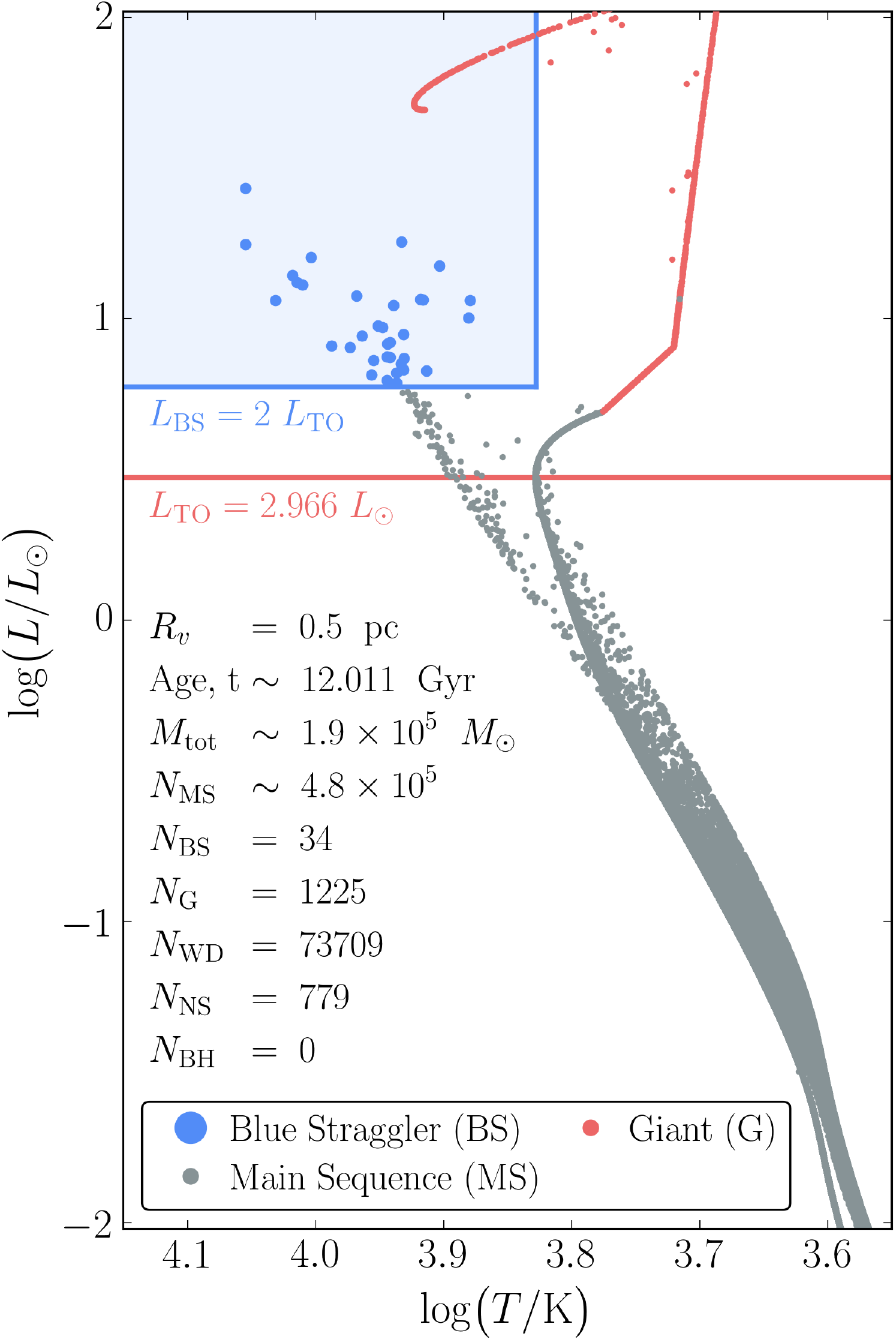}
\caption{\footnotesize Hertzsprung-Russell diagram of the core-collapsed model \textsc{n8-rv0.5-rg8-z0.01} at $t=12\,$Gyr illustrating the blue straggler (BS) selection algorithm. Each dot represents a single or binary star (all binaries are considered unresolved), with BSs colored in blue, giants in red, and the main sequence (MS) in gray. Note that for binaries, the summed luminosity and luminosity-weighted mean temperature are plotted. To select BSs, the turnoff is first defined as the luminosity, $L_{\rm{TO}}$, corresponding to the point on the MS with the highest median temperature, $T_{\rm{TO}}$. $L_{\rm{TO}}$ is indicated by the red horizontal line. BSs are then defined as any single or binary containing a MS star where the luminosity and temperature (defined above for binaries) exceed $L_{\rm{BS}} = 2 L_{\rm{TO}}$ and $T_{\rm{TO}}$, respectively (see the blue-shaded region).}
\label{fig:hrdiag_zoom}
\end{figure}

Blue stragglers (BSs) are hydrogen-burning stars that are photometrically bluer and brighter than the MS turnoff for stars of similar ages \citep{Sandage1953}. A star may become a BS when it undergoes one of several possible stellar interactions that lead to an increase in the star's mass. Such interactions could include accretion of material from a binary companion during Roche-lobe overflow or a physical collision with another star, as described in Section \ref{sec:stellar_collisions}. The latter channel is expected to become important in dense stellar systems like open and globular clusters where stellar collisions are common. BSs are well-observed in nearly all globular clusters in the Milky Way \citep[e.g.,][]{Piotto2003,Ferraro2012} and also in many open clusters \citep[e.g.,][]{MathieuGeller2009}. A number of analyses have explored the formation channels for BSs in clusters \citep[e.g.][]{Ferraro2012,Chatterjee2013b,Hypki2017}. \citet{Chatterjee2013b} noted that in dense GCs (central densities $\gtrsim 10^3\,\rm{pc}^{-3}$), stellar collisions appear to be the dominant formation channel for BSs, while for lower density open clusters, binary mass transfer appears to dominate. This is consistent with recent observational work \citep[e.g.,][]{Geller2011,Gosnell2019} which showed that mass transfer indeed appears to be the dominant formation mechanism in open clusters. \citet{Ferraro2012} noted that BSs can be used as probes of cluster dynamical evolution. Specifically, \citet{Ferraro2012} pointed out that GCs can be grouped into distinct dynamical age families based on their BS radial distributions. Thus BSs can be used as important observational constraints of cluster properties.

We use the following procedure to count the total number of BSs in our models: First, the MS turnoff is defined as the luminosity, $L_{\rm{TO}}$, corresponding to the point on the MS branch with the highest median temperature, $T_{\rm{TO}}$. BSs are then defined as any single or (unresolved) binary containing a MS star where the total luminosity and temperature (luminosity-weighted mean temperature for binaries) exceed $L_{\rm{BS}} = 2 L_{\rm{TO}}$ and $T_{\rm{TO}}$, respectively. Figure \ref{fig:hrdiag_zoom} shows a zoomed-in portion of the HR diagram of simulation \textsc{n8-rv0.5-rg8-z0.01} at $12\,$Gyr, shown earlier in the upper left panel of Figure \ref{fig:hrdiag_multi}. In this figure, the BS selection criterion is visually represented by the blue-shaded region, within which any single or binary containing a MS star is counted as a blue straggler. Column 11 of Table \ref{table:WD-pulsar-BS-full} lists the mean number of BSs in each simulation at late times while Table \ref{table:BSs} lists the total number of BSs in the four characteristic models from previous sections.

\begin{figure}
\begin{center}
\includegraphics[width=\columnwidth]{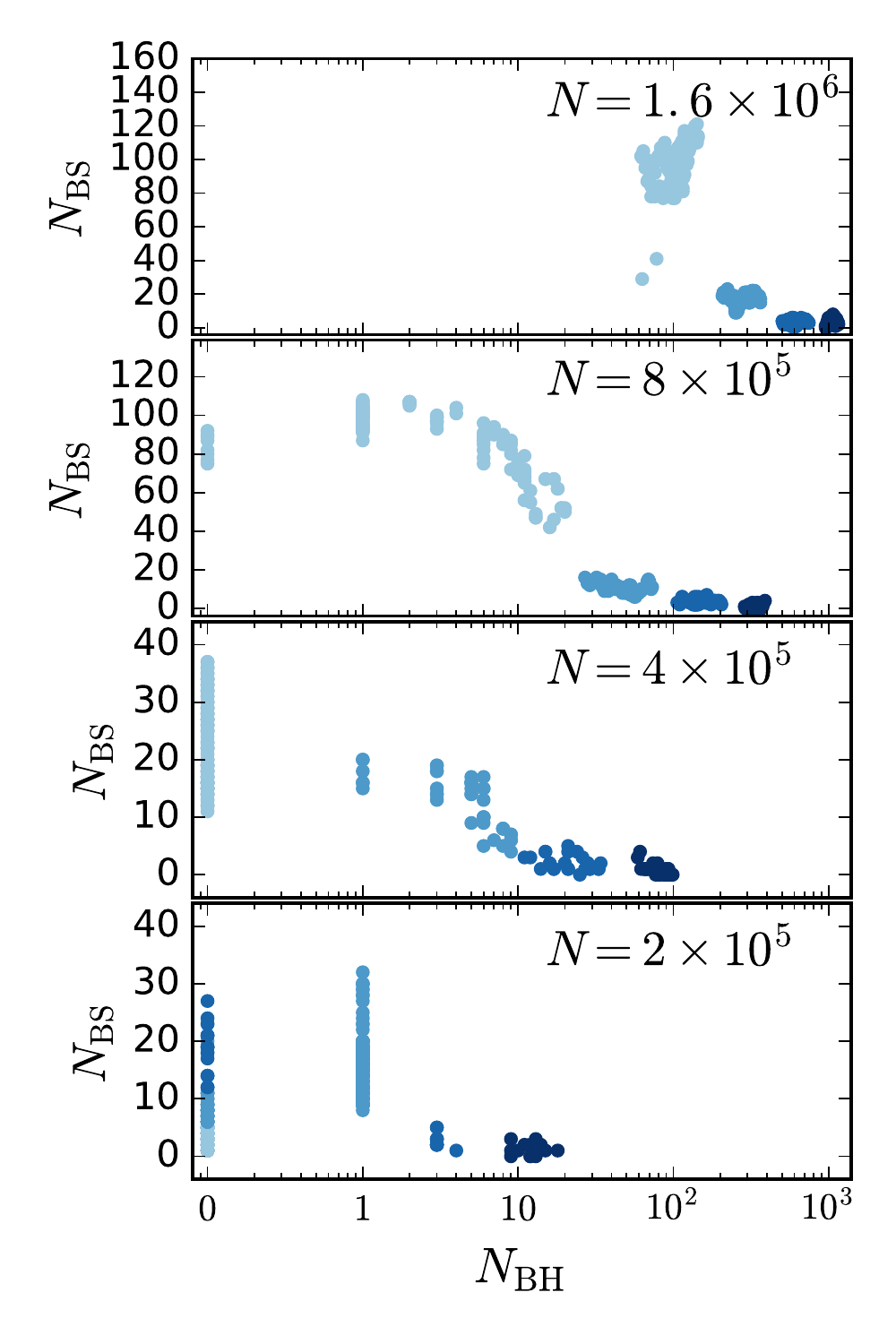}
\caption{ \footnotesize \label{fig:BS_vs_BH} Number of BSs ($N_{\rm{BS}}$) versus total number of retained BHs ($N_{\rm{BH}}$) for snapshots in the age range 10--13 Gyr. Shown here are models with $0.1Z_{\odot}$, the most typical metallicity value for old GCs in the MW (see Figure \ref{fig:Harris}). From top to bottom, we show models of decreasing total particle number. Different shades of blue denote different initial $r_v$, as in previous figures.}
\end{center}
\end{figure}

Figure \ref{fig:BS_vs_BH} shows the total number of BSs versus the total number of BHs for all snapshots with ages in the range 10--13 Gyr from $Z=0.1Z_{\odot}$ models (chosen simply to reflect the median metallicity of MW GCs; see Figure \ref{fig:Harris}). Here, we show only low metallicity models simply because these are most representative of the old GCs observed in the MW. From top to bottom, the different panels correspond to models with $N=1.6\times10^6$, $8\times10^5$, $4\times10^5$, and $2\times10^5$, respectively. A clear anticorrelation exists between the number of BSs and the number of retained BHs. Furthermore, models with smaller initial $r_v$ have, on average, more BSs than models with larger initial $r_v$.

Previous analyses \citep[e.g.,][]{Ferraro2012,Alessandrini2016,Ferraro2019} have noted that BS populations can be used to trace clusters' dynamical ages. Here, we demonstrate this same result with an important addendum: the link between BSs and dynamical age is intertwined with the clusters' evolving BH populations. As discussed in Section \ref{sec:results}, the most dynamically-evolved clusters retain the fewest BHs and are most likely to be found in more centrally-concentrated or even core-collapsed configurations. Through mechanisms identical to those relevant for MSP formation (Section \ref{sec:pulsars}) and WD collisions (Section \ref{sec:WDcollisions}), these dynamically-evolved clusters where BH burning is dynamically insignificant facilitate an increased rate of dynamical interactions of MS stars, and thus, produce more BSs.

\begin{deluxetable}{l|c|c}
\tabletypesize{\scriptsize}
\tablecolumns{3}
\tablewidth{0pt}
\tablecaption{Blue Stragglers in Four Representative GCs\label{table:BSs}}
\tablehead{ & \colhead{$M_{\rm{tot}}\,(M_{\odot})$} & \colhead{Blue Stragglers}}
\startdata
Typical        & $2.3\times10^5$ &   0--2   \\
Core-collapsed & $1.9\times10^5$ &  28--91  \\
Low-mass       & $5\times10^4$   &   0--2   \\
High-mass     & $10^6$   &   2--16
\enddata
\tablecomments{Ranges in total number of blue stragglers for all snapshots in the age range 10--13 Gyr for the four characteristic clusters defined as in Table \ref{table:XRBs}.}
\end{deluxetable}

\section{Binary black hole mergers}
\label{sec:BBHs}

As the catalog of GW detections of merging binary BHs (BBHs) continues to grow \citep{LIGO2018a}, there is mounting evidence suggesting that dynamical interactions in GCs play a prominent role in the formation of merging BBHs in the local universe. In particular, key features of the LIGO/Virgo detections made to date -- including the masses, spins, and detection rates -- point toward dynamical origins for at least some BBH mergers. The existence of a significant cluster dynamics merger channel is further motivated by observational and theoretical evidence showing that GCs similar to those observed in the MW likely host large populations of BHs at present, as discussed in Section \ref{sec:results}. Of course, the story is far from complete and many other BBH merger channels have been proposed, including isolated evolution of high-mass stellar binaries \citep[e.g.,][]{Dominik2012,Dominik2013,Belczynski2016a,Belczynski2016b}, GW capture of primordial BHs \citep[e.g.,][]{Bird2016,Sasaki2016}, secular interactions in hierarchical triple systems \citep[e.g.,][]{AntoniniRasio2016,Antonini2017,Silsbee2017, Hoang2018,Leigh2018,RodriguezAntonini2018,Fragkoc2019,Fragleipern2019}, and dynamical interactions within the disks of active galactic nuclei \citep[e.g.,][]{Bartos2017,Yang2019}.

In this Section, we discuss the formation of merging BBHs in this new set of cluster simulations. We specifically explore how the rates and properties of BBH mergers vary with cluster $r_v$. In Section \ref{sec:merger_channels}, we discuss specific sub-channels for merging BBHs in GCs, including (for the first time in \texttt{CMC} simulations) the contribution of single--single capture mergers. In Section \ref{sec:BBHfits}, we examine how the BBH merger rate scales with cluster mass. In Section \ref{sec:rates}, we calculate the cosmological rates of BBH mergers while we explore in Section \ref{sec:BBH_properties} how BBH properties vary with the simulation parameters considered in this study.

\subsection{Dynamical merger channels}
\label{sec:merger_channels}

As explored in a number of recent analyses, BBH mergers are expected to occur through four distinct dynamical channels in dense star clusters \citep{Samsing2018a,D'Orazio2018,Rodriguez2018b,Zevin2018,Kremer2019b}. Each of these channels is expected to produce BBHs with distinct GW-frequency and eccentricity distributions. We summarize each of these channels and their main features below:

\textit{Ejected mergers:} As BBHs undergo hardening encounters in the core of a cluster, they receive dynamical recoil kicks of magnitudes that scale with the binary orbital velocity \citep[e.g.,][]{Rodriguez2016a}. Thus, as BBHs are hardened to increasingly compact orbital separations, they recoil at higher velocities, until, eventually, the dynamically-attained recoil velocity exceeds the escape velocity of the cluster and the BBH is ejected. Depending upon the orbital parameters at the time of ejection, such a BBH may inspiral and merge due to GW emission within the age of the universe. Such a binary is labeled an ``ejected merger.'' In a typical cluster, these mergers constitute roughly 50\% of all BBH mergers.

\textit{In-cluster 2-body mergers:} Binaries still retained in their host clusters that merge  \textit{between} resonant dynamical encounters (henceforth referred to as the ``two-body merger'' channel). In a typical cluster, two-body mergers consitute roughly 35\% of all BBH mergers.

\textit{In-cluster fewbody mergers:} Binaries that merge through gravitational capture \textit{during} resonant encounters (henceforth referred to as the ``fewbody capture'' channel). In a typical cluster, fewbody capture mergers constitute roughly 10\% of all BBH mergers, with roughly 5\% occuring through binary--single and binary--binary encounters, respectively.

\textit{In-cluster single--single captures:} Most recently, \citet{Samsing2019_singlesingle} pointed out that BBHs also form through GW-capture in single--single BH encounters in clusters. In a typical cluster, ``single--single capture mergers'' constitute roughly 5\% of all BBH mergers.

All of the above dynamical channels produce BBHs with unique and potentially distinguishable properties that should, in principle, be detectable as GW sources by LIGO/Virgo, as well as lower frequency third-generation GW detectors such as LISA \citep{LISA2017}, DECIGO \citep{DECIGO2011,DECIGO2018}, and Tian Qin \citep{TianQin2016}. We list the total number of BBH mergers catalyzed by each of these channels in Table \ref{table:BBH_mergers} in the Appendix. 

\subsection{Average number of mergers per cluster}
\label{sec:BBHfits}

\begin{figure}
\begin{center}
\includegraphics[width=0.9\columnwidth]{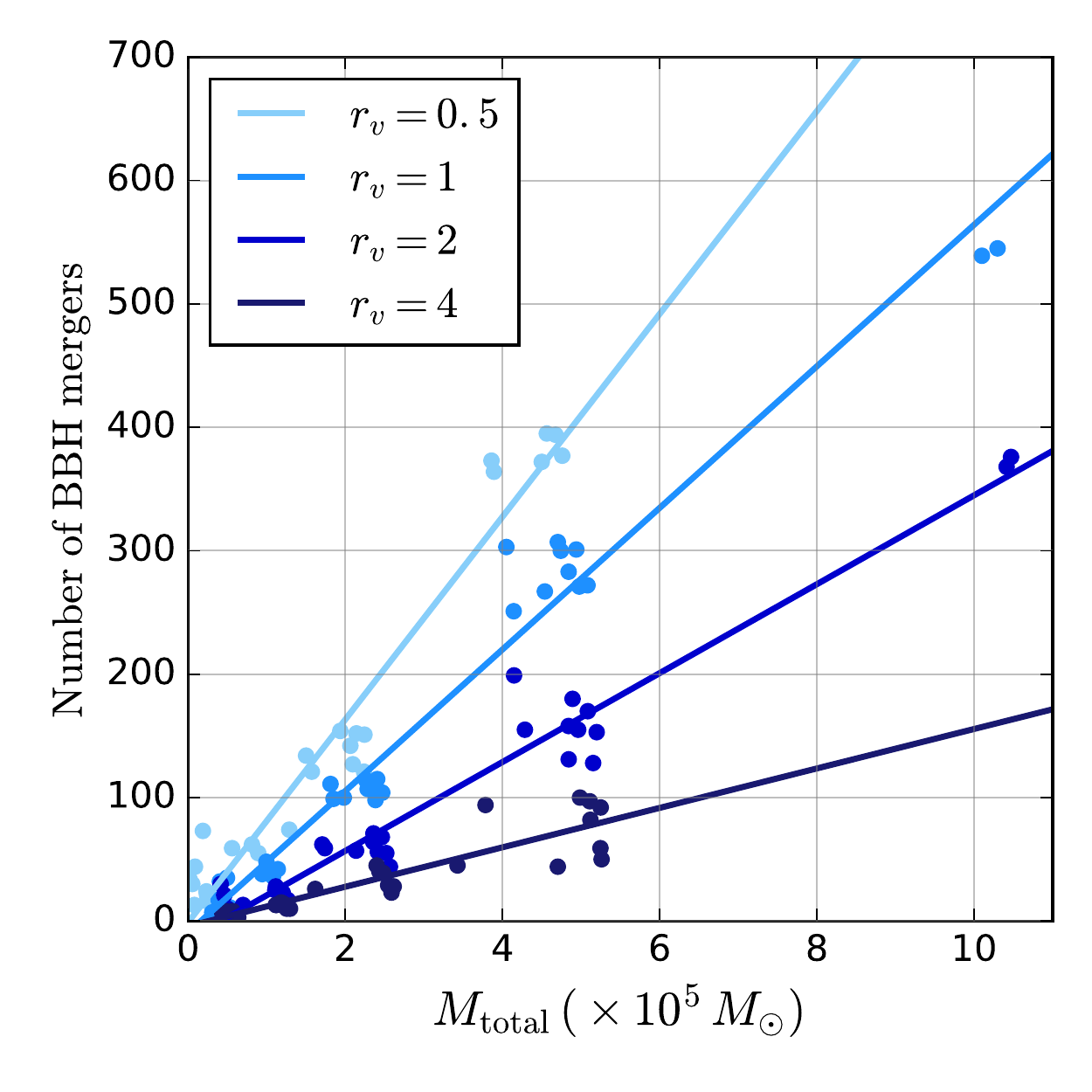}
\caption{\footnotesize \label{fig:BBHfits} Total number of BBH mergers versus final cluster mass for all simulations separated by $r_v$ (various shades of blue as in previous figures). Solid lines show the linear fit to the data for each $r_v$ as described in the text, and as given by Equation \ref{eq:best_fits}.
}
\end{center}
\end{figure}

\begin{deluxetable}{c|ccc|c|c}
\tabletypesize{\scriptsize}
\tablecaption{Average number of BBH mergers per cluster for various initial virial radii computed by integrating over the cluster mass function as described in Section \ref{sec:BBHfits}. \label{table:BBHs}}
\tablehead{
    \colhead{$r_v$} &
    \multicolumn{3}{c}{In-cluster} &
    \colhead{Ejected} & 
    \colhead{Total} \\
    \colhead{} & 
    \colhead{s--s} & 
    \colhead{Fewbody} &
    \colhead{2-body} & 
    \colhead{} & 
    \colhead{}
}
\startdata
0.5 & 26 & 85 & 336 & 299 & 746 \\
1 & 25 & 55 & 233 & 196 & 510 \\
2 & 9 & 35 & 123 & 140 & 308 \\
4 & 4 & 17 & 49 & 70 & 140 \\
\enddata
\end{deluxetable}

The models considered in this study have present-day cluster masses of up to roughly $10^6\,M_{\odot}$. Although this is appropriate for modeling only those clusters observed in the MW (see Figure \ref{fig:Harris}), the cluster mass function for the full population of clusters in the local universe is expected to extend up to larger masses \citep[e.g.,][]{Harris2014,El-Badry2018}. In order to estimate a realistic cosmological rate of BBH mergers in GCs, we must take into account the full cluster mass function. We do so by adopting a method similar to that of \citet{Rodriguez2015a} and \citet{Rodriguez2016a}, as summarized below.

As described in \citet{Rodriguez2015a}, the number of mergers per cluster scales roughly linearly with the total cluster mass. We show this relation for all cluster models in Figure \ref{fig:BBHfits}, separating clusters by initial $r_v$, as in previous figures.
For all models of a given $r_v$, we fit the $N_{\rm{merger}}-M_{\rm{tot}}$ relation shown in Figure \ref{fig:BBHfits} using a linear regression, as in \citet{Rodriguez2015a}.
The best-fit curves for the $N_{\rm{merger}}-M_{\rm{tot}}$ relation shown in Figure \ref{fig:BBHfits} are given for each value of $r_v$ by:

\begin{multline}
\label{eq:best_fits}
    N_{\rm{merger}} = \begin{cases}
    82 \times \Big(\frac{M_{\rm{tot}}}{10^5\,M_{\odot}} \Big) \,;\,r_v=0.5\,\rm{pc}\\
    57 \times \Big(\frac{M_{\rm{tot}}}{10^5\,M_{\odot}} \Big) \,;\,r_v=1\,\rm{pc}\\
    36 \times \Big(\frac{M_{\rm{tot}}}{10^5\,M_{\odot}} \Big) \,;\,r_v=2\,\rm{pc}\\
    16 \times \Big(\frac{M_{\rm{tot}}}{10^5\,M_{\odot}} \Big) \,;\,r_v=4\,\rm{pc}\\
    \end{cases}
\end{multline}

In order to compute the average number of BBH mergers per cluster of a given $r_v$, we then integrate each of the linear relations over a normalized cluster mass function from 0 to $2\times10^7\,M_{\odot}$. As in \citet{Rodriguez2015a}, we assume a log-normal distribution for the cluster mass function with mean $\log M_0 = 5.54$ and width $\sigma_M = 0.52$, based on the GC luminosity functions described in \citet{Harris2014} and assuming a mass-to-light ratio of 2 \citep{Bell2003}. In Table \ref{table:BBHs}, we list the average number of BBH mergers per cluster for each $r_v$. In columns 2--5, we distinguish between BBH mergers occurring through the various channels described in Section \ref{sec:merger_channels}. As shown, the number of mergers occurring through each formation channel (and the total number of mergers) varies inversely with initial $r_v$. Clusters with smaller $r_v$ are denser (see Section \ref{sec:results}), and therefore feature a higher rate of interactions that form BBHs. This is consistent with the results shown in a number of previous analyses \citep[e.g.,][]{Rodriguez2016a,Zevin2018,Choksi2018}. It is clear from Section \ref{sec:results} that an initial $r_v$ of 0.5 pc is appropriate for a fraction of observed clusters in the MW (specifically, those that are most centrally concentrated at present). Thus, in order to estimate a realistic BBH merger rate from clusters, models with $r_v=0.5\,$pc must be incorporated.

\subsection{Merger rates}
\label{sec:rates}

\begin{figure}
\begin{center}
\includegraphics[width=\columnwidth]{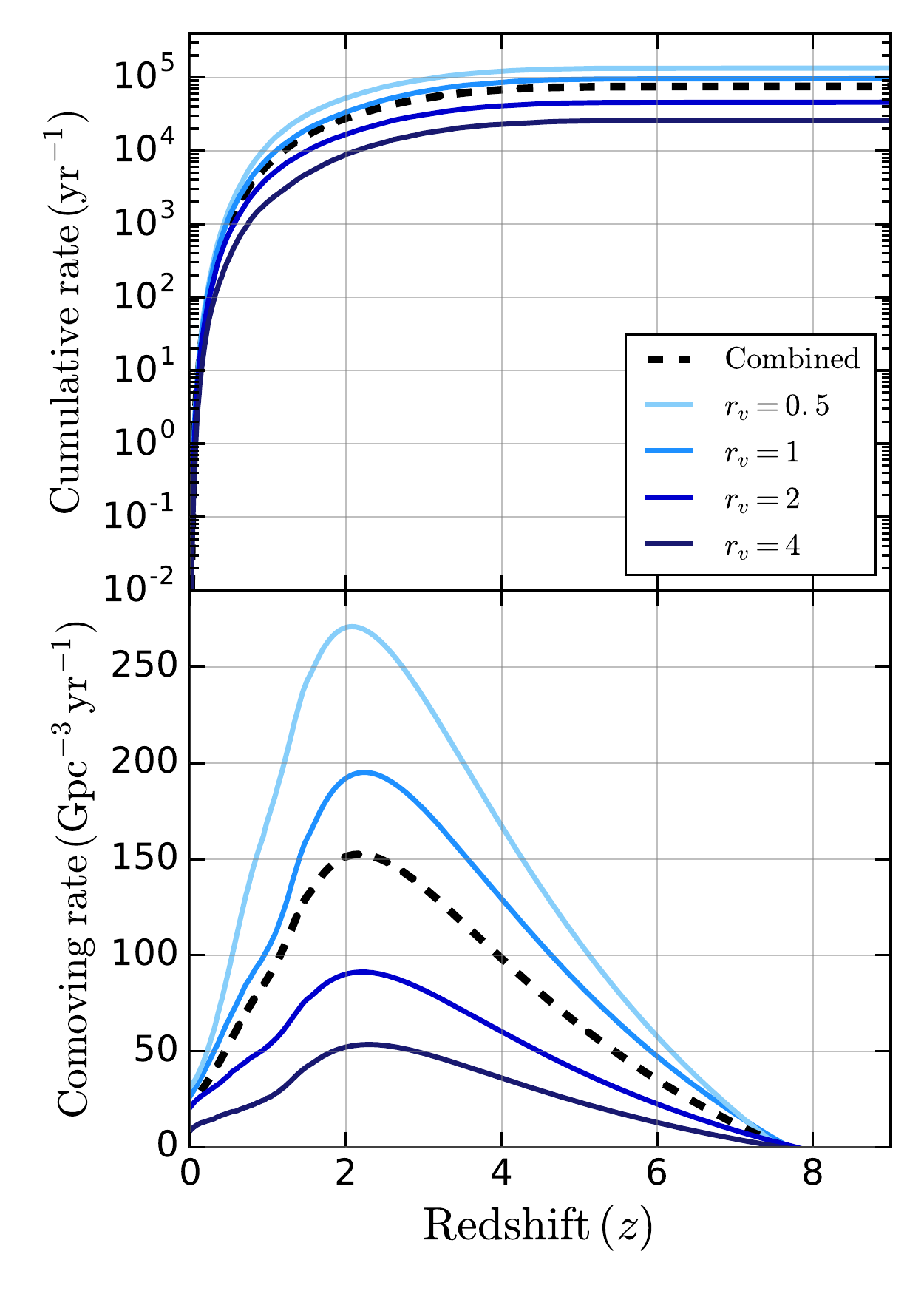}
\caption{\footnotesize \label{fig:BBHrates} Cumulative and comoving BBH merger rates for clusters, colored by initial virial radius (blue curves). The dashed black curve shows the total combined merger rate.}
\end{center}
\end{figure}

In order to calculate the cosmological rate of BBH mergers, we adopt a method similar to \citet{Rodriguez2015a}. The cumulative merger rate is given by:

\begin{equation}
    R(z) = \int_0^z \mathcal{R}(z^\prime) \frac{dV_c}{dz^\prime}(1+z^\prime)^{-1}dz^\prime,
\end{equation}
where $dV_c/dz$ is the comoving volume at redshift $z$ and $\mathcal{R}(z)$ is the comoving (source) merger rate. The comoving rate is given by

\begin{equation}
\label{eq:comoving_rate}
\mathcal{R}(z) = f \times \rho_{\rm{GC}} \times \frac{dN(z)}{dt}.
\end{equation}

Here, $\rho_{\rm{GC}}$ is the volumetric number density of clusters, assuming a constant value of $\rho_{\rm{GC}}=2.31\,\rm{Mpc}^{-3}$ \citep[consistent with][]{Rodriguez2015a,RodriguezLoeb2018}, $f$ is a scaling factor intended to incorporate the contribution of the cluster mass function's high-end tail not covered by our models (see Section \ref{sec:BBHfits}), and $dN(z)/dt$ is the number of mergers per unit time at a given redshift.

We compute $dN(z)/dt$ using the following procedure: first, we generate a complete list of merger times ($t_{\rm{merger}}$) for all BBHs that merge within a Hubble time in our model set (roughly $10^4$ total mergers). For each of these mergers, we draw 10 random ages ($t_{\rm{age}}$) for the host cluster from which the merger originated. We then compute the effective merger time for each BBH merger as $t_{\rm{effective}} = t_{\rm{Hubble}} - t_{\rm{age}} + t_{\rm{merger}}$. As in \citet{Rodriguez2018}, we draw cluster ages from the metallicity-dependent age distributions of \citet{El-Badry2018}. We then compute the number of mergers per time, $dN(z)/dt$, by dividing this list of effective merger times into separate redshift bins. Note that we also scaled down these rates to correct for oversampling -- caused by drawing 10 cluster ages for each merger and by drawing mergers from a large set of cluster models. To correct for the former oversampling, we simply divide the rates by a factor of 10. To correct the latter, we divide by the total number of models sampled (weighting all models equally for simplicity).

We include the scaling factor $f$ in Equation \ref{eq:comoving_rate} to account for the contribution of the cluster mass function's high-mass tail not covered by our models. This factor is calculated as the ratio of the average number of mergers per cluster (computed by integrating over the cluster mass function; see Section \ref{sec:BBHfits} and Table \ref{table:BBHs}) to the average number of mergers per cluster counted from the models sampled. We compute $f$ separately for each $r_v$, as in Table \ref{table:BBHs}. In practice, $f \approx 4$ is typical across all $r_v$ values, so that high-mass clusters ($M \gtrsim 5\times10^5\,M_{\odot}$) contribute roughly four times more mergers than low-mass clusters ($M \lesssim 5\times10^5\,M_{\odot}$).

In Figure \ref{fig:BBHrates}, we show the cumulative rate ($R(z)$) and comoving rate ($\mathcal{R}(z)$) as functions of redshift for the four $r_v$ values considered in this study (blue curves). The combined rate (dashed black curves) is calculated assuming equal contribution from all four values of $r_v$.\footnote{This is likely an oversimplification, but a detailed study of the $r_v$ distribution corresponding to present-day GCs is beyond the scope of this study. See \citet{Choksi2018} for further discussion on this point. We hope to perform a more detailed study on cluster models incorporating more realistic $r_v$-weighting in a later paper.}

For the reasons discussed in Section \ref{sec:BBHfits}, we also see that the merger rate increases significantly as $r_v$ decreases. Thus, as shown in Figure \ref{fig:BBHrates}, if initial virial radii of $r_v=0.5\,$pc or smaller are typical for clusters in the universe, the BBH merger rate from clusters may be roughly twice as high as previous estimates \citep[e.g.,][]{RodriguezLoeb2018}.

Overall, the rates shown in Figure \ref{fig:BBHrates} for models with $r_v=1$ or 2 pc are roughly consistent with those of \citet{RodriguezLoeb2018}, which also used \texttt{CMC} models, but implemented a slightly different rate calculation accounting for metallicity's effect on the cluster mass function. The latest results from the LIGO/Virgo collaboration suggest a local-universe BBH merger rate of $53.2^{+58.5}_{-28.8}\,\rm{Gpc}^{-3}\rm{yr}^{-1}$ \citep{LIGO2018a,LIGO2018b}. Here, we estimate local universe rates ($z=0$) ranging from roughly $9\,\rm{Gpc}^{-3}\rm{yr}^{-1}$ to $30\,\rm{Gpc}^{-3}\rm{yr}^{-1}$, depending on $r_v$, and a combined rate of roughly $22\,\rm{Gpc}^{-3}\rm{yr}^{-1}$.

It is worth noting that some uncertainties are left unexplored here, potentially affecting the estimated rates. For instance, we have focused on only the contribution from clusters that have survived to the present day. However, as noted by \citep[e.g.,][]{Gnedin2014,FragioneAntoniniGnedin2018}, there likely existed a significant population of clusters which did not survive to the present. The remnants of these disrupted clusters may contribute significantly to the BBH merger rate \citep[e.g.,][]{RodriguezLoeb2018,Fragione2018b}. 
Furthermore, in this study we have considered only GCs, however previous analyses \citep[e.g.,][]{Ziosi2014,Banerjee2018,DiCarlo2019} have shown that lower-mass open clusters may also contribute to the BBH merger rate. In this sense, the rates based on this study's particular model set can be viewed as lower limits on the total contribution from clusters.

\subsection{Effect of $r_v$ on BBH masses}
\label{sec:BBH_properties}

\begin{figure}
\begin{center}
\includegraphics[width=0.85\columnwidth]{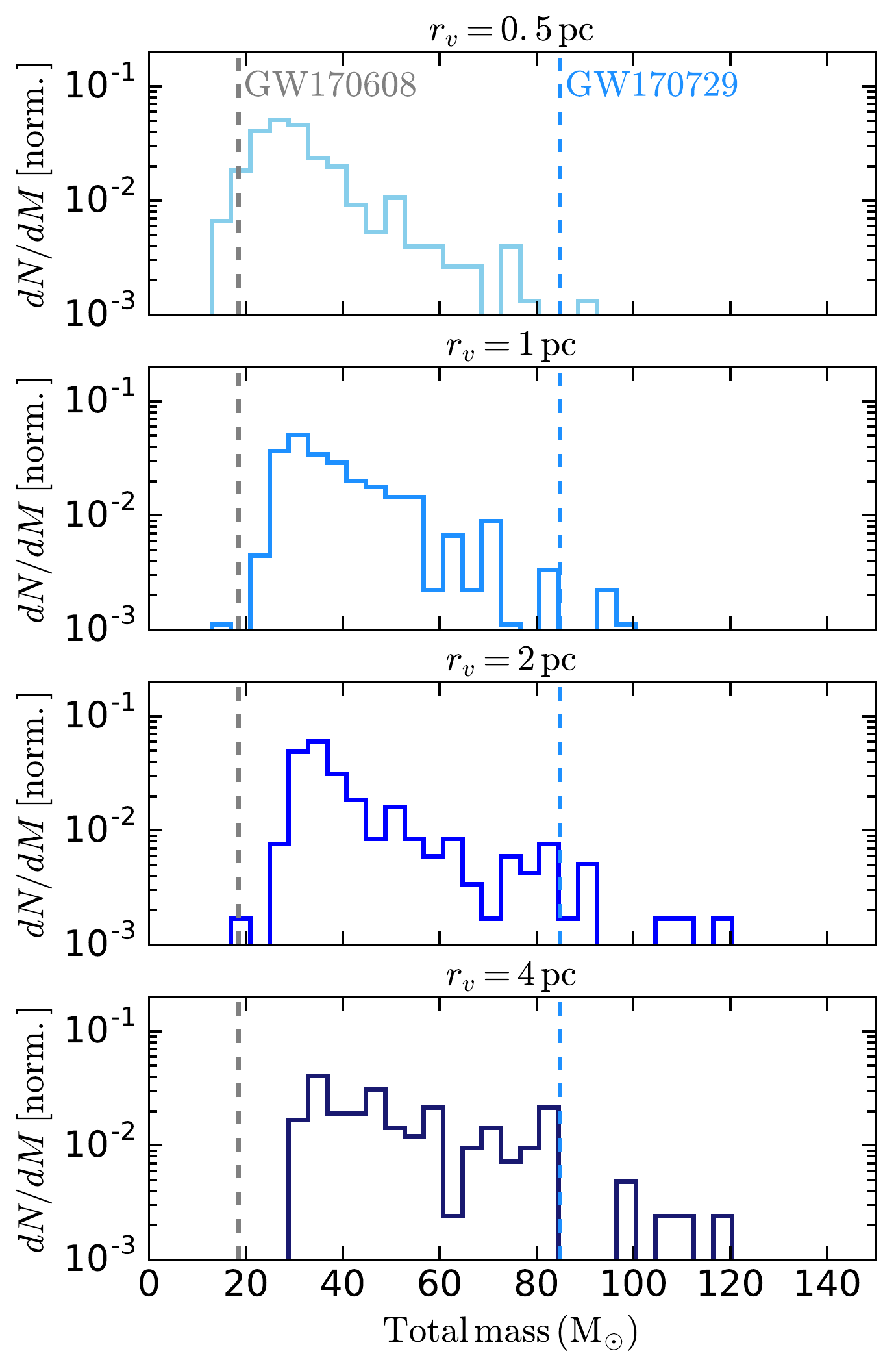}
\caption{\footnotesize \label{fig:BBH_total_mass} Total mass of all BBH mergers occurring at $z<0.5$ ($t_{\rm{mergers}}\gtrsim6\,$Gyr, assuming cluster ages of 12 Gyr. We exclude all solar metallicity models from plot and show only low-metallicity models that are more generally representative of old GCs. From top to bottom, we show models with increasing initital $r_v$. The vertical dashed lines show the total masses of the lowest and highest mass BBH mergers detected to-date by LIGO/Virgo: GW170608 (gray) and GW170729 (blue).}
\end{center}
\end{figure}

As a final point, we discuss the different expected properties of BBHs that originate from clusters of varying $r_v$. In Figure \ref{fig:BBH_total_mass}, we show the total mass of all BBH mergers in low-metallicity cluster models occurring at redshift $<0.5$ (representative of local-universe mergers, which are most relevant to potential detections by LIGO/Virgo). From top to bottom, we show the distributions for models with $r_v=0.5$--$4\,$pc. The BBH mass distribution shifts toward smaller masses as $r_v$ decreases. As shown in Figure \ref{fig:Nbh}, the depletion rate of BHs directly relates to $r_v$, such that clusters with smaller $r_v$ retain fewer BHs at late times. As discussed in, e.g., \citet{Morscher2015}, the most massive BHs in a cluster are generally among the first to be ejected from the cluster and among the first to merge. The most massive BHs sink furthest into the cluster's core, undergoing more frequent dynamical encounters that lead to both ejection and mergers. The lower mass BHs ($M \lesssim 15 M_{\odot}$) become dynamically active only after the most massive BHs have been ejected. In clusters with smaller initial $r_v$, high-mass BHs are dynamically processed and ejected relatively quickly. Therefore, in these clusters, high-mass BHs ($M \gtrsim 30\,M_{\odot}$) tend to merge relatively early (at high redshift).

By late times (low redshift), most of the high-mass BHs in low $r_v$ models have already been ejected or have merged with other BHs, leaving only the least massive BHs in any significant quantity. As a consquence, the mass distribution of BBH mergers shifts to lower masses in clusters with lower $r_v$. For clusters with high $r_v$, the initial relaxation time is longer, so many high-mass BHs still remain at late times. Hence, BBH mergers tend to have higher component masses in these models.

In Figure \ref{fig:BBH_total_mass}, we show the total masses (dashed vertical lines) for the least and most massive BBH mergers observed to date by LIGO/Virgo \citep{LIGO2018a}. By considering cluster models with varied $r_v$, we are able to span the full distribution of observed merger masses. Also note that high-metallicity clusters -- which preferentially form lower-mass BH populations due to the differences in wind mass loss at high metallicities -- provide an alternative way to dynamically form low-mass BBH mergers similar to GW170608. Indeed, \citet{Chatterjee2017b} noted that young high-metallicity clusters may be the \textit{only} way such BBHs can be formed dynamically in GCs. Here, we amend this earlier work and show that low-mass BBHs can also form in low-metallicity clusters, provided that the clusters have sufficiently short initial relaxation times.

\section{Conclusions and Discussion}
\label{sec:conclusions}

\subsection{Summary}

In this paper, we have introduced a set of 148 cluster simulations (computed using the Monte Carlo code \texttt{CMC}) that span wide ranges in initial cluster mass, size, metallicity, and Galactocentric distance. We showed that these models collectively cover nearly the complete range of parameters of the GCs observed in the MW. Specifically, by varying the clusters' initial virial radii (and therefore their initial relaxation time), we showed that our simulations reproduce both clusters that have undergone core-collapse by the present day and those that have not. The onset (or delay) of core-collapse is related to the evolution of a cluster's stellar-mass BH population. When a cluster retains a large fraction of its primordial BH population at late times, the energy generated via ``BH burning'' in the core is sufficient to delay core-collapse. Only when a sufficiently large fraction of primordial BHs have been ejected through dynamical interactions within the BH-dominated core will a cluster be able to reach a core-collapsed state. By examining models with initial star counts ranging from $N=2\times10^5$ to $N=3.2\times10^6$, we demonstrate that the process of BH burning and eventual collapse is relevant for clusters of all realistic present-day masses.

With this model set in hand, we explored the application of these models to the formation of various objects in GCs. We briefly summarize the main results below.

\begin{enumerate}
\item We showed that BH--luminous companion binaries (in both detached and mass-transferring configurations) form at rates consistent with the numbers of accreting and detached BH binaries presently observed in GCs. As shown in previous work, the number of BH binaries does not exhibit strong dependence upon cluster parameters. Meanwhile, the NS binary formation rate anticorrelates with the total number of BHs in the cluster. NS binaries are most likely to form dynamically in core-collapsed clusters with small BH populations.

\item We showed that the number of pulsars (and millisecond pulsars) in a cluster is expected to depend upon the BH population in a manner similar to NS binaries. In line with the previous results of \citet{Ye2018}, the number of pulsars anticorrelates with the number of BHs. We demonstrate here that, for clusters of similar total mass, those with smaller initial $r_v$ generally host more pulsars at late times compared to clusters with initially larger $r_v$.

\item We demonstrated that up to dozens of accreting WD binaries can form in typical GCs. These binaries may be observed as CVs or AM CVn, which are well observed in a number of GCs. Unlike NS binaries and millisecond pulsars, the numbers of accreting (and detached) WD binaries in clusters do not vary in an obvious manner with initial $r_v$ nor with the total numbers of BHs. Previous analyses have suggested dynamical interactions may actually lead to \textit{disruption} of CV-progenitors. Furthermore, WD binary formation is particularly sensitive to assumptions about binary evolution. 

\item We also explored the collision rates of WDs with other stellar remnants and discussed implications for potential high-energy transients. Because WD collisions are primarily driven dynamically (unlike CVs, for example), the number of WD collisions per cluster exhibits a clear dependence upon initial $r_v$ and BH number; core-collapsed clusters hosting few BHs at present are ideal candidates for WD collisions.

\item We discussed the number of luminous star collisions in our models. We showed that, at early times, massive MS stars and giants most frequently undergo collisions with low-mass ($M\sim0.1\,M_{\odot}$) M-dwarfs, simply because these low-mass stars dominate the assumed cluster initial mass function. This result may be sensitive to assumptions regarding the amount of primordial mass segregation in clusters. These various early-time stellar collisions may have important implications for BH formation, which will be explored in later work.

\item Additionally, we explored luminous star collisions that occur at late times ($t>8\,$Gyr) and discussed the implications for blue straggler star formation. We showed that clusters can contain up to 100 blue stragglers or more. Generally, the number of blue stragglers is anticorrelated with the number of stellar-mass BHs. This shows that the link between blue straggler populations and cluster dynamical age \citep[see, e.g.,][]{Ferraro2012} is connected to stellar-mass BH populations, not independent of, as asserted in previous work \citep{Ferraro2019}.

\item Finally, we examined the number of BBH mergers in our model set. We explored the total number of mergers that occur through four distinct dynamical channels and discussed how the relative rates from these channels may depend upon various cluster features. We computed cosmological BBH merger rates in our models and showed that if a large number of clusters form with small initial $r_v$ ($r_v \lesssim 1\,$pc), the BBH merger rate may be higher than previous estimates by a factor of a few. From all of our models together, we estimate a BBH merger rate of roughly $20\,\rm{Gyr}^{-3}\rm{yr}^{-1}$ in the local universe. 

\end{enumerate}

\subsection{Discussion and Future work}

There are of course several complexities pertaining to the evolution of GCs that are not captured in the present work. In this section, we describe several such complexities and discuss avenues for future work.

In the present version of \texttt{CMC}, we assume fixed circular orbits within the Galatic potential. This determines the influence of the Galactic tidal field on the cluster evolution. In reality, true GC orbits are not circular \citep[for a recent review, see][]{Baumgardt2019} and thus the mass loss of GCs is affected also by e.g., disk shocking and passages close to the Galactic center. The mass loss prescriptions in the simulations presented in this work are at least moderately affected in the absence of the effect of their orbits' eccentricities and inclinations with respect to the Galactic disk \citep[see, e.g.,][for a discussion of some of these effects]{BaumgardtMakino2003}. We intend to incorporate such effects in future work.

In this paper, we examined the evolution of clusters for three distinct metallicities ($0.01, 0.1,$ and $1\,Z_{\odot}$), which overall span the range in observed metallicities of the MW GCs (see Figure \ref{fig:Harris}). As discussed in Section \ref{sec:results}, the average mass, the maximum mass, and the nature of the mass distribution of the BHs in clusters govern the dynamical evolution of both the BH subsystem and the cluster as a whole. As illustrated in Figure \ref{fig:BH_vs_ZAMS}, the mass distribution of BHs vary with cluster metallicity. In particular, the BH distribution varies significantly as the metallicity approaches $0.1Z_{\odot}$ from $Z_{\odot}$ and, since (pulsational) pair instability supernovae are assumed, the retained BH mass distribution varies relatively little between $0.1Z_{\odot}$ and $0.01Z_{\odot}$ \citep[see also, e.g.,][]{Belczynski2016b,Banerjee2019}. Thus, an intermediate metallicity grid point (i.e., $0.5Z_{\odot}$, similar to the observed metallicity of clusters in the Large Magellanic Cloud), may yield interesting extensions of the results presented in this study. Indeed, such intermediate metallicity \texttt{CMC} models were computed in \citet{Chatterjee2017b} where it was shown that the metallicity can have a substantial effect upon the mass distribution of BBH mergers. In the future, we plan to expand the present set of \texttt{CMC} models to include finer grid points that will capture such details.

Throughout this work, we demonstrated that our set of cluster models match well the bulk features of the MW GC population (see Figure \ref{fig:Harris}). However, it is also worth asking if specific models from our set are able to effectively match individual observed clusters. Such an exercise was performed in \citet{Kremer2019a} using a much smaller set of \texttt{CMC} models (with generally identical physics). In that analysis, the \texttt{CMC} models were used to effectively match small number of MW clusters with similar total mass (NGC 3201, M22, M10, and NGC 6752). In a forthcoming study \citep{Rui2019}, we will perform a similar exercise for this complete set of cluster models and specifically demonstrate the techniques one may use to identify which model in a large set best fits any particular observed cluster. Furthermore, recent work by \citet{Weatherford2018} used observed measurements of mass segregation to predict the number of stellar-mass BHs retained in three MW GCs with known BH candidates (M10, M22, and 47 Tuc). A follow-up analysis \citep{Weatherford2019} implements the complete set of cluster models introduced here along with observed mass segregation measurements for 50 GCs in the ACS Survey for MW GCs \citep{Sarajedini2007}. This work further constrains the number of BHs retained in specific GCs.

We have touched briefly here upon a wide range of stellar sources that are observed in GCs, including blue stragglers, LMXBs, millisecond pulsars, and CVs. The goal of this analysis is to simply demonstrate that, to ``zero-th order,'' these various sources are formed in our models at rates roughly consistent with what is observed. However, more detailed analyses are necessary to explore a number of questions. For example, in the case of blue stragglers, how do various formation channels (e.g., stellar collisions versus binary mass-transfer) contribute to the overall blue straggler population observed in realistic clusters? How do these channels depend upon various initial conditions, such as $r_v$ and metallicity, as well as BH properties? Pertaining to millisecond pulsars, what are the binary companions expected for dynamically-formed pulsars and are the companions found in our models consistent with the array of companions (e.g., Helium WDs, redbacks, black widows) identified for observed pulsars in GCs? Furthermore, what is the role of tidal capture in pulsar formation? The list of more focused topics motivated by this study's results is extensive. Indeed, several of these more detailed analyses are already underway.

\acknowledgments
We thank the anonymous referee for many helpful comments and suggestions that have improved the manuscript. We also thank Scott Coughlin for assistance setting up the \texttt{CMC} web platform. This work was supported by NSF Grant AST-1716762 and through the computational resources and staff contributions provided for the {\tt Quest} high performance computing facility at Northwestern University. {\tt Quest} is jointly supported by the Office of the Provost, the Office for Research, and Northwestern University Information Technology.
This work also used computing resources at CIERA funded by NSF PHY-1726951.
K.K.\ acknowledges support by the National Science Foundation Graduate Research Fellowship Program under Grant No. DGE-1324585.
C.S.Y.\ acknowledges support from NSF Grant DGE-0948017.
S.C.\ acknowledges support from NASA through Chandra Award Number TM5-16004X issued by the Chandra X-ray Observatory Center
(operated by the Smithsonian Astrophysical Observatory for and on behalf of NASA under contract NAS8-03060). 
M.S.\ acknowledges funding from the European Union's Horizon 2020 research and innovation programme under the Marie-Sklodowska-Curie grant agreement No. 794393.
G.F.\ acknowledges support from a CIERA Fellowship at Northwestern University.

\bibliographystyle{aasjournal}
\bibliography{mybib}

\appendix

We include in the Appendix five tables containing more detailed information for each simulation. In Table \ref{table:models}, we list all initial cluster properties and various features at the end of the simulations. In Table \ref{table:XRBs_full}, we include information concerning BH and NS binaries at late times. In Table \ref{table:stellar_collisions}, we list the total number of stellar collisions occurring in each simulation. In Table \ref{table:WD-pulsar-BS-full}, we list the mean total number of WD binaries, pulsars, and blue stragglers in each simulation. In Table \ref{table:BBH_mergers}, we list the total number of binary BH mergers occurring in each simulation.

\setcounter{table}{0}
\renewcommand{\thetable}{A\arabic{table}}

\startlongtable
\begin{deluxetable*}{l|l||c|c|c|c||cccccccccc}
\centerwidetable
\tabletypesize{\scriptsize}
\tablewidth{0pt}
\tablecaption{Initial cluster parameters and various populations at end of simulation for all model GCs\label{table:models}}
\tablehead{
\colhead{} &
\colhead{Simulation} &
\colhead{$N$}&
\colhead{$r_v$}&
\colhead{$R_{\rm{gc}}$}&
\colhead{$Z$}&
\colhead{$M_{\mathrm{tot}}$}&
\colhead{$r_{c,\mathrm{theoretical}}$}&
\colhead{$r_h$}&
\colhead{$\sigma_v$}&
\colhead{$N_{\mathrm{MS}}$}&
\colhead{$N_{\mathrm{G}}$}&
\colhead{$N_{\mathrm{WD}}$}&
\colhead{$N_{\mathrm{NS}}$}&
\colhead{$N_{\mathrm{BH}}$}&
\colhead{$N_{\mathrm{final}}$}\\
\colhead{}&
\colhead{}&
\colhead{$\times10^5$}&
\colhead{pc}&
\colhead{kpc}&
\colhead{$Z_{\odot}$}&
\colhead{$\times10^5$ $M_\odot$}&
\colhead{pc}&
\colhead{pc}&
\colhead{km s$^{-1}$}&
\colhead{}&
\colhead{}&
\colhead{}&
\colhead{}&
\colhead{}&
\colhead{}
}
\startdata
1 & \textsc{n2-rv0.5-rg2-z0.01} & 2 & 0.5 & 2 & 0.01 & \multicolumn{9}{c}{$\cdots$ \textit{disrupted} $\cdots$} & 15719 \\
2 & \textsc{n4-rv0.5-rg2-z0.01} & 4 & 0.5 & 2 & 0.01 & \multicolumn{9}{c}{$\cdots$ \textit{disrupted} $\cdots$} & 15719 \\
3 & \textsc{n8-rv0.5-rg2-z0.01} & 8 & 0.5 & 2 & 0.01 & 1.10 & 0.31 & 2.04 & 9.32 & 221131 & 808 & 54524 & 593 & 0 & 265985 \\
4 & \textsc{n16-rv0.5-rg2-z0.01}$^\dagger$ & 16 & 0.5 & 2 & 0.01 & - & - & - & - & - & - & - & - & - & - \\
\hline
5 & \textsc{n2-rv0.5-rg2-z0.1} & 2 & 0.5 & 2 & 0.1 & \multicolumn{9}{c}{$\cdots$ \textit{disrupted} $\cdots$} & 15719 \\
6 & \textsc{n4-rv0.5-rg2-z0.1} & 4 & 0.5 & 2 & 0.1 & \multicolumn{9}{c}{$\cdots$ \textit{disrupted} $\cdots$} & 15719 \\
7 & \textsc{n8-rv0.5-rg2-z0.1} & 8 & 0.5 & 2 & 0.1 & 1.40 & 0.31 & 1.62 & 6.09 & 316698 & 1612 & 62257 & 248 & 1 & 365516 \\
8 & \textsc{n16-rv0.5-rg2-z0.1} & 16 & 0.5 & 2 & 0.1 & 3.70 & 0.29 & 1.13 & 9.92 & 949522 & 4145 & 153586 & 729 & 20 & 1062282 \\
\enddata
\tablecomments
{The final cluster mass $M_{\mathrm{tot}}$, core radius $r_c$, half-light radius $r_h$, central velocity dispersion $\sigma_v$, and number of main sequence stars $N_{\mathrm{MS}}$, giants $N_{\mathrm{G}}$, white dwarfs $N_{\mathrm{WD}}$, neutron stars $N_{\mathrm{NS}}$, and black holes $N_{\mathrm{BH}}$ for each model. Models marked with a dagger ($^\dagger$) indicates the model was stopped due to onset of collisional runaway (see Section \ref{sec:definitions}).}
\end{deluxetable*}

\startlongtable
\begin{deluxetable*}{l|l||c|ccc|ccc||c|ccc|ccc}
\centerwidetable
\tabletypesize{\scriptsize}
\tablewidth{0pt}
\tablecaption{Mean numbers of black hole and neutron star binaries in all models at late times. \label{table:XRBs_full}}
\tablehead{
    \colhead{} &
    \colhead{} &
    \colhead{} &
    \multicolumn{3}{c}{Detached} &
    \multicolumn{3}{c}{Accreting} &
    \colhead{} &
    \multicolumn{3}{c}{Detached} &
    \multicolumn{3}{c}{Accreting}\\
    \colhead{} &
    \colhead{Simulation} &
    \colhead{BH} &
    \colhead{BH-MS} &
    \colhead{BH-G} &
    \colhead{BH-WD} &
    \colhead{BH-MS} &
    \colhead{BH-G} &
    \colhead{BH-WD} &
    \colhead{NS} &
    \colhead{NS-MS} &
    \colhead{NS-G} &
    \colhead{NS-WD} &
    \colhead{NS-MS} &
    \colhead{NS-G} &
    \colhead{NS-WD}
}
\startdata
1 & \textsc{n2-rv0.5-rg2-z0.01} & $\cdots$ & \multicolumn{3}{c|}{$\cdots$ \textit{disrupted} $\cdots$} & \multicolumn{3}{c||}{$\cdots$} & $\cdots$ & \multicolumn{3}{c|}{$\cdots$ \textit{disrupted} $\cdots$} & \multicolumn{3}{c}{$\cdots$} \\2 & \textsc{n4-rv0.5-rg2-z0.01} & $\cdots$ & \multicolumn{3}{c|}{$\cdots$ \textit{disrupted} $\cdots$} & \multicolumn{3}{c||}{$\cdots$} & $\cdots$ & \multicolumn{3}{c|}{$\cdots$ \textit{disrupted} $\cdots$} & \multicolumn{3}{c}{$\cdots$} \\3 & \textsc{n8-rv0.5-rg2-z0.01} & 0 & 0.00 & 0.00 & 0.00 & 0.00 & 0.00 & 0.00 & 658 & 0.75 & 0.00 & 2.64 & 0.04 & 0.04 & 2.70\\
4 & \textsc{n16-rv0.5-rg2-z0.01}$^\dagger$ & - & - & - & - & - & - & - & - & - & - & - & - & - & - \\
\hline
5 & \textsc{n2-rv0.5-rg2-z0.1} & $\cdots$ & \multicolumn{3}{c|}{$\cdots$ \textit{disrupted} $\cdots$} & \multicolumn{3}{c||}{$\cdots$} & $\cdots$ & \multicolumn{3}{c|}{$\cdots$ \textit{disrupted} $\cdots$} & \multicolumn{3}{c}{$\cdots$} \\6 & \textsc{n4-rv0.5-rg2-z0.1} & $\cdots$ & \multicolumn{3}{c|}{$\cdots$ \textit{disrupted} $\cdots$} & \multicolumn{3}{c||}{$\cdots$} & $\cdots$ & \multicolumn{3}{c|}{$\cdots$ \textit{disrupted} $\cdots$} & \multicolumn{3}{c}{$\cdots$} \\7 & \textsc{n8-rv0.5-rg2-z0.1} & 0 & 0.00 & 0.00 & 0.00 & 0.00 & 0.00 & 0.00 & 262 & 1.09 & 0.00 & 1.31 & 0.01 & 0.04 & 3.04\\
8 & \textsc{n16-rv0.5-rg2-z0.1} & 50 & 2.15 & 0.00 & 0.75 & 1.21 & 0.02 & 0.00 & 732 & 2.98 & 0.00 & 0.79 & 0.00 & 0.23 & 6.65\\
\enddata
\tablecomments{Average number of BH and NS binaries of with various companion types for all simulations at late times ($10$--$13\,$Gyr). 
We distinguish here between binaries that are found in accreting and detached configurations. Accreting binaries can be viewed as low-mass X-ray binary candidates. See Section \ref{sec:XRBs} for further discussion.}
\end{deluxetable*}

\startlongtable 
\begin{deluxetable*}{l|l|ccccc|cccc|ccc}
\tabletypesize{\scriptsize}
\tablewidth{0pt}
\tablecaption{Total number of stellar collisions per model\label{table:stellar_collisions}}
\tablehead{
    \colhead{} &
	\colhead{Simulation} &
    \colhead{MS-MS}&
    \colhead{MS-G}&
    \colhead{MS-WD}&
    \colhead{MS-NS}&
    \colhead{MS-BH}&
    \colhead{G-G}&
    \colhead{G-WD}&
    \colhead{G-NS}&
    \colhead{G-BH}&
    \colhead{WD-WD}&
    \colhead{WD-NS}&
    \colhead{WD-BH}
}
\startdata
1 & \textsc{n2-rv0.5-rg2-z0.01} & 650 & 346 & 72 & 1 & 136 & 33 & 62 & 1 & 44 & 0 & 0 & 0 \\
2 & \textsc{n4-rv0.5-rg2-z0.01} & 1738 & 1309 & 299 & 35 & 625 & 134 & 267 & 24 & 48 & 4 & 0 & 1 \\
3 & \textsc{n8-rv0.5-rg2-z0.01} & 2133 & 796 & 1221 & 152 & 100 & 3 & 294 & 37 & 34 & 51 & 12 & 7 \\
4 & \textsc{n16-rv0.5-rg2-z0.01} & - & - & - & - & - & - & - & - & - & - & - & - \\
\hline
5 & \textsc{n2-rv0.5-rg2-z0.1} & 792 & 732 & 57 & 1 & 60 & 35 & 54 & 3 & 43 & 0 & 0 & 0 \\
6 & \textsc{n4-rv0.5-rg2-z0.1} & 1489 & 1207 & 553 & 28 & 87 & 28 & 253 & 14 & 39 & 19 & 4 & 0 \\
7 & \textsc{n8-rv0.5-rg2-z0.1} & 2628 & 3388 & 1138 & 93 & 131 & 35 & 350 & 18 & 81 & 27 & 7 & 1 \\
8 & \textsc{n16-rv0.5-rg2-z0.1} & 4170 & 12172 & 452 & 14 & 249 & 35 & 127 & 7 & 207 & 2 & 0 & 0 \\
\enddata
\tablecomments{Total number of stellar collisions of various combinations of stellar types occurring in each model.}
\end{deluxetable*}


\startlongtable 
\begin{deluxetable*}{l|l|c|ccc|ccc|cc|c}
\tabletypesize{\scriptsize}
\tablewidth{0pt}
\tablecaption{White dwarf binaries, pulsars, and blue stragglers \label{table:WD-pulsar-BS-full}}
\tablehead{
    \colhead{}&
    \colhead{Simulation}&
    \colhead{$N_{\rm{WD}}$}&
    \multicolumn{3}{c}{Detached WD binaries}&
    \multicolumn{3}{c}{Accreting WD binaries}&
    \colhead{$N_{\rm{pulsar}}$} &
    \colhead{$N_{\rm{MSP}}$}&
    \colhead{$N_{\rm{BS}}$}\\
    \colhead{}&
    \colhead{}&
    \colhead{}&
    \colhead{MS}&
    \colhead{G}&
    \colhead{WD}&
    \colhead{MS}&
    \colhead{G}&
    \colhead{WD}&
    \colhead{}&
    \colhead{}&
    \colhead{}
}
\startdata
1 & \textsc{n2-rv0.5-rg2-z0.01} & - & -& -& -& -& -& - & 0 & 0 & - \\
2 & \textsc{n4-rv0.5-rg2-z0.01} & - & -& -& -& -& -& - & 0 & 0 & - \\
3 & \textsc{n8-rv0.5-rg2-z0.01} & 57663 & 46.18 & 0.0 & 10.93 & 5.91 & 0.09 & 3.46 & 5.43 & 2.84 & 32.5 \\
4 & \textsc{n16-rv0.5-rg2-z0.01} & - & -& -& -& -& -& - & 0 & 0 & - \\
\hline
5 & \textsc{n2-rv0.5-rg2-z0.1} & - & -& -& -& -& -& - & 0 & 0 & - \\
6 & \textsc{n4-rv0.5-rg2-z0.1} & - & -& -& -& -& -& - & 0 & 0 & - \\
7 & \textsc{n8-rv0.5-rg2-z0.1} & 63257 & 79.83 & 0.0 & 11.0 & 14.48 & 0.31 & 9.96 & 6.96 & 3.75 & 79.0 \\
8 & \textsc{n16-rv0.5-rg2-z0.1} & 148248 & 184.31 & 0.0 & 19.0 & 15.85 & 1.94 & 15.48 & 9.14 & 0.36 & 114.0 \\
\enddata
\tablecomments{Average number of WD binaries (see Section \ref{sec:WDs}), pulsars (see Section \ref{sec:pulsars}), and blue stragglers (see Section \ref{sec:BSs}) for all late time snapshots for each model.}
\end{deluxetable*}

\startlongtable 
\begin{deluxetable*}{l|l|c|cccc|c}
\tabletypesize{\scriptsize}
\tablewidth{0pt}
\tablecaption{Binary BH mergers\label{table:BBH_mergers}}
\tablehead{
    \colhead{} &
	\colhead{Simulation} &
    \colhead{$N_{\rm{BH}}$ at 14 Gyr}&
    \multicolumn{3}{c}{In-cluster}&
    \colhead{Ejected}&
    \colhead{All Mergers}\\
    \colhead{} &
    \colhead{} &
    \colhead{} &
    \colhead{single--single} &
    \colhead{Fewbody} &
    \colhead{2-body} &
    \colhead{} &
    \colhead{}
}
\startdata
1 & \textsc{n2-rv0.5-rg2-z0.01} & - & 4 & 0 & 5 & 12 & 21 \\
2 & \textsc{n4-rv0.5-rg2-z0.01} & - & 9 & 1 & 16 & 42 & 68 \\
3 & \textsc{n8-rv0.5-rg2-z0.01} & 0 & 26 & 3 & 45 & 0 & 74 \\
4 & \textsc{n16-rv0.5-rg2-z0.01} & - & - & - & - & - & -  \\
\hline
5 & \textsc{n2-rv0.5-rg2-z0.1} & - & 7 & 0 & 2 & 9 & 18 \\
6 & \textsc{n4-rv0.5-rg2-z0.1} & - & 5 & 0 & 11 & 28 & 44 \\
7 & \textsc{n8-rv0.5-rg2-z0.1} & 0 & 16 & 2 & 44 & 72 & 134 \\
8 & \textsc{n16-rv0.5-rg2-z0.1} & 53 & 46 & 13 & 151 & 163 & 373 \\
\enddata
\tablecomments{Column 3 shows the total number of BHs retained at end of each simulation (models marked with a ``-'' disrupted before 14 Gyr). Columns 4--7 show the total number of BBH mergers occuring through single--single capture, fewbody mergers, 2-body mergers, and ejected mergers, respectively. See Section \ref{sec:merger_channels} for descriptions of these various channels. Column 8 shows the total number of BBH mergers per model.}
\end{deluxetable*}
\listofchanges

.

\end{document}